\documentclass[final, aps, pra, twocolumn, superscriptaddress]{revtex4}

\usepackage{doi} 
\usepackage{graphicx}
\usepackage[color]{showkeys}
\usepackage{ifdraft}
\usepackage[usenames,dvipsnames]{xcolor} 
\usepackage{amsmath}

\usepackage{bm}
\renewcommand {\vec} [1] {{\bm #1}}

\usepackage{listings}
\usepackage[]{changes}

\ifdraft{
\usepackage[paperwidth=210mm,paperheight=297mm,centering,hmargin=2.5cm,vmargin=2cm, marginparwidth=2cm]{geometry}
}{}

\usepackage[footnote=true]{snotez} 
\definechangesauthor[name={Alfredo A. Correa}, color=blue]{AC}
\definechangesauthor[name={Xavier Andrade}, color=Maroon]{XA}

\usepackage{xstring}
\usepackage{xparse}
\usepackage{xspace}

\newcommand{\defineabbrevchanges}[1]{%
\expandafter\NewDocumentCommand\csname #1\endcsname{sogd<>}{%
\IfBooleanTF##1{%
\IfNoValueTF{##2}{\IfNoValueTF{##3}{\IfNoValueF{##4}{\deleted[id=#1]{##4}}}{\IfNoValueTF{##4}{\added[id=#1]{##3}}{\replaced[id=#1]{##3}{##4}}}}{\IfNoValueTF{##3}{\IfNoValueTF{##4}{\added[id=#1,comment={##2}]{}}{\deleted[id=#1, comment={##2}]{##4}}}{\IfNoValueTF{##4}{\added[id=#1,comment={##2}]{##3}}{\replaced[id=#1,comment={##2}]{##3}{##4}}}}%
}{%
\IfNoValueTF{##2}{\IfNoValueTF{##3}{\IfNoValueF{##4}{\deleted[id=#1]{##4}}}{\IfNoValueTF{##4}{\added[id=#1]{##3}}{%
\replaced[id=#1]{##3}{##4}%
}}}{\IfNoValueTF{##3}{\IfNoValueTF{##4}{\added[id=#1,comment={##2}]{}}{\deleted[id=#1,comment={##2}]{##4}}}{\IfNoValueTF{##4}{\added[id=#1,comment={##2}]{##3}}{\replaced[id=#1,comment={##2}]{##3}{##4}}}}%
}%
}%
}%
\defineabbrevchanges{AC}
\defineabbrevchanges{XA}

\def\inq{\textsc{inq}\xspace}
\def\Inq{\textsc{Inq}\xspace}

\usepackage{ifthen}
\newcommand{\CC}[1][]{$\text{C\hspace{-.25ex}}^{_{_{_{++}}}}
                      \ifthenelse{\equal{#1}{}}{}{\text{\hspace{-.625ex}#1}}$}

\begin{document}
\title{INQ, a modern GPU-accelerated computational framework for (time-dependent) density functional theory}

\author{Xavier Andrade}
\email{xavier@llnl.gov}
\affiliation{Quantum Simulations Group, Lawrence Livermore National Laboratory, Livermore, California 94551, USA}

\author{Chaitanya Das Pemmaraju}
\affiliation{Stanford Institute for Materials and Energy Sciences, SLAC National Accelerator Laboratory, Menlo Park, CA, 94025, USA}

\author{Alexey Kartsev}
\affiliation{Stanford Institute for Materials and Energy Sciences, SLAC National Accelerator Laboratory, Menlo Park, CA, 94025, USA}

\author{Jun Xiao}
\affiliation{Stanford Institute for Materials and Energy Sciences, SLAC National Accelerator Laboratory, Menlo Park, CA, 94025, USA}

\author{Aaron Lindenberg}
\affiliation{Stanford Institute for Materials and Energy Sciences, SLAC National Accelerator Laboratory, Menlo Park, CA, 94025, USA}

\author{Sangeeta Rajpurohit}
\affiliation{The Molecular Foundry, Lawrence Berkeley National Laboratory, Berkeley, California 94720, USA}

\author{Liang Z. Tan}
\affiliation{The Molecular Foundry, Lawrence Berkeley National Laboratory, Berkeley, California 94720, USA}

\author{Tadashi Ogitsu}
\affiliation{Quantum Simulations Group, Lawrence Livermore National Laboratory, Livermore, California 94551, USA}

\author{Alfredo A. Correa}
\affiliation{Quantum Simulations Group, Lawrence Livermore National Laboratory, Livermore, California 94551, USA}

\begin{abstract}
We present \inq, a new implementation of density functional theory (DFT) and time-dependent DFT (TDDFT) written from scratch to work on graphical processing units (GPUs).
Besides GPU support, \inq makes use of modern code design features and takes advantage of newly available hardware.
By designing the code around algorithms, rather than against specific implementations and numerical libraries, we aim to provide a concise and modular code.
The result is a fairly complete DFT/TDDFT implementation in roughly 12,000 lines of open-source \CC code representing a modular platform for community-driven application development on emerging high-performance computing architectures.
\end{abstract}


\maketitle 



\section{Introduction}

The density functional theory (DFT) framework~\cite{Hohenberg_1964,Kohn_1965} provides a computationally tractable way to approximate and solve the quantum many-body problem for electrons~\cite{Kohn_1999}, both in the ground-state and in the excited-state~\cite{Runge_1984}.
In the last decades, DFT has been extremely successful, to the point of becoming the standard method for first principles simulations in computational chemistry, solid state physics and material science~\cite{Hafner2006,Burke2012,Mardirossian2017,Hasnip2014}.
This success has been possible in large part due to computer programs than can solve the DFT equations efficiently, accurately and reliably~\cite{Hehre1969,Chelikowsky1994,Briggs1996,Kresse1996,Soler2002,Castro2006,Gygi2008,Blum2009,Enkovaara2010,Shao2015,Genova2017,Draeger2017,Giannozzi2017,Noda2019,Apra2020,Kuhne2020,Gonze2020,Seritan2021}.

However, there are still many challenges that DFT faces in terms of theory and computer codes to be applicable to new kinds of problems~\cite{Verma2020}.
One of these challenges is the application of data science to electronic structure, for example through high-throughput material screening~\cite{Curtarolo2013, Oba2018,Marzari2021}, structure discovery~\cite{Oganov2006} or machine learning techniques~\cite{Lemonick2018,Schmidt2019,Tkatchenko2020,Li2021,Shaidu2021}, where thousands or even millions~\cite{Hachmann2014} of DFT calculations are performed.
For this, it is important to manage input parameters, output results, executions and error exceptions of simulation codes as simply as possible.

Another issue is that the standard approximations for exchange and correlation (XC) functionals fail for many types of systems~\cite{Cohen2008,Gori2010,Jain2016}.
This is particularly important in time-dependent DFT (TDDFT)~\cite{Runge_1984,Ullrich2014} where the usual adiabatic approximation for the XC functional misses part of the physics involved in the description of excited states~\cite{Onida2002,Castro2009,Elliott2011,Ullrich2014,Lian2018}.
For example, the adiabatic and semi-local functionals in TDDFT cannot describe excitons~\cite{Onida2002},
which are important to simulate from first principles as they control the optical gaps, energy relaxation and transport pathways of semiconductors, molecular, and nanostructured systems.  

Recently it has been shown that it is possible to describe excitons in TDDFT using hybrid functionals and better XC approximations~\cite{Izmaylov2008,Ullrich2016,Refaely-Abramson2015a,Kummel2017,Pemmaraju2018b,Sun2021}.
As expected, these improvements come with considerable additional computational costs.

Fortunately, computing power has increased at an rapid pace in the last few years, with exascale-level supercomputers coming in the next few years~\cite{Kothe2019}.
To use this new hardware to simulate new physics in more realistic systems, we require new software that can run efficiently on these modern computing platforms.

The superscalar central processing units (CPUs) that have dominated high-performance computing for a long time have been largely replaced by graphics processing units (GPUs).
GPUs have a much more parallel architecture that offers a considerably larger numerical throughput and memory bandwidth than a CPU with similar cost and power consumption.
Unfortunately, GPUs cannot directly run codes written for the CPU as they need code that is written in parallel and the GPU memory needs to be managed explicitly.

This means that existing DFT codes, many started more than two decades ago, need to be adapted to this new paradigm.
This is not an easy task as the solution of the DFT/TDDFT equations is quite complex.
It involves a combination of algorithms and computational kernels that need to work efficiently together.
Many of the DFT software packages contain hundreds of thousands of lines of code.
Due to these factors, the adoption of GPUs in DFT simulations has been slow when compared with other areas like classical molecular dynamics~\cite{Stone2007,Biagini2019}.
Right now only a few DFT codes offer production-ready GPU support, and with limited performance gains for specific cases~\cite{Genovese2009,Andrade2013,Hacene2012,Romero2017,Huhn2020}.

When faced with the necessity of an efficient DFT implementation that can make use of large GPU-based supercomputers, we decided that it was more practical to start an implementation from scratch, instead of modifying an existing code.
We took this opportunity to create a modern code that profits from current design ideas and programming tools.
At the same time, it is based on years of experience that our team has in the development of the codes \textsc{octopus}~\cite{Castro2006,Andrade2010thesis,Andrade2012,Andrade2013,Andrade2015,Tancogne2020} and \textsc{qball}~\cite{Schleife2012,Schleife2015,Draeger2017} (the TDDFT implementation branch from \textsc{Qbox}~\cite{Gygi2005,Gygi2008}).
This allowed us to write \inq, a streamlined implementation of DFT and TDDFT that is very compact, only 12,000 lines of code, and that is easy to maintain and further develop.

This article gives a complete overview of the design and implementation of \inq.
We show how modern approaches to coding can be applied to DFT problems with advantageous results.
We then present results of some calculations performed with \inq and how they compare with other DFT codes in terms of accuracy.
The numerical performance of the code will be the subject of a future publication.

The current version of \inq uses innovative coding practices to implement conventional DFT algorithms, including both plane-wave and real-space strategies that have been used in other DFT codes.
This approach will allow newer algorithmic improvements to be incorporated into \inq in a systematic and fail-safe way, building upon a solid foundation of tried-and-tested methods.

\section{Code development}

\begin{figure}
	\includegraphics[width=\columnwidth]{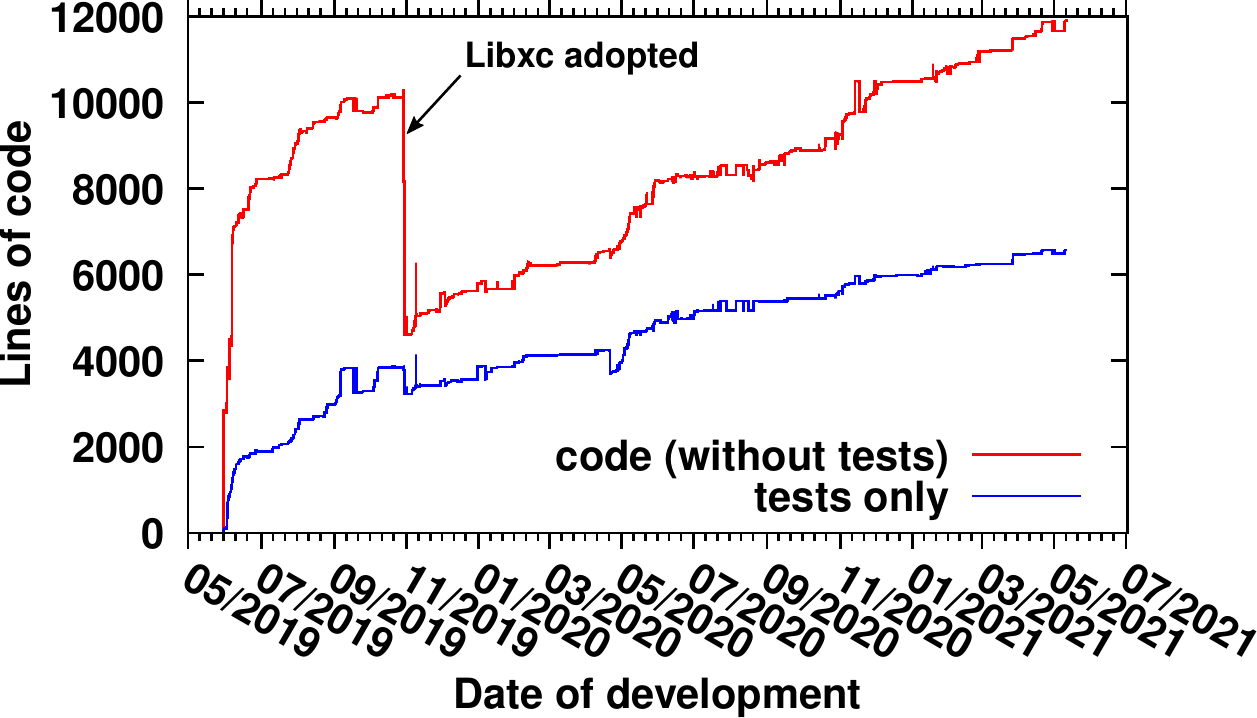}
	\caption{
		\label{fig:lines}
		\Inq source-code lines \emph{vs.} time since the start of the project. 
		Note that test and code are developed side by side, and the amount of lines of code is comparable.
		We find testing fundamental for continuous development and maintenance of the code.
		The dip near 11/2019 reflects the removal of internal exchange and corrections (XC) functional computation, which was replaced by the external library \textsc{libxc}.
		This illustrates the importance of delegating functionality to other high quality libraries when possible.
	}
\end{figure}

We start the article by discussing our general development strategy for \inq.
This is one of the most fundamental aspects of  code development  that can make the difference between a successful piece of software or an abandoned project.

A scientific program is not a static entity that is written once and used for years without modification.
Besides the usual problems in the code that are regularly discovered and need to be fixed, modification is essential in software.
More precise or more computationally efficient algorithms and theories appear and need to be tested or implemented.
And finally, computational platforms change and codes need to be adapted for them.

This is why our main objective when designing and developing \inq is that it can be easily modified and improved, while providing consistent results and behavior.
At the same time, it should offer all the features expected of a modern electronic structure code and the highest performance possible.
We combine several elements to achieve our objective, that we briefly discuss next and that are expanded in the following sections.

First, \inq does not follow the traditional code interface based on input files.
Instead, \inq is a library where the user input is a computer program.
We explain our approach in section~\ref{sec:interface}.

Second, \inq has a highly modular structure where the code is divided into several components with well defined tasks.
In this way, the different components  can be developed and tested independently.
They can also be shared with other codes, to avoid duplication of work and enable collaborative development.
In fact, we use third party components when they are available, and when possible contribute to their development (as in the case of \textsc{libxc}).
This is discussed in detail in section~\ref{sec:structure}.

A third aspect is the use of modern programming in \CC, explained in section~\ref{sec:c++}, that allows for a high level of abstraction while retaining the efficiency of a low level language.
The result is a code that is simple to write and read, and that resembles as much as possible the underlying mathematics.
Details like memory management and parallelization are hidden as much as possible and most programmers do not need to worry about them.
A particularly advantageous application of modern \CC is to facilitate GPU programming.
We developed a simple and general model to write GPU code that we describe in section~\ref{sec:gpu}.

An important aspect of the code design of \inq is that we recognize that it is very hard to determine what is the best code design \emph{a priori}, so we do not attempt to do it.
When implementing something our priority is to have the simplest implementation that works, write tests, and only then figure out what it is the best way of coding it.
And even then we are always willing to change it, since the ``optimal'' solution might change over time depending on other factors.

Finally, for a scientific code the reliability of the results, and their consistency after code modification is essential.
With this idea in mind, we rely heavily on a systematic, exhaustive, and continuous testing of the code.
Tests are written at the same time, or even before, a new component or feature is developed~\cite{Martin2009}.
They ensure that the implementation always gives the expected results.
Once the initial results are validated, the tests verify that the results do not change when the code is modified or when it is run on a different platform.
In particular, they check that the CPU and GPU versions of the code give the same results.

We use two types of tests: unit tests and integration tests.
Unit tests are small tests written for every individual component of the code, a class or function.
They verify the components are giving proper results under all conditions.
The reference values for these tests are normally obtained from analytical results for known conditions, including corner cases.
Integration tests in \inq are calculations for particular systems whose results are verified analytically or with other codes.
These examples are designed to span all the possible types of calculations and features.

All the development of \inq is tracked through a source control system.
For this we use \textsc{git}~\cite{Loeliger2012} and keep our central repository on the \textsc{GitLab} service.
As a set of changes is made to the code, all the tests are automatically executed by a `continuous integration' (CI) system provided by \textsc{GitLab}.
We configured this system to compile and run the code in a variety of platforms (CPU/GPU, serial/parallel, and different compilers and libraries).
A failure in compilation, or any of the tests, blocks the changes from being included into the code until they are fixed.

The CI also uses the \textsc{codecov} tool to evaluate how well the tests are checking all the lines of code.
The code coverage in \inq is approximately 95\%, which means that almost every line of code is executed during testing.

In Fig.~\ref{fig:lines} we show the number of lines of code in \inq as a function of time, since the start of the project.
It illustrates several things.
In the first place, it shows that \inq offers an implementation of DFT and TDDFT in roughly 12,000 lines of code (not considering tests).
This is quite an achievement considering that a code that offers similar functionality like \textsc{qball}, has more than 60,000 lines.
Other electronic structure codes can reach much higher counts, in the hundreds of thousands of lines of code.
The figure also shows how the tests in \inq are developed concurrently with the code, and that constitute a significant part of the total amount of the code.
While this might seem a waste of effort, in fact it makes development easier and faster by reducing the time spent on code debugging.
The tests make errors much less likely, and when errors do appear they are much easier to detect and fix.

\section{The \inq user interface}\label{sec:interface}

\begin{lstlisting}[language={[11]C++},float=*, numbers=left, showstringspaces=false, captionpos=b, label={lst:input_file}, caption={
	Example of an \inq program (equivalent to an input file) for the DFT calculation of the nitrogen molecule ground state.
	The input is a regular \CC source code file that is compiled normally.
	Up to line 9, we initialize the code to use \inq.
	In line 12 we define the interatomic distance for our molecule, note that the user needs to explicitly give the units, in this case \r{A}ngstrom.
	Lines 14 to 16 create the geometry of the molecule inside a standard \CC dynamic array (\texttt{std::vector}).
	In line 18, the ionic configuration geometry and the cell are brought together in an ionic subsystem.
	In line 19, an electronic subsystem (KS electrons) is created and allocated across all \textsc{MPI} tasks available.
  Line 20 initializes the electronic plane-wave coefficients with a guess for the ground-state calculation.
	In line 23, a convenience function called `calculate', optimizes the KS system to the ground state for fixed ion configuration.
	Finally, the calculated total energy is printed to the screen.
}]
#include <inq/inq.hpp>
  
int main(int argc, char * argv[]){

  using namespace inq;
  using namespace inq::input;  
  using namespace inq::magnitude;
  
  inq::environment env(argc, argv);
  auto comm = mpi3::environment::get_world_instance();

  auto distance = 1.06_angstrom;

  std::vector<atom> geo(2);
  geo[0] = "N" | coord(0.0, 0.0, -distance/2);
  geo[1] = "N" | coord(0.0, 0.0,  distance/2);
  
  system::ions ions(cell::cubic(10.0_bohr) | cell::finite(), geo);
  system::electrons electrons(comm, ions, basis::cutoff_energy(80.0_Ry));

  ground_state::initialize(ions, electrons);

  auto result = ground_state::calculate(ions, electrons, interaction::pbe());

  std::cout <<"N2 energy = "<< result.energy.total() <<" Hartree\n";
}
\end{lstlisting}

\begin{figure}
	\includegraphics[width=\columnwidth]{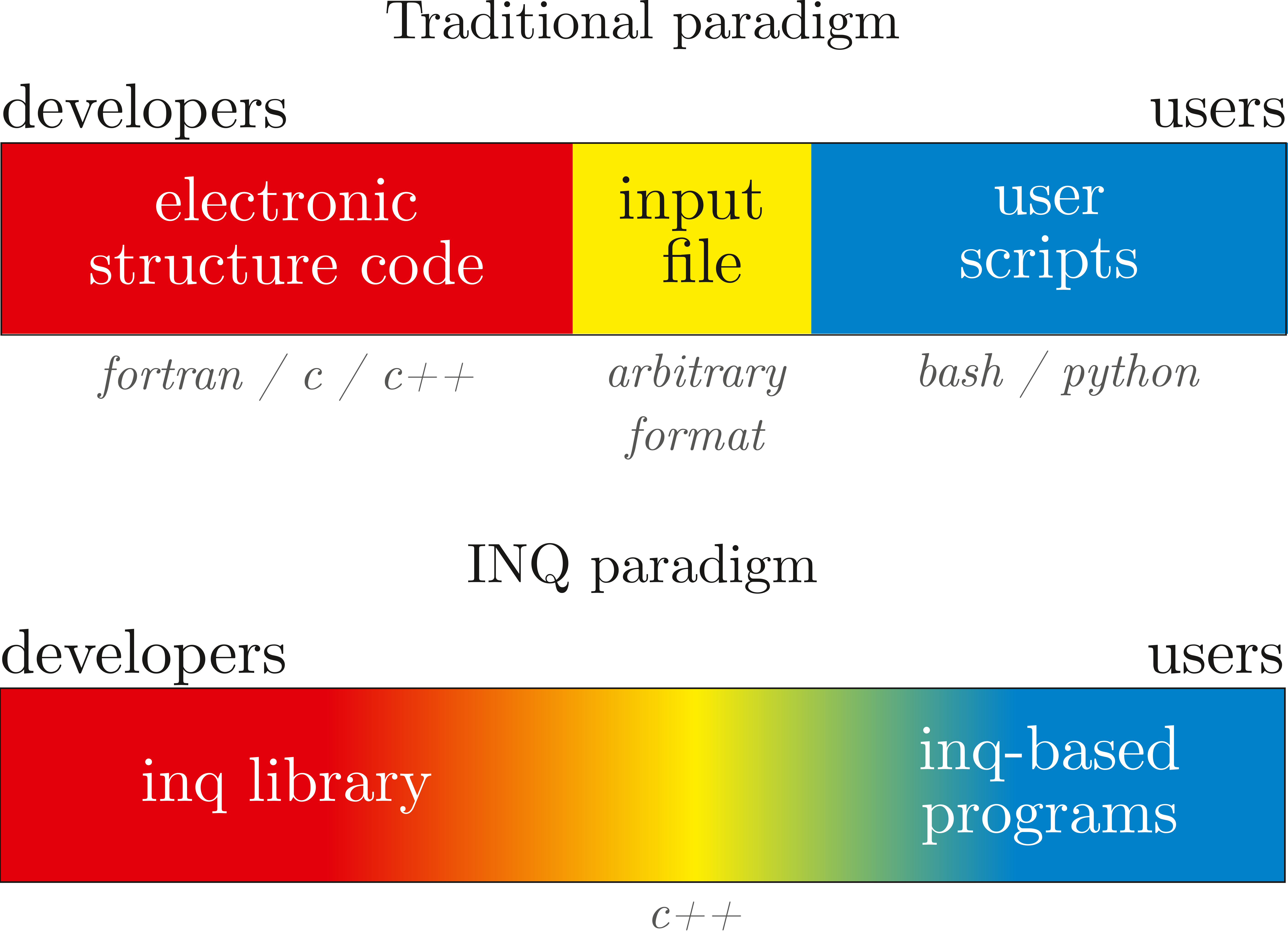}
	\caption{
		Schematically we develop \inq with the idea of blurring the distinction between input and main program. 
		We offer all the available opportunities to drive simulations in the same framework that the low level functionality is implemented in.
		This paradigm obviates the need the for input files, output files for postprocessing and the potential need to deal with three different languages.
		A coherent organization of the information of a simulation is accessible in memory at any level of the framework.
    Simple or complex programs can be written in the same language than the rest of the library;
    there is no need to invent an ad-hoc input format or language to specify the desired simulation.
		Simple simulation tasks (e.g. plain ground-state calculation, time-dependent propagation) corresponds to simple programs that use the library.
		Complex uses of the system may need to access advanced levels of functionality, further development of the library or interaction with other third party libraries (e.g. other molecular codes, machine learning libraries).
	}
	\label{fig:paradigm}
\end{figure}

\Inq follows a modular philosophy: the code is split into different components with well defined tasks.
Following that idea, \inq itself is designed as a component that can be used by other programs.
This means that in practice, \inq is not a program that is executed by users but a \CC library that provides all the functionality of a DFT/TDDFT code.
Instead of using an ad-hoc input-file format, \inq input files are directly written in standard \CC and compiled, possibly taking advantage of information specific to the desired calculation.
A sample input file is shown in Listing~\ref{lst:input_file}.
While this may require a bit of effort from users, this approach has several key advantages in comparison to traditional codes.
A similar approach is used by the \textsc{gpaw} code that uses Python scripts as input~\cite{Enkovaara2010} and by the Atomic Simulation Environment ~\cite{HjorthLarsen2017} in an attempt to control uniformly different existing codes with a single Python interface.

In the past, quantum simulations were quite expensive and researchers could only afford to make a few simulations in a research project.
Today, fast computers, efficient and reliable codes, and data-analysis techniques have made it feasible for a single user to perform large numbers of DFT/TDDFT simulations.
In particular, this has lead to the concept of high-throughput computational materials screening.

The library approach of \inq offers a considerable advantage in this scenario.
In the first place, users need not learn an additional syntax for writing input files.
With \inq, since the input file is already a program in a general programming language, all the automation can be done directly and the code can be easily integrated with other libraries.
For example, we have written a \CC program that can connect to the materials project~\cite{Jain2013} to download a structure and use it as an input for \inq.

A particularly complicated part of writing scripts that call a third party code is the parsing and post-processing of results from the output files.
In \inq instead, users can directly access the results as data structures instead of having to write parsing routines for output files.
Any post-processing of the data can be also done directly and make use of the functionality already provided by \inq.

\begin{figure}
	\includegraphics[width=\columnwidth]{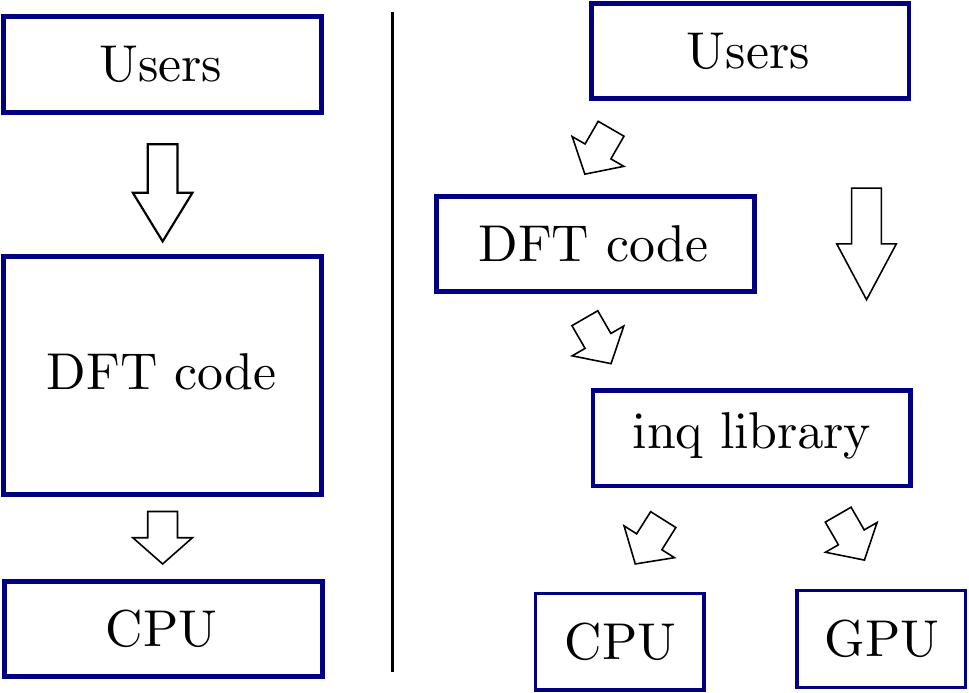}
	\caption{
		Different paradigms of numerical simulation in relation with the usage and development.
		On the left, a traditional paradigm: The users interact with a monolithic code.
		The code defines its own interface and uses underlying hardware.
		Even if some users become developers themselves, the task of developing and the task of usage are clearly separated.
		Each new development has to be reflected in the interface (e.g. input format) for the users.
		This has the advantage that the monolithic code have stability and give well defined access to functionality, at the cost of development turnaround and efficiency, e.g. postprocessing and interaction with other codes is always a separate task.
		On the right, the proposed paradigm: The system is prepared in such a way that users have access to low, intermediate, or high level functionality.
		The resulting library offers a trade off between level of access and level of difficulty (knowledge on the part of the user).
		The library hides for the most part specialized interaction with the hardware (e.g. CPU or GPU), but no more.
		Simple or complex programs can be written in the same language than the rest of the library.
		Optionally a more traditional `DFT code' can be adapted on top of the \inq library, allowing the DFT code to use the high-performance components and GPU support of \inq while retaining the traditional interface and capabilities.
	}
	\label{fig:user_flow}
\end{figure}

An additional advantage is that users do not have to face a barrier when they need to modify or extend \inq.
Since they are already using \CC, it is natural to explore the code and its lower-level interfaces to tailor \inq for their specific needs.
For example, if a user needs to implement a new observable, they can directly access the data structures and use \inq operations over them.

The difference between the usual paradigms of calculations and \inq is illustrated in Fig.~\ref{fig:paradigm}.
\Inq blurs the boundaries between the electronic structure code, the input file and the calculation processing scripts, and merges them all into a continuum with different levels of abstraction and detailed access.
This continuum gives more power to the users, and developers of other programs, to adapt the code to their specific needs.

Another purpose of \inq is to provide highly-efficient GPU accelerated routines that can be used by other DFT codes that do not support GPUs natively.
As shown in Fig.~\ref{fig:user_flow}, in this modality users can run \inq through another code that acts as a front end, instead of running it directly.
The advantage would be that higher level functionality can be directly implemented on top of \inq, and that users can use \inq through a familiar interface.

All the source code for \inq is freely accessible online from our \textsc{GitLab} website \url{https://gitlab.com/NPNEQ/inq/}.
The code is released under the Lesser General Public License version 3 (LGPLv3).
This is an open source license that allows anyone access to the source code of \inq and works that derive from it.
We think that this is important to ensure the reproducibility and transparency of the results, and to ensure open access to science~\cite{Ince2012,Smart2018}.
Note, however, that this specific license allows other codes to link with \inq as a library independently of their license.

As developers, we are aware that the installation of a scientific code can become cumbersome.
To make it easier for users to compile \inq we aim to provide an easy to use build system, that works out of the box in most cases.
The build system for \inq is based on \textsc{CMake}, but we use a wrapper script to offer the familiar \texttt{configure} script.
We limit the number of library dependencies to the standard ones for scientific codes, and include in the software package libraries that would be difficult for the users to obtain and compile.

\section{The modular structure of \inq} \label{sec:structure}

\begin{figure*}
	\includegraphics[width=\textwidth]{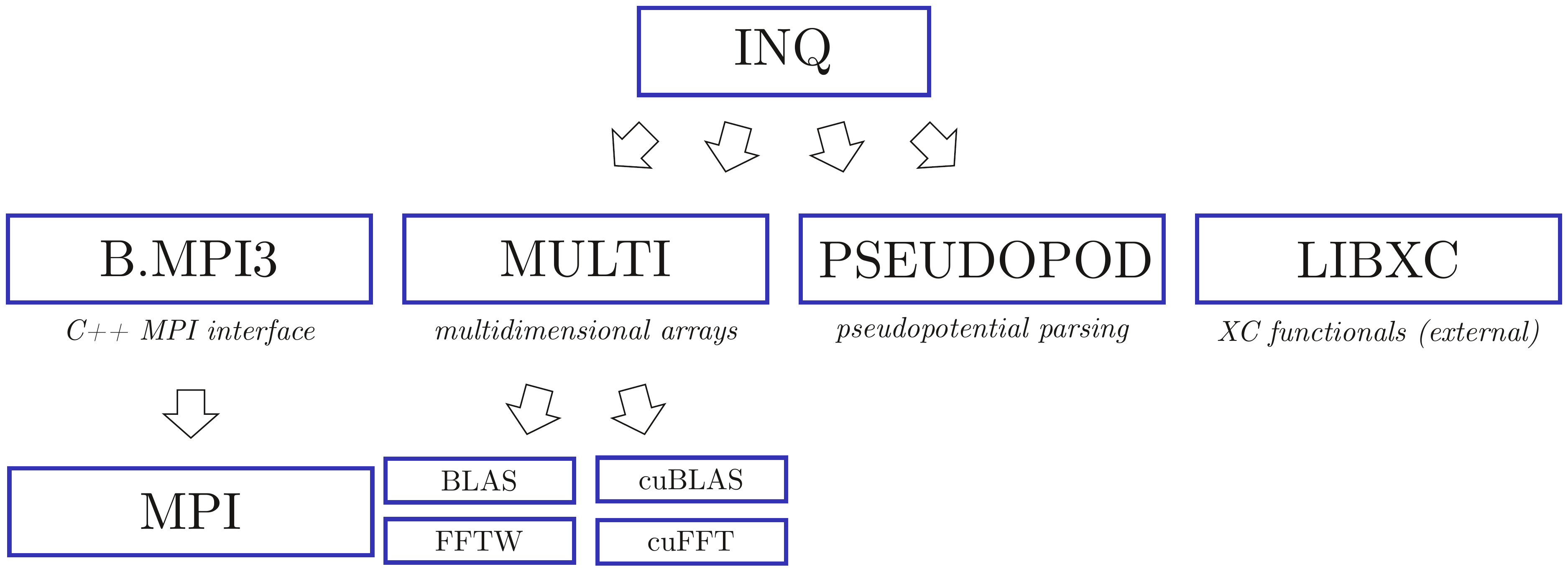}
	\caption{
		Simplified architecture of the \inq code and the main library dependencies.
		The first layer of libraries correspond to facilities directly related to the implementation of DFT and TDDFT.
		The facilities are understandable for a professional researcher in the theory of electronic structure:
		access to different exchange and correlation functionals and pseudopotential information.
		Various array representation of states, and spatial discretizations, high-level linear algebra operation, like orthogonalization, application of Hamiltonian and representation of potential fields and the communication patterns.
		At the lowest level we have very specialized libraries, usually vendor specific that can be switched depending on the platform.
	}
	\label{fig:structure}
\end{figure*}

The implementation of \inq is based on a modular design.
We have identified from the code some components that perform different tasks and turned them into independent components.
It is remarkable how modern coding techniques can make several \emph{separate} libraries work together with absolutely no code interdependence.
This is achieved by writing code against `concepts' or interfaces rather than leaning on specific implementations~\cite{Stepanov2009}.

This movement towards libraries rather than monolithic codes is slowly picking up in the community.
For example the ESL (CECAM Electronic Structure Library)~\cite{Oliveira2020} is a collection of modules for electronic structure calculations written in several languages to help reduce duplicate effort.

The main components of \inq are shown in Fig.~\ref{fig:structure}.
In the following subsections we briefly describe these components.
In the future we plan to integrate other components, like \textsc{spglib}~\cite{Togo2018} to handle symmetries and \textit{k}-point generation.
We also plan to split other parts of the code into independent libraries once they have a mature enough interface.

\subsection{\textsc{Multi}: multi-dimensional arrays} \label{sec:multi} 

An essential component needed to implement DFT in a simple and elegant way are multidimensional arrays.
The type of basis which define regular grids in real and reciprocal space, and the utilization of massive GPU parallelism makes regular contiguous arrays the ideal data structure to perform numeric operations.
Historically, \CC doesn't provide any particular kind of multidimensional arrays, leaving to the user the option to implement or use different libraries that emulate multidimensional indexed access.
We found no pre-existing library was particularly suited for use in the GPU and with a sufficiently modern design that matched our needs and therefore choose to implement it ourselves.

\textsc{Multi} provides multidimensional array access and description of the data layout in memory, both in CPUs and GPUs.
Its key goal is to abstract as much as possible operations over arrays and make it easier to write algorithms without compromising maximum performance.
The library also provides interoperability with existing libraries, in particular numerical ones such as linear algebra (BLAS-like) and fast Fourier transforms (FFT)~\cite{Cooley1965}, and the \CC Standard Library (STL) algorithms.

\textsc{Multi} is currently heavily used in the \inq implementation and it is its main internal \emph{data structure}.
Specifically, electronic states data (complex coefficients) is \emph{viewed} simultaneously, depending on the part of the code, as 2-dimensional or 4-dimensional.
3-dimensional sub-arrays can represent the spatial (volumetric) structure of the problem either in real or reciprocal space.
This referential manipulation of arrays (also known as `views') allow to minimize copies and especially avoid copies between GPU and CPU.

In addition, arrays can support different memory spaces (e.g. GPUs) through custom pointers.
The use of custom pointers allows to syntactically separate (at compilation-time) GPU and CPU memory and their corresponding operation (e.g. choose between a GPU and CPU code to dispatch an abstract operation).
Finally, the library is responsible for memory allocations which are notoriously slow in the GPU.
In \inq, the library allows reducing the number of allocations by optionally managing internal blocks of preallocated memory.

The fact that this library is used in a completely different HPC code, \textsc{QMCpack}~\cite{Kim2018} is witness to the general purpose aspect of this library.

\subsection{\textsc{B-MPI3}: Message passing for \CC}\label{sec:mpi3} 

\Inq achieves distributed memory parallelism (parallelism between several nodes) by using the Message Passing Interface (\textsc{MPI}) infrastructure, including extensions that allow direct GPU to GPU data transfer.
The problem of standard \textsc{MPI} is that its interface can be quite cumbersome to use, since it requires a large amount of user provided information information (explicit types, data pointers, layout) even for constructing simple messages.
To avoid this issue, the low-level \textsc{MPI} calls in \inq are wrapped by a higher level library called \textsc{B-MPI3}.

\textsc{B-MPI3} is a \CC library wrapper for version 3.1 of the \textsc{MPI} standard that simplifies the its utilization maintaining a similar level of performance.
We find this \CC interface more convenient, powerful, and less error prone than the standard \textsc{MPI} C-based interface.
For example, pointers are not utilized directly and it is replaced by an iterator-based interface and most data, in particular arrays and complex objects are serialized automatically into messages by the library.
\textsc{B-MPI3} interacts well with the \CC Standard Library and containers, and can take advantage of the aforementioned \textsc{Multi} array library. 
The library is general purpose and, as a standalone open source project, it can be reutilized in other scientific projects, also including \textsc{QMCpack}~\cite{Kim2018}.

\subsection{The \textsc{pseudopod} pseudopotential parser}

Pseudopotentials are an essential part of electronic structure codes that rely on uniform representations like plane-waves or real-space grids.
They provide two benefits: make the wave-functions smoother around the nuclei and avoid the explicit simulations of the core electrons.
There are many types of pseudopotentials, and even within each type there are multiple ways of generating them.
This means that users normally have to carefully select the pseudopotentials they want to use from dozens of options.

To make things worse, pseudopotentials come in many different file formats that are not compatible between each other.
The idea of \inq is to support as many formats as possible instead of introducing a new one, as this gives the largest possible flexibility for users.
This means we need to know how to parse most pseudopotential formats.
This is a task that is fairly disconnected from the rest of a DFT code, that only needs to access the pseudopotential information in memory.
Based on this, we decided to make an independent library, named \textsc{pseudopod}, that could take care of this task and other pseudopotential-related functionality.

The goal of \textsc{pseudopod} is to provide a fully standalone library that can be used for other electronic structure codes.
\textsc{Pseudopod} is based on code written for \textsc{octopus} and \textsc{qball}, and as such, it has been tested quite extensively.
It provides \inq, and any code that uses it, a unified interface, independent of the file format, to access the information in pseudopotential files.
This interface is, in addition, GPU-aware, so that the construction of the potential and projectors by the calling code can be done directly from a GPU kernel.

\textsc{Pseudopod} can parse several popular formats like \texttt{UPF1} and \texttt{2} (\textsc{Quantum Espresso}), \texttt{PSML}~\cite{Garcia2018} (\textsc{siesta}), \texttt{psp8} (\textsc{Abinit}), and \texttt{QSO} (\textsc{Qbox}).
This includes Optimized Norm Conserving Vanderbilt pseudopotentials~\cite{Hamann2013}.
The formats based on XML files are parsed using the \textsc{RapidXml} library.
(\textsc{RapidXml} is a simple library that is distributed with \textsc{pseudopod}, so the users need not install it separately.)

\textsc{Pseudopod} also has additional functionality, as it includes routines to \emph{filter} pseudopotentials to remove high-frequency components and avoid aliasing effects~\cite{Briggs1996,Tafipolsky2006}, something that \inq uses and that is essential for real-space DFT codes.
It also provides auxiliary routines like Bessel-transforms, range separation and spherical harmonics that codes might need to process and apply pseudopotentials.

Finally, a very important functionality of \textsc{pseudopod} is to handle pseudopotential sets.
Pseudopotential sets are a relatively recent development that makes much simpler and reliable to run DFT calculations.
Several research groups have provided a collection of pseudopotentials for most of the elements in the periodic table~\cite{Willand2013,DalCorso2014,Garrity2014,Kucukbenli2014,Topsakal2014,Schlipf2015,Prandini2018,Van2018}, that have been curated and validated.
This means that users do not need to select individual pseudopotentials but they can just pick a consistent set of them.

\textsc{Pseudopod} contains as part of its files two of these sets: Pseudodojo~\cite{Van2018} and SG15~\cite{Schlipf2015}; so they can be used directly by the code without needing to be download separately by the user.
The calling code only needs to select the set, and requests the pseudopotential for the specific element.

\subsection{\textsc{Libxc} and GPU support}

\textsc{Libxc} is a standalone library of exchange and correlation (XC) functionals initially developed by M. Marques~\cite{Marques2012,Lehtola2018}.
We rely on \textsc{libxc} to calculate the XC energy and potential for DFT calculations.
This refers to a number of programmed analytic functions that map values of the density and density gradients to energies (and higher derivatives there of).
This is a part of an electronic structure code that requires a lot of code, and that is quite cumbersome and error-prone to implement.
The effect of adopting \textsc{libxc} can be seen in Fig.~\ref{fig:lines}; 
it shows a big drop in \inq's line of code count early on when we started to rely on this external library instead of a few internal functionals.

One previous limitation of \textsc{libxc} is the lack of native GPU support.
In general, the cost of the XC evaluation is comparatively small in a DFT code so the computational cost difference is not very large.
However, a CPU-only implementation forces the code to copy data back and forth between the CPU and the GPU, which is expensive.
So, for an efficient GPU code we need to evaluate the XC functional on the GPU.

As part of the development on \inq, we implemented a GPU version of \textsc{libxc} based on \textsc{cuda} and contributed it back to the official version of the library.
Our implementation makes minimal changes to the library and relies on \textsc{cuda} `unified memory' to store the internal data structures on \textsc{libxc}.

\section{Modern \CC programming} 
\label{sec:c++}

Ideally a programming language should allow us to express the physics clearly, without exposing too much the implementation details or the underlying data structures.
At the same time, it should be efficient so that the programmer has control of what the low-level code is doing and it doesn't introduce spurious operations.
Additionally, availability of compilers, portability and compatibility with GPU computing are key to this project.
Considering all these factors we decided to use \CC for \inq.

\CC is an evolving programming language that is specially attractive to create high-performance applications.
Due to its legacy, it is a system language that allows to control low-level hardware operations while at the same time offering various abstraction mechanisms to represent complex problems.
Separate compilation and linking allows interoperability with hardware, vendor-specific libraries as well as abstraction libraries.
The combination is ideal for simulation problems that require high-performance but also require expressing simulation steps with reasonable simplicity.

In \CC there is no single way to use the language, a rather complex syntax allows paradigms to coexist in a single unit of code, usually correlated with the level of abstraction.
The absence of a single paradigm requires certain discipline which we describe below as \emph{modern techniques}.
Techniques change as the language evolves and the community continues research.
Yet, some characteristics tend to stick as they are deemed to produce good code.
Here we mention some of these techniques that are relevant to this simulation code in particular.

A simulation code usually deals with the manipulation of a simulated system, whose \emph{state} is represented by a set of program variables stored in memory.
Historically, academic codes tend to make this state \emph{global} to the program since usually only one system is simulated at a time.
This state is usually represented by global variables or variables that are designed to live for the whole duration of the program execution.
A global state makes it easy to access the data of the simulation from any place in the program, it makes the code very easy to change and reduces the number of arguments taken by functions that otherwise run in the tens of arguments.
In addition the memory associated with the system does not need to be managed by any special mechanism;
in the worst case, memory is released by the operating system itself once the program shuts down.

Global state has a long term maintenance cost though; 
the program becomes hard to reason about because changes to the state of the simulation can be made by any part of the code in an inconsistent way, by parts that can be far removed from each other.
Global state also implies the existence of a \emph{single} simulation object at a time, generalizing the code to simulate several systems simultaneously is hard to achieve or requires use of isolated instances of the program.

As we explained, we designed the code to be run as a standalone code or as a library.
As a library, the constraint of a single simulated object is not natural and overly restrictive.
Besides, a library needs to interact with other unknown-in-advance parts of the program and therefore cannot take over all the resources or memory to itself.
Therefore, we rejected the idea of a single global simulated system from the start in our design.

It turns out that this fundamental decision led us to use other interesting techniques.
First, since simulated systems had to be manipulated by functions, and there is no single instance of it, functions have to take the simulation variables explicitly.
Therefore sets of variables associated with a system are naturally bunched together as objects or user-defined \CC classes.
Also, objects can be more easily protected against inconsistent modifications (or against modifications at all) by certain functions if they do not naturally need to modify the state.

In all realistic settings, computers have limited resources.
Each object or subobject requires controlled access to resources.
Since an object's lifetime is not indefinite in general in our programming paradigm, the request for resources should be accompanied by release when the resource is no longer needed.
It is key to recognize that resources include more than just memory; 
it includes file handles, \textsc{MPI} communicators, threads, mutexes, GPU memory, or anything that is limited in a computer.
Any resource can \emph{leak} if not properly managed.
Ideally this management should be automatic and not explicitly coded.

\CC can emulate automatic and deterministic management, by which any resource that needs explicit control can be requested on construction of a manager object and released upon its destruction; after that, most resources are managed automatically by scopes or logical units of the code (such as functions or other higher level classes).

Finally, in scientific programming, the code is an expression of equations and mathematical operations of a certain model or theory.
Ideally the code should represent as closely as possible the mathematics, to make the software simple to understand, write, and verify.
Unfortunately this is not always possible to achieve as programming languages offer limited expressiveness and when they do, they usually come with a considerable performance overhead.
The key point is that performance (or scalability or computational complexity) itself cannot be abstracted away or hidden from the user.
In modern \CC, it is possible to design the code such that simple expressions can be written without penalizing performance.
For example in \inq the application of the Hamiltonian, and other operators, in the code looks as simple as
\begin{lstlisting}[language={[11]C++}]
auto hphi = hamiltonian(phi);
\end{lstlisting}
while in most DFT codes the Hamiltonian operator (if it exists as such) is a complicated function with many arguments.
Moreover the above syntax achieves optimal performance without unnecessary copies, an issue that plagued old ways of utilizing the \CC language.

Such simplicity allows the developers to focus on writing algorithms and equations and not with the intricacies of the specific data structures in the code.
Additionally, it is a polymorphic code that can work with any type of Hamiltonian operator.
The previous line is an example of code that is written with a `concept' in mind, rather than a specific implementation of the \lstinline{phi} or \lstinline{hamiltonian} objects.
For example, in \inq many algorithms are tested with a simple ``Hamiltonian'' given by simple dense Hermitian matrix, using exactly the same routine that works for any type of operator.
The idea is that as long as the operation makes sense syntactically and that reasonable developer(s) of the code agree in the semantic meaning of an operation a function can be written simultaneously for different classes that share the same conceptual meaning.
This type of code becomes a \emph{template}, a fundamental \CC feature, that can be compiled into very different machine code, even using different underlying numerical libraries, but still representing the same conceptual operation.
\CC templates not only make code more generally applicable but also have the side effect of `late binding'.
This is a feature by which actual machine code is compiled only when all information about the code is available.
More importantly template functions can have injected code, allowing the compiler to optimize code across function boundaries, including \emph{inlining}.
These ideas are part of a very productive trend in \CC called generic programming, which is based on well defined mathematical reasoning~\cite{Stepanov2014}.

\section{Graphic Processing Units} \label{sec:gpu}

Programming on GPU presents some unique challenges for code developers.
The main ones are that the GPU has its own memory space where data must be stored prior to use, and that the code has to be explicitly parallelized.
Because of this, it is challenging to adapt an existing code to run on the GPU, as extensive modifications to the whole code are needed.
This is one of the reasons we decided to start a new code, designed from the ground up to run on GPUs (but that also runs on CPUs).
This effort is guided by our previous experience in the GPU port of the code \textsc{octopus}~\cite{Andrade2012c,Andrade2013,Andrade2015}.

Since GPUs from different vendors are available or will be available in the near future, it is essential to design a code that is portable, and that ideally is performance portable.
This is, the code can run on different platforms and moreover it can run efficiently on all of them without extensive re-tuning.

\begin{figure}
  \includegraphics[width=\columnwidth]{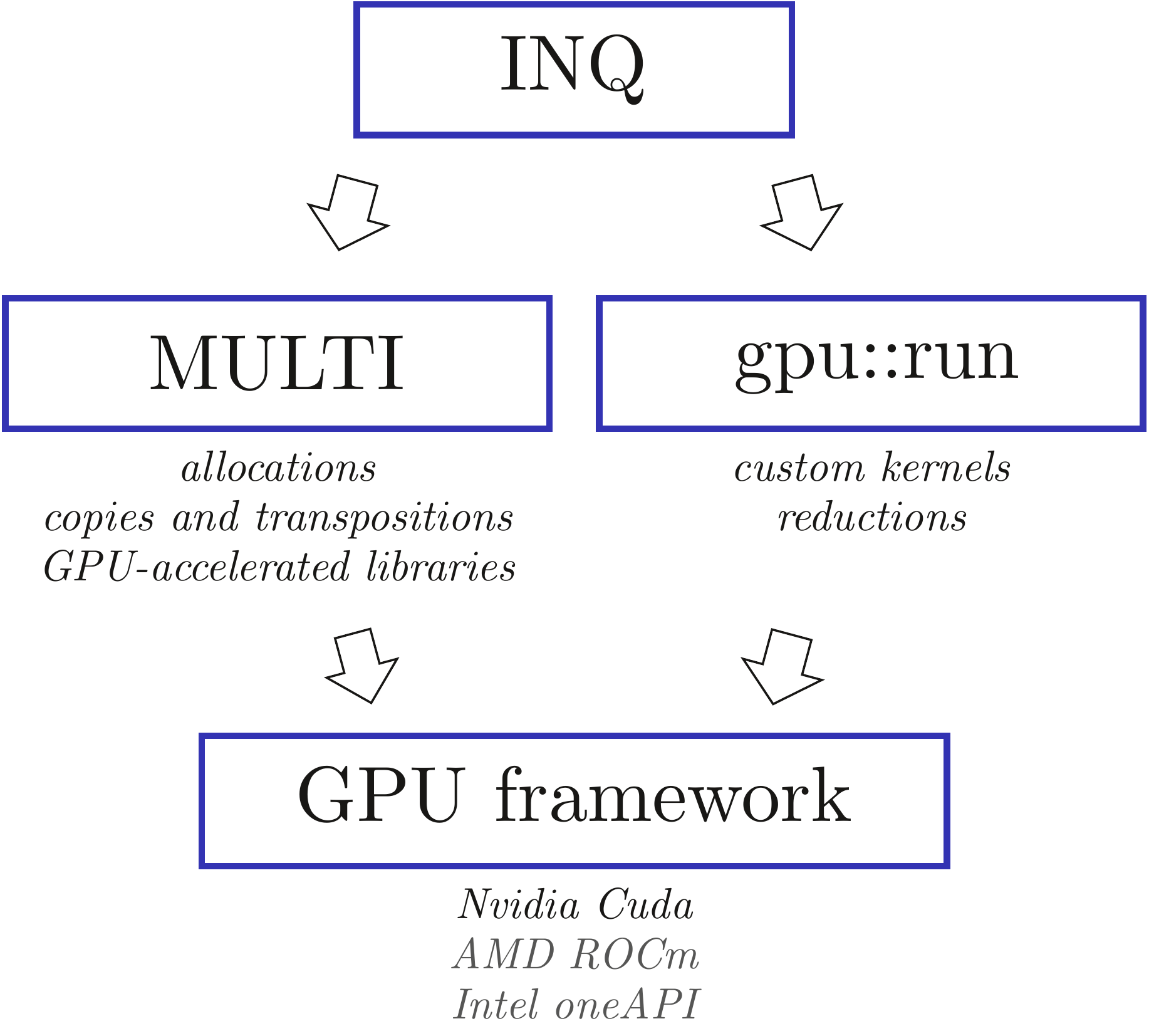}
  \caption{The two components that \inq uses to run GPU code.
    \textsc{Multi} is an array library, that besides the multidimensional indexing abstraction, takes care of memory allocations, memory copy and provides an interface for lineal algebra and fast Fourier transforms.
    \texttt{gpu::run} is a simple abstraction layer to run kernels and perform reductions on the GPU.
    This layer of abstraction allows us to write code that is agnostic to the specific processors that are used.
    Right now we are based on \textsc{cuda}, but in the near future we plan to support the code on AMD and Intel GPUs.
    Both \textsc{multi} and \texttt{gpu::run} also support execution on the CPU, so we do not need to write duplicated code for each processor type.
  }
  \label{fig:gpu_structure}
\end{figure}

With this objective in mind we studied the different available platforms for high level GPU programming like \textsc{raja}~\cite{Beckingsale2019}, \textsc{Kokkos}~\cite{CarterEdwards2014}, \textsc{OpenMP}~\cite{Lee2010} and \textsc{SyCL}~\cite{Alpay2020}.
Unfortunately we could not find one that was directly suitable for the operations needed for DFT/TDDFT and that was mature enough at the moment we started the project (mid-2019).
So we decided to design our code using the \textsc{cuda} \CC extensions with a thin layer of abstraction on top, that allows us to make most of the code independent of the GPU backend.
This layer has two components that take care of different tasks, a scheme of this approach is shown in \ref{fig:gpu_structure}.
The first component is the \textsc{Multi} library that takes care of allocations, array copies and transpositions, and GPU-accelerated libraries for linear algebra and FFTs.
The second one is a routine we wrote called \texttt{gpu::run}.
It executes kernels on the GPU and adds some extra functionality like the calculation of reductions.
To adapt \inq to other GPUs we will just need to extend these two components, and not the whole code.

To simplify the code and its development we use what Nvidia calls \emph{managed memory}, a type of memory that can transparently be accessed from the CPU and GPU.
However, a large performance price is paid when the memory is accessed (even inadvertently) from the wrong processor.
Our memory management philosophy is to assume that all the data will be kept in the GPU memory.
We think this is reasonable since the amount of memory on the GPU is increasing substantially, and that the individual GPU memory requirements can be reduced by using more \textsc{MPI} nodes with GPUs.
This contrasts with the upgrade strategy followed by many CPU legacy codes, in which some identified tasks are delegated the GPU together with a two-way copy of data from CPU to GPU and back.

In the code we use \textsc{Multi} arrays that are allocated in \textsc{cuda} managed memory, so that they can be accessed both from CPU and GPU code.
This completely hides both the allocation and memory location in most of the high level code.
However, our experience in the optimization of the code shows that managed memory can slow down significantly GPU kernels accessing large memory blocks, memory that has been just allocated.
To avoid this problem, it is necessary to selectively use a `prefetch' call after allocation to speed up kernel execution considerably~\cite{Chien2019}.

The \textsc{Multi} library offers generic operations like matrix multiplications and FFTs that behind the scenes are dispatched by the compiler to the corresponding library.
For example, \textsc{BLAS} for the CPU data and \textsc{cuBLAS} for (Nvidia) GPU data.
This allows us to abstract a large part of the operations and make them run as efficiently as possible using vendor optimized libraries under a common syntax.

There are, however, several operations in a DFT code that are quite specific and are not available from external libraries.
For them we have to write our own GPU kernels in the \textsc{cuda} language, which is an extension of the \CC language.
We use modern \CC to do this in a simple, readable and portable way.

\CC offers a nice way of defining functions locally (inside another function) called \emph{lambdas}. 
These lambdas can `capture' local information and they can be passed to other functions, such as GPU kernels.
Based on this functionality we wrote a simple function to execute lambdas on the GPU called \texttt{gpu::run}.
The idea behind this function is that it can replace loops in a way that is similarly readable.
Unlike the \CC \textsc{STL}, \texttt{gpu::run} is oriented to the index execution pattern in multiple dimensions and not to one-dimensional data structures.

This is a simple example of \texttt{gpu::run}.
Consider the following double loop that is how a standard CPU code would be written:
\begin{lstlisting}[language={[11]C++}]
for(int i = 0; i < n; i++){
  for(int j = 0; j < m; j++){
    c[i][j] = a[i][j] + b[i][j];
  }
}
\end{lstlisting}
With \texttt{gpu::run} we would write this code like this:
\begin{lstlisting}[language={[11]C++}]
gpu::run(
  m, n, 
  [...](auto j, auto i){
    c[i][j] = a[i][j] + b[i][j];
  }
);
\end{lstlisting}
(Where the lambda `capture' arguments \lstinline{[...]} are omitted for brevity.)

The number of size arguments (2 in this case) determines how many loops there are and what are the corresponding index ranges (0 to \lstinline{m} and 0 to \lstinline{n} in this case).
The last argument is a lambda-function that indicates the work that each iteration does.
When compiling for the CPU, the code above is interpreted as normal loops, just like the original code.
When compiling for the GPU, the lambda is executed in parallel inside a GPU kernel instead of loops.

Many operations can be written in this fashion, such as element-by-element transformations or copies.
The main exception happens when there are sums (or in general reductions) inside some of the loops.
So we have implemented a special case of \texttt{gpu::run} for this cases.

Take for example the case of a matrix multiplication.
This can be written in a loop like this
\begin{lstlisting}[language={[11]C++}]
for(int i = 0; i < n; i++){
  for(int j = 0; j < m; j++){
    c[i][j] = 0.0;
    for(int l = 0; l < k; l++){
      c[i][j] += a[i][l]*b[l][j];
    }
  }
}
\end{lstlisting}
Notice that there is a sum in the loop over \lstinline{k}.
In that case, to use the version of \texttt{gpu::run} we just showed, we would need to include the loop with the reduction inside the lambda, like this:
\begin{lstlisting}[language={[11]C++}]
gpu::run(m, n,
  [...](auto j, auto i, auto l){
    c[i][j] = 0.0;  
    for(int l = 0; l < k; l++){
      c[i][j] += a[i][l]*b[l][j];
    }
});
\end{lstlisting}
The problem with this strategy is that it is not very efficient on a GPU in the case when \lstinline{m} and \lstinline{n} are relatively small compared to \lstinline{k}.
So we need a more general solution that also parallelizes the loop over \texttt{k} on the GPU.

In general, performing reductions in parallel over GPU code is not simple, since horizontal communication between threads is not straightforward.
The main strategy for reductions is to do a tree-based approach~\cite{Harris2007}, even though more advanced hardware-dependent strategies exist.
Our approach in \inq was to implement this reduction algorithm and expose it through a special \texttt{gpu::run} interface.
This would make the matrix multiplication look like this:
\begin{lstlisting}[language={[11]C++}]
c = gpu::run(m, n, gpu::reduce(k),
  [...](auto j, auto i, auto l){ 
      return a[i][l]*b[l][j];
  }
);
\end{lstlisting}
As before, the \lstinline{m} and \lstinline{n} are normal iterations, while the \lstinline{gpu::reduce} tag around \lstinline{k} indicates that a reduction should be performed over this dimension (third index).
The result of the reduction is returned as an array \lstinline{c} of dimensions \lstinline{n} by \lstinline{m}.
Note that in this case the lambda must \emph{return} a value for each iteration, this is the value that will be accumulated.

This strategy allows us to easily write operations that otherwise are complicated to implement on the GPU.
Plus, they hide the reduction algorithm that could be optimized depending on the specific hardware.

Please note that we used the matrix multiplication just as an example in this section.
In \inq, we always use \texttt{gemm} for matrix multiplications, as it is much more efficient as it is optimized by the vendor for each platform (e.g. as part of \textsc{cuBLAS}).
The main objective of \lstinline{gpu::run} is to allow us to quickly write GPU code for many operations.
Some kernels that are critical for performance might need further optimization.

It is well known that allocation (explicit request of memory) and deallocation (release of memory) of GPU memory is relatively time-consuming.
Therefore, it is paramount to avoid these type of requests in performance-critical parts of the code.
A naive approach, such as making the variables live in larger scopes, or even making them global, can spoil the architecture of the program.
Instead, \inq adds another layer of abstraction by utilizing custom memory \emph{allocators} that can optimize memory management for particular use patterns.
Specifically, large chunks of memory can be recycled across different objects avoiding the costly system-level release of memory.

\section{Distributed-memory parallelization (\textsc{mpi})}

If we want to model large systems it is essential to use distributed-memory parallelization, where the code runs simultaneously over multiple nodes at the same time (up to hundreds of thousands in some cases).
For this parallelization we use the standard message-passing paradigm, where each process has its own data and can exchange information with other processes.

The simplest and more efficient method of parallelization for message passing is to distribute the data among processors, with processors executing mostly the same code (single program multiple data)~\cite{Pacheco1997}.
It allows us to increase the number of processors as the problem size increases.
The challenge is to use an optimal data distribution to avoid communication as much as possible.
In the case of \inq, this means a distribution of the arrays that is optimal for parallel FFTs (this is explained in detail in section~\ref{sec:fields}).

DFT and TDDFT have traditionally performed very well in CPU-based parallel supercomputers~\cite{Gygi2005,Andrade2012,Jornet2015,Hasegawa2011,Draeger2017}.
However, GPUs present additional challenges for parallel programming.
While, GPUs can compute much faster than a CPU, network communication hardware has not seen a similar jump in performance.
This means that the relative cost of communication with respect to computation has increased in GPU-based supercomputers.
On top of that, communicating data between GPUs through \textsc{MPI} can be challenging to implement for vendors.

In \inq we directly call \textsc{MPI} over data in GPU memory.
Ideally, most modern \textsc{MPI} implementations are GPU-aware and can recognize the GPU memory space and they can directly access data.
In the best scenario they can also communicate the data to and from GPU memory without passing through main memory, but this is not always the case.
For non-GPU-aware implementations, \emph{managed memory} residing on the GPU would be automatically copied to main memory before \textsc{MPI} calls.
Unfortunately this last case has a non-negligible overhead in the communication cost.

\section{DFT and TDDFT implementation}

In this section we describe the physical model that \inq simulates, the solution algorithms, the main data representation, and the computational kernels we need to use.

The density functional framework provides a way to calculate the density, and other observables, of an interacting many-body system by using a non-interacting system as reference~\cite{Runge_1984}.
The main quantities are then the density \(n\) and the set of Kohn-Sham (KS) orbitals (or states) \(\varphi_k\) of the reference non-interacting system.
Mathematically these objects are fields, or functions: they have a value associated to each point in space.
The density is generated from the occupied orbitals by the formula
\begin{equation}
	\label{eq:density}
	n(\vec{r}, t) = \sum_k\left|\varphi_k(\vec{r}, t)\right|^2 \ .
\end{equation}

The operator that generates the dynamics of \(\varphi_k\) is the KS Hamiltonian (in atomic units)
\begin{multline} \label{eq:hamiltonian}
	\hat{H} = 
		-\frac12\hat\nabla^2 
		+ \sum_I\hat{v}_I(\vec{r}-\vec{R}_I)  + v_\text{pert}(\vec{r}, t) \\
		+ v_\text{h}[n](\vec{r}) + v_\text{xc}[n](\vec{r})\ .
\end{multline}

The first term is the Laplacian operator that represents the kinetic energy.
The second term is the sum of the potentials of the ions, at positions \(R_I\), represented by a non-local pseudopotential.
\(v_\text{pert}\) represents a possible time-dependent perturbation, for example an external electric field.
The two last terms \(v_\text{h}[n](\vec{r})\) and \(v_\text{xc}(\vec{r})\) mimic the electronic interaction in the density functional approach.
They are in principle functionally dependent on the density at all points and, for the time-dependent case, at all past times~\cite{Runge_1984}. 

The actual dynamics of the electrons is obtained from the time-dependent KS equation
\begin{equation} \label{eq:tdks}
	i\frac{\mathrm{d}}{\mathrm{d} t}|\varphi_k(t)\rangle = \hat{H} |\varphi(t)\rangle \ .
\end{equation}
Note that because of the density dependence in $\hat{H}$, this equation is coupled with Eq.~\ref{eq:density} so both must be solved self-consistently.

In the stationary case Eq.~\ref{eq:tdks} yields the ground-state KS equation~\cite{Kohn_1965}
\begin{equation}
  \label{eq:ks}
  \hat{H} |\varphi_k\rangle = \varepsilon_k |\varphi_k\rangle\ .
\end{equation}
that minimizes the energy
\begin{multline}
  E = \sum_k\varepsilon_k - \frac12\int\mathrm{d}\vec{r}\,n(\vec{r})v_\mathrm{h}(\vec{r})\\
  + E_\mathrm{xc}[n] - \int\mathrm{d}\vec{r}\,n(\vec{r})v_\mathrm{xc}(\vec{r})\ .
\end{multline}

The real-time equation for the electrons (Eq.~\ref{eq:tdks}) can be coupled with a set of classical equations of motions for the ions, under the force generated by the electrons.
This produces Ehrenfest dynamics~\cite{Ehrenfest1927,Andrade2009} that approximates the non-adiabatic dynamics of the system.

Our idea is to provide a computational toolkit that allows, not only to solve eqs.~\ref{eq:tdks} and \ref{eq:ks}, but also to perform all the operations that appear in the density functional approach as distinct code routines.
All of these routines are well tested and implemented so that they can execute efficiently and transparently on CPUs, GPUs and \textsc{MPI} parallelization.
This will allow \inq developers and users to to implement new functionalities and test new theories, in a simple to use and fast framework.

The main choice for a density functional code is to select a basis where the orbitals, density and other fields are going to be represented by a finite amount of data.
In \inq we use the popular plane-wave approach, that despite the name, actually uses two representations.
One is the basis of plane-waves or Fourier space, while the second one is a real-space grid.
This is advantageous because the Laplacian operator is diagonal in Fourier space while the potential terms are local or semi-local in real-space, and because we can go efficiently between the two representations using FFTs.
Of course, other alternatives exist for the discretization of the equations.
In chemistry the dominant approach is to expand the fields into a set of atomic orbitals, usually represented by Gaussian wave-functions~\cite{Szabo2012,Olsen2021}.
In real-time TDDFT it is quite common to use real-space grids where the Laplacian operator is approximated by high-order finite-differences~\cite{Chelikowsky1994}.

\subsection{Ground-state solution} \label{sec:ground-state}

To obtain the ground state of an electronic system in DFT we need to solve the KS equation (Eq.~\ref{eq:ks}) numerically.
We follow the standard approach of directly solving the non-linear eigenvalue problem through a self-consistent iteration, that deals with the non-linearity which appears through the density dependency~\cite{Kresse1996}.
Each iteration we need to solve a linear eigenvalue problem for the Hamiltonian operator for a given guess density.
The solution gives a new density, that is mixed with the previous guess density as a guess for the next iteration.
When the input and output densities are sufficiently similar the solution is converged.

The mixing of the density in each step is relatively simple to implement and does not require much computation,
while, at the same time, stabilizing the convergence of the iterative solution.
In \inq we use the Broyden method~\cite{Broyden1965} by default.
The code also implements (static) linear mixing and Pulay mixing~\cite{Pulay1980,Kresse1996} as alternatives.

The computationally expensive part comes from the solution of the eigenvalue problem.
Since the dimension of our space can be very large, it is not practical to use a method that directly works over a dense matrix (e.g. the basis representation of \(\hat H\) ).
Instead, we use eigensolver algorithms that only need the application of the operator over trial vectors~\cite{Saad2011}.
These methods are usually called \emph{iterative} since they progressively refine approximate eigenvectors until convergence is achieved.
(However, it must be noted that as a corollary of the Abel-Ruffini theorem all eigensolvers, including the dense-matrix ones, must be iterative in some sense~\cite{Axler2014}).

In practice it is not necessary or convenient to fully solve the eigenvalue problem in each self-consistency iteration.
Instead, a few iterations of the eigensolver are done at each step, so eigenvector and self-consistency convergence is achieved together.

The eigensolver we use in \inq by default is the preconditioned steepest (SD) descent algorithm.
This is a simple method that has the advantage that it can work simultaneously and independently on all eigenvectors, which is particularly useful for GPUs, as it makes a lot of data parallelism available.
(We are working on the implemention of the RMM-DIIS method~\cite{Kresse1996} that converges faster that SD and is also highly parallelizable.)
We also implement the conjugate gradient~\cite{Payne1992} and Davidson~\cite{Davidson1975} eigensolvers, however they do not perform as efficiently as SD at the moment.
For preconditioning we use the method of Teter~\emph{et al}.~\cite{Teter1989} that is applied in Fourier space.

As part of the eigensolver process, we need two operations that involve \emph{rotations} in the space of eigenvectors: orthogonalization and subspace diagonalization~\cite{Kresse1996}.
The cost of these procedures is dominated by linear algebra operations that scale cubically with the number of atoms, while other operations are quadratic.
As this linear algebra becomes dominant for large systems, it is important to do them efficiently.
In the serial case this is not complicated to do, since highly optimized versions of \textsc{blas}~\cite{Blackford2002} and \textsc{lapack}~\cite{Anderson1999} are available for the CPU and GPU.
For the parallel case, the situation is more complicated at the moment.
While the \textsc{scalapack}~\cite{Blackford1997} library provides a set of parallel linear algebra routines, it only runs on CPUs.
The \textsc{slate} library~\cite{Gates2019}, currently under development, in the near future will provide a modern replacement for \textsc{scalapack}.
Until \textsc{slate} is complete or other library becomes available, \inq cannot fully run in parallel for ground-state calculations.
Only domain parallelization is available.

An important aspect of the solution of the ground state is selecting an appropriate initial guess for the density and the orbitals.
The density is initialized as a sum of the atomic density of the atoms, obtained from the pseudopotential files.
The orbitals are initialized as random values for each coefficient in real-space, uniformly distributed in a symmetric range around 0.
We have found that this produces orbitals that are linearly independent and that contain components of the ground-state orbitals.

However, special care must taken when running in parallel, both in the GPU and \textsc{MPI}, with the generation of random numbers.
If each parallel domain uses a different random number generator, the initial guess would depend on the number of processors used, making it difficult to get consistent and repeatable results.
(And this is without even considering the issue of correlation between generators, that can happen in parallel.)
Our solution was to use a permuted congruential generator (PCG)~\cite{Oneill2014} that it can be fast-forwarded by skipping steps in logarithmic time.
So, when running in parallel, for each point we can forward the generator to obtain the same number it would have in the serial case.
This approach produces random orbitals that are exactly the same independently of the number or processors used, and independently of whether we are running on the CPU or GPU.
The overall cost of the randomization is quasi-linear, and in practice we see it is negligible compared to other operations.
For the implementation we use the small library provided by the author of PCG, that we modified to run on GPUs.

Having a consistent starting point that does not depend on the parallelization is essential for testing.
Otherwise, it would be very hard to determine whether differences in the results come from errors in the \textsc{MPI} or GPU parallelization or from the differences in the starting guess.

\subsection{Real-time propagation}

Time propagation is used to calculate excited state properties by following the real-time dynamics of the electrons under external perturbations.
It can be used to calculate a large number of linear and non-linear response properties.
It requires the integration in time of the time-dependent KS equation (Eq.~\ref{eq:tdks}).

How to do the real-time propagation efficiently is an area of active research, and many methods have been proposed~\cite{Castro2004,Schleife2012,Kidd2017,Gomez2018,Jia2018,Bader2018,Zhu2018}.
We use the enforced time-reversal symmetry (ETRS) propagator with the exponential approximated by a 4th order Taylor expansion~\cite{Castro2004}.
This is a quite popular method due to its efficiency, numerical stability and relatively simple implementation.
However, we have introduced an additional ``trick'' with respect to previous implementations that allows us to reduce the computational time by 33\%.

In ETRS, the propagator is
\begin{multline} \label{eq:etrs}
	|\varphi_n(t+dt)\rangle 
	= \exp\left(-i\frac{\delta t}2 H(t + dt)\right) \times \\
	  \times \exp\left(-i\frac{\delta t}2 H(t)\right)|\varphi_n(t)\rangle\ .
\end{multline}
Since \(H(t+dt)\) is not known due the non-linearity of the equations, we need to approximate it.
This approximation is usually obtained from doing a full-step propagation of the states
\begin{equation}
  \label{eq:etrs0}
  |\varphi^\text{approx}_n(t+dt)\rangle = \exp\left(-i\delta t H(t)\right)|\varphi_n(t)\rangle\ .
\end{equation}
From \(|\varphi^\text{approx}_n(t+dt)\rangle\) we then get an approximation for \(n(\vec{r}, t + dt)\), and from there \(H(t + dt)\).

Hence, for each step of ETRS we need to calculate three exponentials, two in Eq.~\ref{eq:etrs} and one in Eq.~\ref{eq:etrs0}.
The important detail is that two of these exponentials, \(\exp\left(-i\frac\delta{2} t H(t)\right)|\varphi_n(t)\rangle\) and \(\exp\left(-i\delta t H(t)\right)|\varphi_n(t)\rangle\), have a very similar form.
They only differ by a factor of \(2\) in the exponent.

Let's consider the truncated Taylor approximation of the exponential of an operator multiplied by a scalar
\begin{equation}
\exp(\lambda A)|v\rangle = \sum_{k=0}^4\frac1{k!}\lambda^kA^k|v\rangle\ .
\end{equation}
When we evaluate this expression numerically, the expensive part is \(A^k|v\rangle\).
It is easy to see then, that we can evaluate this expression for several values of \(\lambda\) for almost the same cost of a single value, since \(\lambda\) only appears as a multiplicative coefficient.
In practice, this means we can calculate \emph{together} two of the three exponentials needed for ETRS, \emph{reducing the cost to only two exponentials per step}.


For Ehrenfest dynamics, we propagate the ions with the velocity Verlet~\cite{Verlet1967} algorithm using the forces given by Eq.~\ref{eq:force_grad}.
Both electrons and ions must be propagated consistently, to ensure that the ionic potential is evaluated at the correct time in Eq.~\ref{eq:etrs}.
Otherwise the time-reversal symmetry is broken and a drift appears in the total energy.

The propagation of the ion coordinates itself  is not numerically expensive.
However, it is the calculation of the forces from the electronic state and the recalculation of the ionic potential that can be the most time consuming  when including ion dynamics.
For these reasons these operations must be implemented efficiently and run on the GPU.

One significant advantage of TDDFT is its potential for parallelization.
The real-time propagation in TDDFT conserves orthogonality of the KS states mathematically and also numerically~\cite{Andrade2009}.
The first consequence of this property is that, unlike the ground state, we do not need the orthogonalization operation which makes the overall cost of the time-propagation quadratic with the size of the system instead of cubic.
The second consequence is that the propagation of each state is independent from the rest, with the only ``interaction'' between them coming from the self-consistency of the equations.
This means that it is very efficient to parallelize the propagation by distributing states among processors~\cite{Andrade2012}.

\subsection{Fields}
\label{sec:fields}

\begin{figure}
\includegraphics[width=\columnwidth]{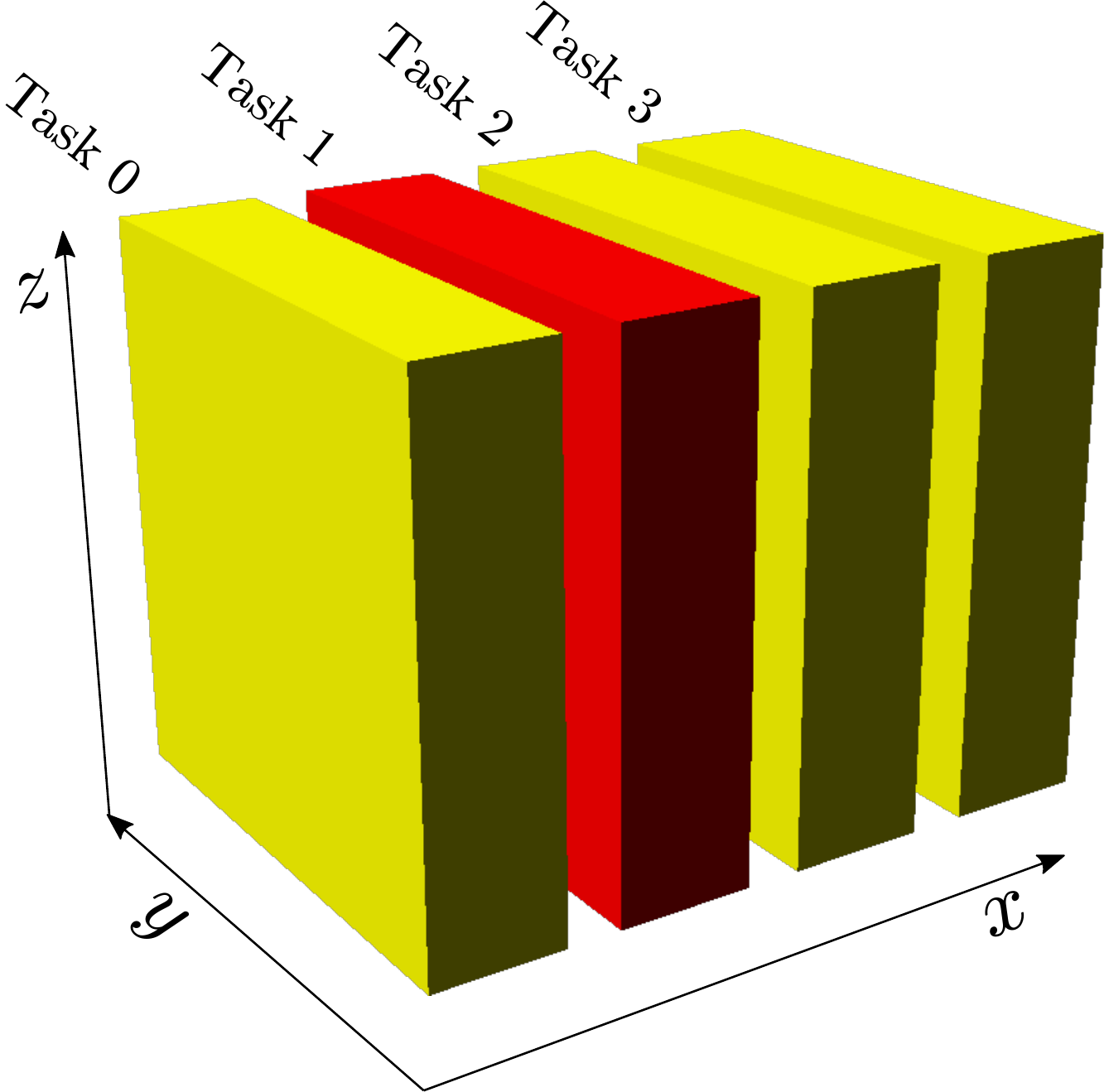}
\caption{ \label{fig:slabs}
	Distribution of the 3-dimensional grid into slabs for optimal communication when doing FFTs in parallel.
	The order of spatial axis corresponds to a real-space field; in reciprocal space \(x\) and \(z\) are interchanged.
}
\end{figure}

A fundamental data structure in \inq is the \texttt{field} type, that represents a mathematical function or field in a finite domain.
The information it contains is a basis and the coefficients of the field in that basis.
Quantities represented by \emph{field} are for example the density and the local potential.

A property of the basis used inside a \emph{field} is the type of representation, for example real space (RS) or Fourier space (FS).
This type of basis is represented through \CC types, which allows us to define polymorphic functions that work over a \texttt{field}, independently of the basis on which it is represented.
For example the \texttt{gradient} function can accept both a \texttt{field} in RS or FS as input, and do the correct operation for each case.
Of course, \inq provides functions to transform a field from one representation to the other by a FFT.

In both RS and FS, the coefficients in a field have a geometrical structure as a 3-dimensional (3D) array.
However, for many operations that do not rely on a geometric context, it is more convenient to access the coefficients as a 1-dimensional (1D) array.
Thanks to \textsc{multi}, a \texttt{field} object offers both a linear and 3D view of the coefficient data.
Both representations share the underlying data so no copies are done, and the access is done without any overhead in both cases.

When running in parallel, a field is distributed by giving each processor a part of the coefficients.
A large part of the operations we need to perform are local, they do not mix information from different coefficients.
Integrals can be calculated efficiently by computing the local sums on each processor and then performing a reduction operation over all the processors.
For this the field also includes a communicator, that allows communication between all processors that contains parts of the field.
The operations that really mix information from all points, and use the larger amount of communication, are the FFTs required to move from RS to FS and vice versa.
It is the FFT operation then that determines they way we divide our points among processors so that communication is minimized.

A 3D FFT is calculated as a sequence of 1D FFTs in each direction.
To minimize communication we distribute the 3D grid into \emph{slabs} by dividing one dimension evenly among processors, as show in Fig.~\ref{fig:slabs}.
For a RS-field the \(x\)-dimension is the one that is split.
So we Fourier-transform the \(y\) and \(z\)-dimensions first as those transforms can be done locally.
Now we redistribute the array among nodes, so that the \(z\) dimension is split.
Of course this is the step that involves communication as all nodes need to exchange information.
Once this is done we can do the FFT in the remaining \(x\)-dimension that now is local.
Note than in the case of FS-field, we end up with a 3D grid that is split along the \(z\)-direction.

The approach we describe above is the standard approach for the parallelization of 3D FFTs and it is implemented for the CPU in \textsc{FFTW} and other libraries.
For the GPU unfortunately there isn't a general \textsc{MPI} library yet that provides all the required functionality for \inq, so we need to do use our own implementation.
The HeFFTe~\cite{Ayala2020} library promises to fill this gap and we expect to use it in the near future.

\subsection{Field sets and Kohn-Sham states}

One of the fundamental objects we need to represent in DFT are the KS states.
As they are just a group of several fields, it would be natural to represent them as an array of \emph{field} objects.
However, this is not practical for numerical performance as we will end up with separated blocks of memory.
In code, it is much more efficient to operate over data all at once instead of one by one~\cite{Wadleigh2000}, especially on the GPU~\cite{Andrade2012c,Andrade2013}.

Instead, we define a specialized \emph{field-set} data structure, that represents a group of several fields.
As well as the basis information, it contains the coefficients for the whole set stored in a single array.
This array can be seen as 2-dimensional (2D), or equivalently a matrix, where one index corresponds to the coefficient index and the other to the index in the set.
This representation is particularly useful for linear algebra operations that can be done directly using \textsc{Multi}.
The array of coefficient can also be accessed as a 4-dimensional array with indices corresponding to the 3 spacial dimensions of the coefficients plus the set index.
We use this representation for operations with geometrical context including FFTs.

\begin{figure}
\includegraphics[width=\columnwidth]{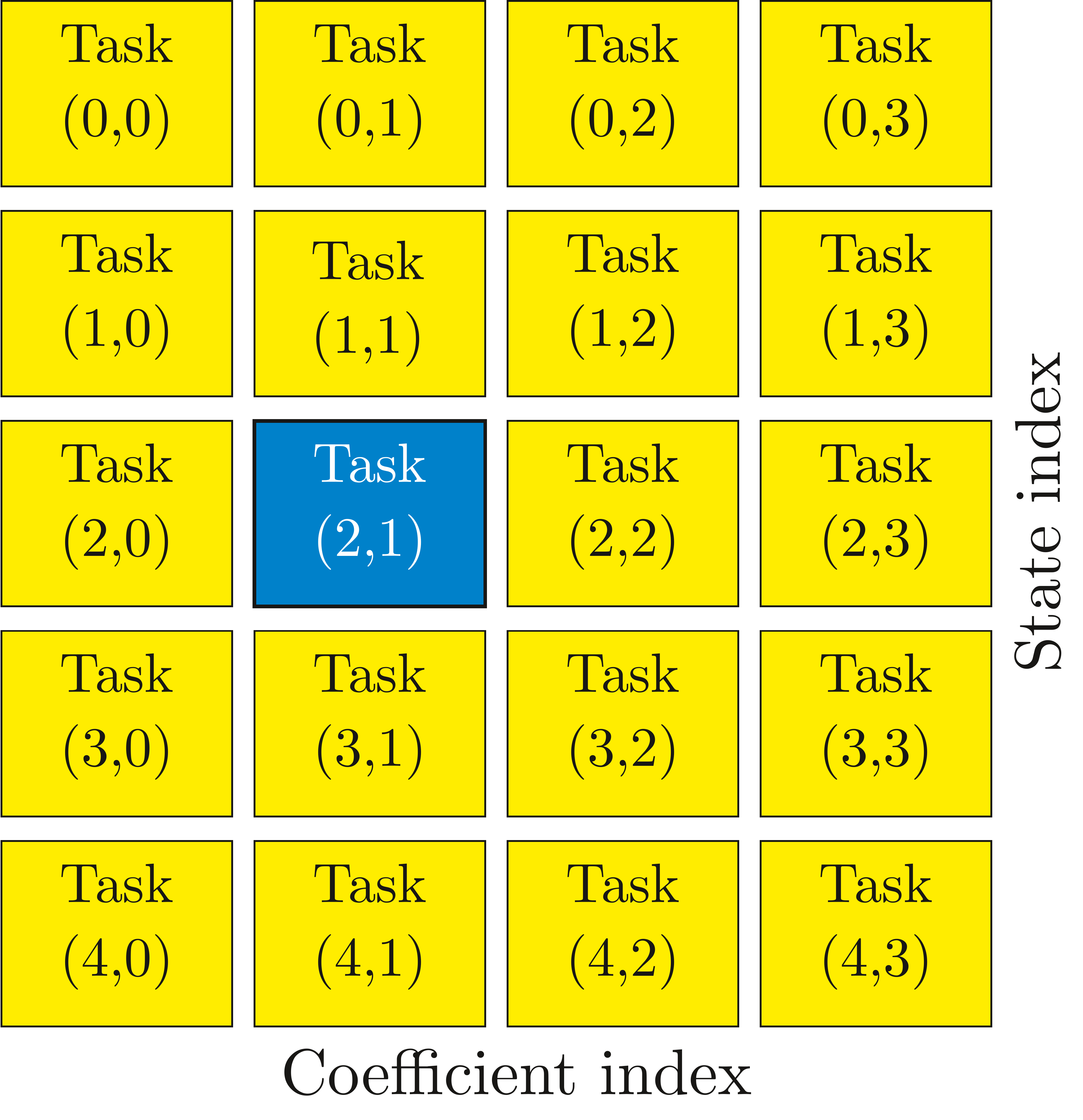}
\caption{
	\label{fig:2dparallel}
	Representation of parallel distribution of the Kohn-Sham states in a 2-dimensional (2D) data decomposition.
  We can consider the states as a 2D matrix where one dimension is the state index and the other is the basis coefficient index (where the 3-dimensional indices have been flattened).
  In parallel, this matrix is distributed in blocks to the different processors.
  For convenience we can think of the processors as arranged in a 2D grid, where each one is labeled by two coordinates that indicate which range of states they have and which range of coefficients.
  For example, if there are 400 coefficients and 50 states, task (2,1) (shown in blue) would have coefficients 200 to 299 from the states 10 to 19.
}
\end{figure}

The parallelization strategy for a field set is to use a 2D decomposition of the coefficient array.
The coefficient dimension is divided in the same way done for \texttt{field}.
The second division is done along the set indices, distributing the states among processes.
In practice, we are giving each processor a 2D block of the orbitals, as shown in Fig.~\ref{fig:2dparallel}.

Using such a decomposition has several advantages.
In first place, as the system size grows both the number of states and the number of basis coefficients grows linearly.
So a two-level parallelization can keep the amount of data local to each processor constant, if the number of processors is grown accordingly (this is the weak-scaling scenario).
Second, each parallelization level is naturally limited either by cost of communication or Ahmdal's law.
By combining multiple levels of parallelization we can use a larger number of processors than with a single one (this would be the strong scaling case).

Most parallel operations are trivially parallelizable in one of this two levels, while requiring communication in the other level.
For a example, a parallel FFT requires the mixing of coefficients which implies communication, as we discuss in section~\ref{sec:fields}, however it is trivially parallelizable for the different states.
On the other hand, the calculation of the density, Eq.~\ref{eq:density}, it is trivially parallelizable in RS points but requires a parallel reduction over states.
The case where we need communication across both levels of parallelization are linear algebra operations, that only appear for the ground-state case, and that need to be handled by linear algebra libraries (as mentioned in sec.\ref{sec:ground-state}).

\subsection{Implementation of the KS Hamiltonian}

The KS Hamiltonian, Eq.~\ref{eq:hamiltonian}, is the main operation in the density functional formalism.
Even though it is a linear operator (for a given density, in each self-consistency iteration) it would not be convenient to store it explicitly as a dense, or even as a sparse, matrix.
Instead, we use it in operator form, by having a routine that applies the Hamiltonian over a \emph{field set}.
This way each one of the terms that appear in Eq.~\ref{eq:hamiltonian} can be calculated in the optimal way.

The kinetic energy operator, given by the Laplacian in Eq.~\ref{eq:hamiltonian}, is calculated in Fourier space, where it is diagonal.
The local potential is diagonal in real-space instead.
And the non-local part of the pseudopotential is also applied in real-space, where the projectors are localized (this is discussed in detail in the next section).
This means that the main operations inside the Hamiltonian are two Fourier transforms to switch back and forth between real-space and Fourier representations using FFTs.

\subsection{Pseudopotentials}\label{sec:pp}

The current version of \inq uses Optimized norm-conserving Vanderlbit (ONCV) pseudopotentials~\cite{Hamann2013}.
In comparison to ultrasoft pseudopotentials~\cite{Vanderbilt1990} and projector-augmented waves (PAW)~\cite{Blochl1994}, ONCVs are simpler to apply, do not introduce additional terms in the equations, and still produce accurate results~\cite{Willand2013,Lejaeghere2016,Van2018,Tancogne2020}.

In \inq the pseudopotentials are fully applied in real space in Kleinman-Bylander (KB) form.
In this form there are two terms for the pseudopotential, the local part and a non-local part.

The local part is, in RS, a simple multiplicative potential that is applied together with the Hartree and XC potentials.
Since the local part is long range, it needs to take into account the effect of all the periodic replicas of the atoms when the system is periodic.
To calculate it, we separate the local term into two terms, short range and long range, using the standard error function separation.
The short range potential is calculated directly in the grid.
The long range potential is given by the error-function, that is generated by a Gaussian charge distribution.
So the solution of the Poisson equation for the Gaussian charges of all atoms gives us the long range part with the correct boundary conditions.

The non-local part requires a bit more attention in order to apply it efficiently.
The KB separation yields a non-local potential of the form
\begin{multline} \label{eq:vnl}
	v_\text{nl}\varphi_n(\vec{r}) = \sum_I\sum_{\ell m}\beta^{I}_{\ell m}(\vec{r} - \vec{R}_I)\\
	\int_{\left|\vec{r}' - \vec{R}_I\right| <R_c}\mathrm{d}\vec{r}'\beta^{I}_{\ell m}(\vec{r}' - \vec{R}_I)\varphi_n(\vec{r}')\ ,
\end{multline}
where \(\beta^{I}_{lm}\) are the projector functions for each angular momentum component \(\ell m\) for each atom \(I\).
The important property is that the projectors are localized in space, so the integral in Eq.~\ref{eq:vnl} can be done over a sphere around each atom instead over the whole RS grid.
This means that the cost of applying the pseudopotential is proportional to \(N_\mathrm{atoms}\), a scaling similar to the other components of the KS Hamiltonian.

The problem of applying the pseudopotentials in real space is that there is a spurious dependency of the energy with respect to relative position of the atoms (point particles) and grid points.
This is known as the egg-box effect.
The cause of the problem is the aliasing of the high frequency components of the pseudopotential that cannot be represented in the grid.
When the pseudopotential is applied in Fourier space these components are naturally filtered out.

To control the egg-box effect, most real-space codes use some sort of filtering that removes those high frequency components~\cite{Briggs1996}.
A hard cutoff in Fourier space would not work, as it introduces ripples in real space that destroy the localization of the projectors.
So a more sophisticated approach is needed that yields a localized and soft pseudopotential in real space.
In \inq we use the approach by Tafipolsky and Schmid~\cite{Tafipolsky2006}, that in our experience with \textsc{octopus} has produced good results~\cite{Varsano2009,Andrade2010thesis}.
The filtering process is done once per run and calculated directly in the radial representation of the pseudopotential by the \textsc{pseudopod} library, so it does not add any additional computational cost.


\subsection{Poisson solver}\label{sec:poisson}

The Poisson solver is needed in \inq to obtain the long range ionic potential and the Hartree potential (produced by the electron density itself).
We solve the equation in FS, where the Poisson equation becomes a simple algebraic equation.
To solve the equation in RS, additionally two FFTs are needed to go to FS and back.

In principle the solution obtained in Fourier space has natural periodic boundary conditions.
For finite systems, where neither the density nor the potential is periodic, this method introduces spurious interactions between cells, slowing the convergence with the size of the supercell.
In this case we use a modified kernel that exactly reproduces the free boundary conditions~\cite{Rozzi2006}.
Unfortunately this approach needs the simulation cell to be duplicated in each direction, making the cost of the Poisson solution 8 times more expensive for finite systems.
Even with this factor of 8, this method is quite competitive in comparison with other approaches for free boundary conditions~\cite{Garcia2014}.
Both kinds of boundary conditions are implemented in \inq and can be chosen by the user depending on the physical system.

\subsection{Forces}

An accurate and efficient calculation of the forces is essential for adiabatic and non-adiabatic molecular dynamics.
In the density functional framework the forces are given by the formula
\begin{equation} \label{eq:force}
	\vec{F}_I = -\sum_k\langle\varphi_k|\frac{\partial \hat v_I(\vec{r}-\vec{R}_I)}{\partial \vec{r}}|\varphi_k\rangle\ ,
\end{equation}
Note that for the adiabatic forces in ground-state DFT this expression comes from applying the Hellman-Feynman theorem.
While in the non-adiabatic case, this is directly the expression for the forces since \(\varphi_k\) and \(R_I\) are independent variables~\cite{Andrade2009}.

The problem of Eq.~\ref{eq:force} is that the gradient of the ionic potential appears.
This gradient has higher energy components than the potential itself.
This means that the force calculated using Eq.~\ref{eq:force} would be the bottleneck in the convergence with the basis.
The gradient can also be complicated to calculate since it can have many terms, in particular in the case of spin-orbit coupling.
We can avoid this problem by transforming Eq.~\ref{eq:force} into an expression that contains the gradients of the orbitals instead~\cite{Hirose2005}.
\begin{equation} \label{eq:force_grad}
	\vec{F}_I = \sum_k\langle\nabla\varphi_k|\hat v_I(\vec{r}-\vec{R}_I)|\varphi_k\rangle\ + \mathrm{c.c.}\ .
\end{equation}
Since the orbitals are smoother than the ionic potential, this last expression has better numerical properties as it converges much faster with the basis set.
It is also much simpler to implement, since the derivatives of the ionic potential are not needed.
It only needs the calculation of the gradient of a field that, just like the Laplacian, is a local operation in Fourier space.
The approach has also been extended for second-order derivatives~\cite{Andrade2015}, that will be needed in the calculation of phonon frequencies~\cite{Baroni2001}.

\section{Results}

In this section we show some of the results obtained with the \inq code.
The main objective is to validate the results and show that \inq can be reliably used for production runs.
For these reason all the calculations shown were done on GPUs, still the CPU version of the code gives the same results to a high precision.
In a future work we will show in detail numerical performance benchmarks, our focus here is on the physical predictions.
We start by comparing the results of \inq with other codes using the same simulation parameters, to show the results are practically the same.
Additionally, RT-TDDFT based electronic stopping power which investigates the system response to time-dependent spatially localized perturbations is simulated and compared to previously published results.

\subsection{Validation with other codes}
\begin{figure*}
	\includegraphics[width=\textwidth]{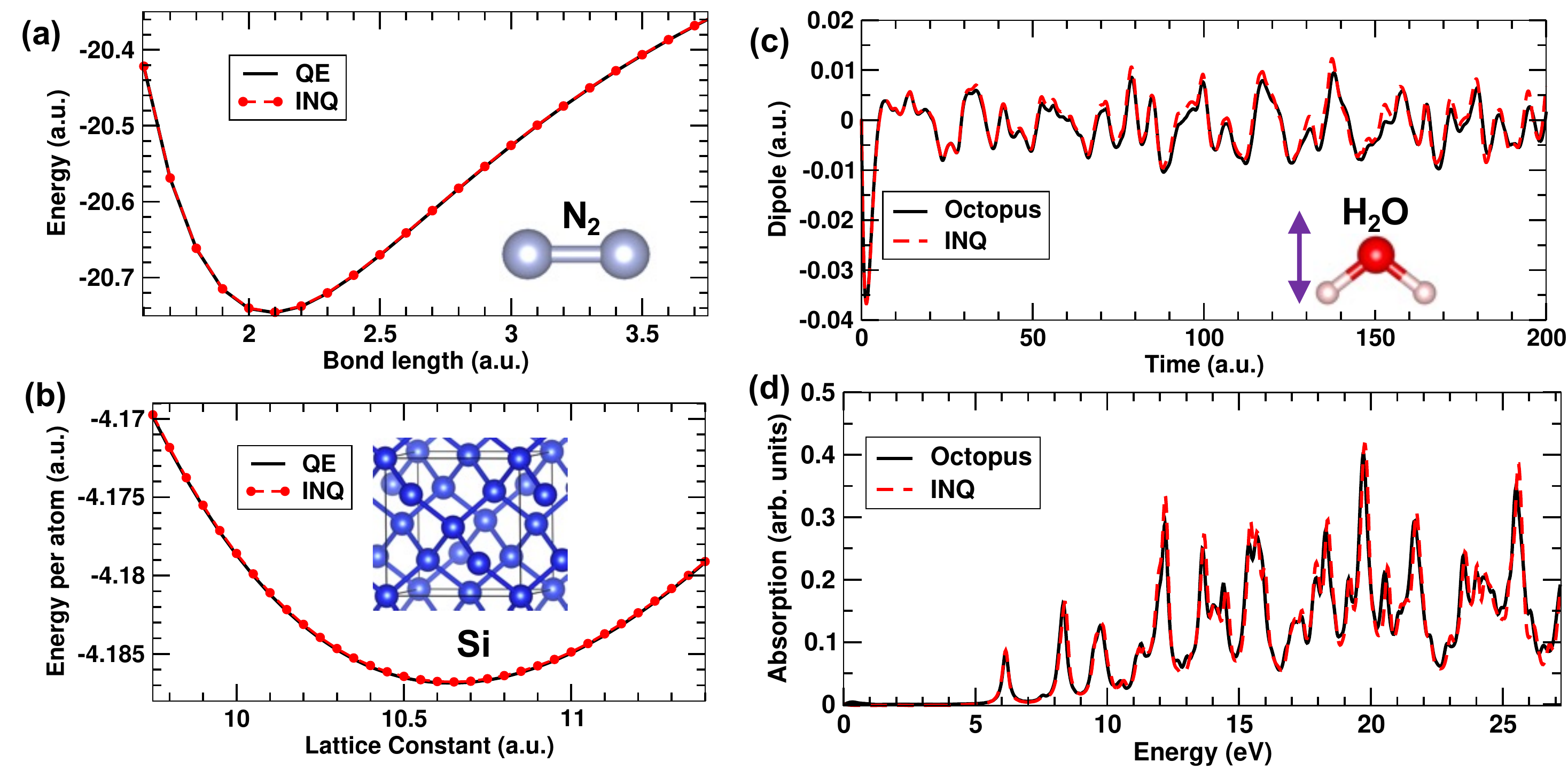}
	\caption{
		\label{fig:comparison}
		Comparison between \inq and the established electronic structure codes \textsc{quantum espresso} (\textsc{QE}) and~\textsc{octopus}.
    Results for ground state total energies and real-time TDDFT optical response are shown.
    (a) Total energy vs bond length in N$_2$ (b) Total energy vs lattice constant in bulk Silicon (c) Time evolution of the dipole moment in gas phase H$_2$O following a ``\textit{kick}" perturbation (d) Linear optical absorption of molecular H$_2$O. In all cases there is a high-level of agreement between the codes.
	}
\end{figure*}

For a scientific code it is fundamental to produce results that are consistent with other approaches and codes.
For that, in this section we compare the results of \inq with other established codes for a few cases.
They illustrate a wider validation work that has been consistently done since we started the development on \inq.
Our ultimate goal is to use the \emph{Delta factor}~\cite{Lejaeghere2016} to assess the reliability of the codes for a large number of systems, however some missing functionality prevents us from doing that at the moment.

\Inq currently implements DFT total energies, forces and real-time propagation of the time-dependent KS equations for both molecular and solid-state systems within a periodic supercell framework.
In the following, results from \inq are compared with corresponding quantities obtained from established plane-wave and real-space grid codes: \textsc{Quantum Espresso} (\textsc{QE})~\cite{Giannozzi2017} and \textsc{octopus}~\cite{Castro2006}.
The quantities we compare are molecular bond lengths and crystal lattice parameters based on total-energy minimization as well as linear optical response from real-time propagation.
The total energy and optical response simulations employed standard norm-conserving scalar relativistic PBE pseudopotentials from Pseudodojo~\cite{Van2018} which are compatible with all of the codes used in the validation. 

Fig~\ref{fig:comparison}(a) shows the total energy of the \(\mathrm{N_2}\) molecule as a function of the bond length as obtained from \inq and QE at a plane-wave cutoff of \(80~\mathrm{a.u.}\).
Fig~\ref{fig:comparison}(b) shows the total energy dependence on the lattice constant for an 8-atom conventional unit cell of bulk Si. A plane-wave cutoff of \(40~\mathrm{a.u.}\) was used in this instance and the Brillouin zone was sampled only at the zone center. 
In each case we find that absolute total energies from \inq are within \(0.3~\mathrm{mHartree/atom}\) of the QE reference. 
The residual difference in total energies is attributed to the real-space treatment of the non-local pseudopotential term within \inq in contrast to the traditional reciprocal space treatment in plane-wave codes such as \textsc{QE}. 
As mentioned in Section~\ref{sec:pp}, the real-space approach is expected to scale more favorably for larger system sizes which \inq aims to target for applications.

In Fig~\ref{fig:comparison}(c),(d), the electronic response of a gas phase water molecule to a time-localized `kick' electric field perturbation is plotted as obtained from \inq and \textsc{octopus}. 
The simulation cell consists of a water molecule enclosed in a \(10~\text{\AA}\times 10~\text{\AA}\times 10~\text{\AA}\) box at a plane-wave cutoff of \(40~\mathrm{a.u.}\) 
Within \inq following the perturbation applied at the initial time \(t = 0\), the time-dependent dipole moment of the system is recorded as given by the first  moment of the time-evolving charge density distribution. 
In \textsc{octopus} the time-dependent current is first calculated within the velocity gauge RT-TDDFT approach~\cite{Yabana2012} and the dipole is then estimated as the time integral of the current. 
It is clear from Fig~\ref{fig:comparison}(c) which shows the real-time evolution of the dipole moment for a perturbation along the \(C_2\) symmetry axis of \(\mathrm{H_2O}\) that the \inq and \textsc{octopus} results are consistent as expected for a finite system. 
The above procedure is repeated for perturbations applied along three orthogonal coordinate directions and subsequently, from the Fourier transform of the time-dependent dipole moments, the optical absorption spectrum is calculated and plotted in Fig~\ref{fig:comparison}(d). 
It is apparent that excitation frequencies predicted by \inq and \textsc{octopus} are in good agreement over a wide frequency range, validating the real-time evolution algorithm implemented in \inq. 
 
\subsection{Electronic stopping}

Stopping power in materials is a fundamental physical process by which energetic particles penetrating matter are slowed down by generating excitations of the material media, such as atomic displacements, phonons, electron-holes, secondary electrons, or plasmons~\cite{Sigmund2006}.
If these fast particles move at velocities that are at least a fraction of that of the electrons in the materials (e.g. the Fermi velocity) the dissipative process is dominated by a continuum of electronic excitations.
For example, this can happen as a result of an energetic nuclear event or by artificial ion accelerators, 
Electronic stopping power results are typically condensed into curves that relate dissipation rate, energy loss per unit distance, versus projectile velocity.
Extensive databases of experimental curves exists in the literature~\cite{Ziegler1985}.
Besides being a quantity of great importance in nuclear technology~\cite{Zhang2016}, ion implantation~\cite{Haussalo1996}, and medicine~\cite{Caporaso2009}, the modeling of electronic stopping power is deeply intertwined with the development of the atomistic theory of matter~\cite{Thomson1912, Darwin1912, Bohr1913} and the electron gas~\cite{Bethe1930, Fermi1947, Lindhard1964}.

\begin{figure}
\includegraphics[width=\columnwidth]{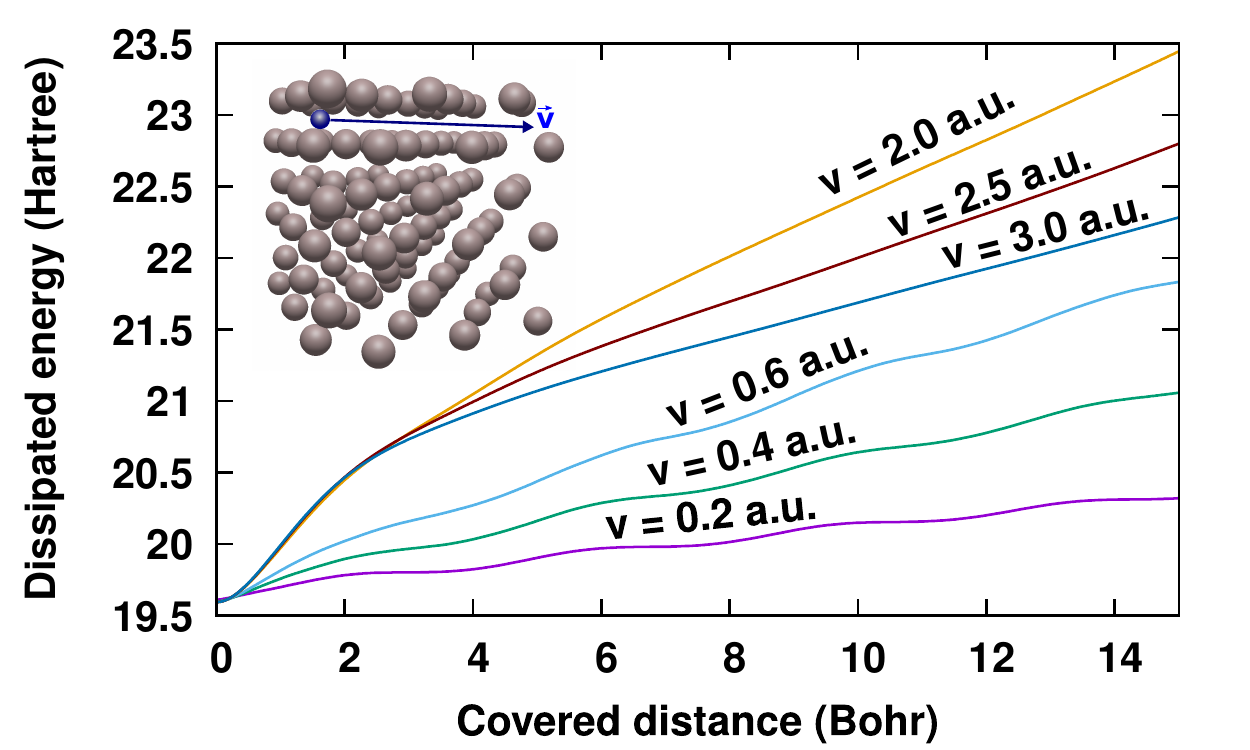}
\caption{
	\label{fig:al_td_overlay}
	Energy dissipation as a function of distance covered by a proton in a rectilinear trajectory in the \(\langle 100\rangle\) channel for different velocities in an aluminum fcc supercell as calculated by a real-time calculation with the \inq code.
	Note that at low velocity the dissipation rate is proportional to the velocity; instead, above a certain velocity the dissipation rate decays.
	Since the ions in the host material are fixed the dissipation rate in the steady state is defined as the electronic stopping power for proton in Aluminum shown in Fig.~\ref{fig:al_stopping}.
}
\end{figure}

\begin{figure}
\includegraphics[width=\columnwidth]{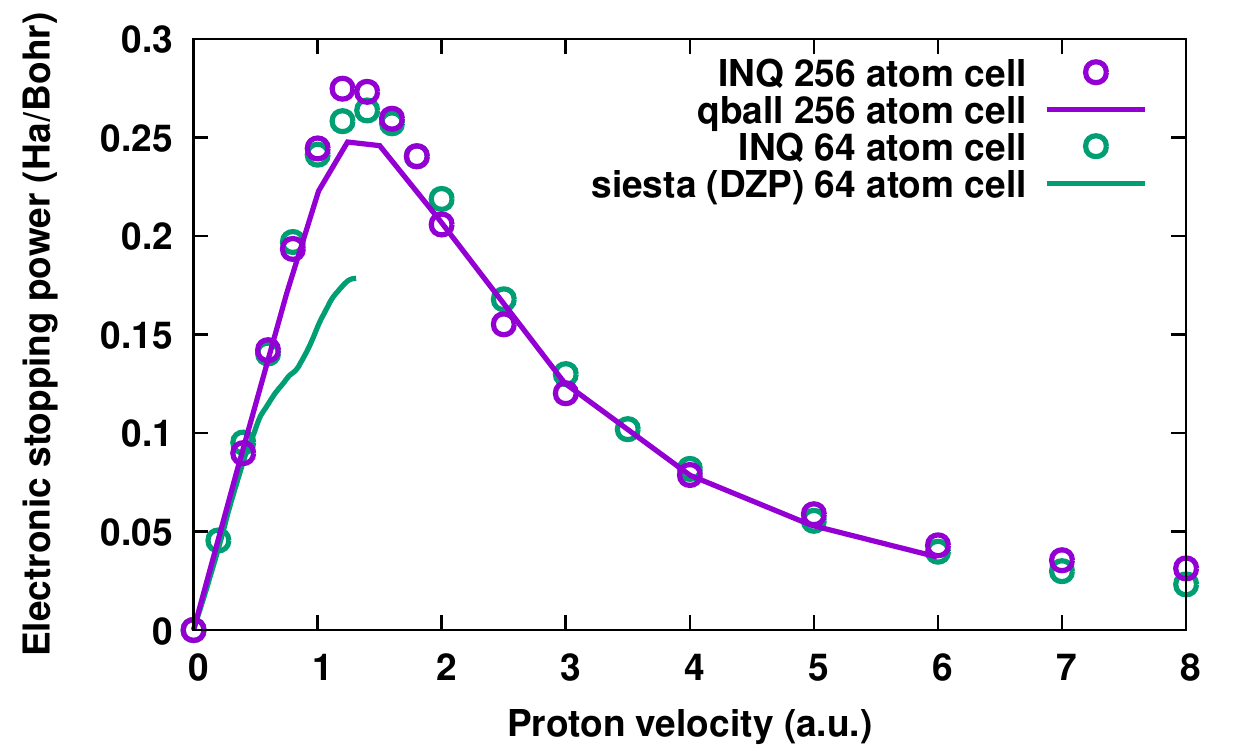}
\caption{
	\label{fig:al_stopping}
	Electronic stopping power obtained by direct simulation of a proton projectile moving in aluminum atom fcc super cell with different codes: \inq, \textsc{qball} (from Ref.~\cite{Schleife2015}) and \textsc{siesta} with atom-centered double-zeta plus polarization basis (DZP) (from Ref.~\cite{Correa2012}).
	Each point is obtained by evaluating the steady energy rate dissipation, e.g from Fig.~\ref{fig:al_td_overlay}.
	All simulations are with 3 valence (explicitly simulated) electrons per aluminum which is enough for channeling stopping power.
}
\end{figure}

Electronic stopping power is, from the point of view of the ion dynamics, a fundamentally non-adiabatic and dissipative process.
As such, the process can in principle be tackled by a direct simulations based on real-time TDDFT.
If we imagine the process as a specific realization of a particle (projectile) traversing tens or hundreds of lattice parameters in a well defined trajectory, we realize that the simulation naturally calls for a large supercell.
The interaction between the projectile and the electron gas is simultaneously an aperiodic localized perturbation in space and also it is extended spatially as the simulation progresses.
Therefore simulations of stopping power had become one of the most important application of \emph{large scale} TDDFT simulations, 
both for its predictive power and as a benchmark of different approximations of TDDFT.
By varying a single parameter (projectile velocity) different technical limitations of the methods can be reached;
for example:
i) the position of the maximum of the stopping curve is sensitive to the completeness of the basis set~\cite{Correa2012}, 
ii) the asymptotic behavior is affected by supercell size effects~\cite{Schleife2015}
iii) accuracy of the low velocity limit is affect by the ability of simulating long times~\cite{Quashie2016}
and iv) highly charged projectiles (high \(Z\)) probe deep core electrons and the pseudopotential approximation~\cite{Ullah2018}.
For a full review and details on this type of calculations see Ref.~\cite{Correa2018}.

In this section we present a benchmark calculation of stopping power in the prototypical case of a proton projectile in FCC aluminum.
Briefly, stopping power is extracted from simulation by forcing the projectile ion to move across a well defined direction in the supercell and monitoring the additional energy gained by the system in the process.
The average rate of increase in the energy corresponds to the kinetic energy loss by the projectile to the electronic system.
The rate is generally obtained by linear fitting a energy-distance curve resulting from the real-time simulation~\cite{Pruneda2007}.

We use a plane-wave cutoff of \(25~\mathrm{a.u.}\) to simulate 3 valence electrons per aluminum ion.
Figure~\ref{fig:al_td_overlay} shows how the energy transfer achieves a steady state whose slope corresponds to the stopping power at that velocity.
This information is collected as an electronic stopping curve (Fig.~\ref{fig:al_stopping}) of energy dissipation per unit length as a function of velocity.

These electronic stopping results illustrate that \inq is ready for state-of-the-art large-scale time-dependent computations, retaining accuracy for high-energy excitation processes  such as in stopping power and other particle radiation-related applications thanks to a robust plane-wave basis.

\section{The future of \inq}

\inq is designed to be extensible.
Not by any means a finished code, \inq allows new features to be added easily so as to tackle scientifically challenging problems that cannot be addressed with current codes. 
These problems might require a unique combination of functionalities not found in a single existing code base, or might require modification of current algorithms or implementation of new ones. 

The support of \emph{exact exchange}~\cite{Lin2016,Jia2019,Carnimeo2019} is planned for a future release of \inq. 
Exact exchange is needed for the accurate description of many transition metal oxides. 
Manganite perovskites~\cite{Ehrke2011,Li2013,Rajpurohit2020} and vanadium oxides \(\mathrm{VO_2}\) \cite{Morrison2014, Otto2019} undergo metal-insulator transitions following optical excitation, likely associated with the emergence of ferromagnetic correlations on sub-picosecond timescales.
Similarly, the melting of antiferromagnetic order in photo-excited nickelates within a few picoseconds~\cite{Caviglia2013} is suggested to be directly linked to the melting of charge order.
The combination of exact exchange and GPU scalability is needed for real-time TDDFT simulations at sizes and timescales required to investigate coherent dynamics in these systems. 
These simulations will clarify the roles of defects, dynamic disorder, and phonons beyond the accuracy of simplified models~\cite{Filippis2012,Werner2015,Kohler2018,Stolpp2020,Rajpurohit2020a}. This will enable comparisons with a broad range of experimental techniques that now exist for probing these coupled electronic and structural dynamics including time-resolved optical and terahertz spectroscopy~\cite{Stoica2019,Guzelturk2020}, as well as ultrafast x-ray and electron scattering approaches~\cite{Trigo2013,Sie2019}.

Support of relativistic effects will also be included in a future release of \inq. 
The spin-orbit interaction is important in topological materials and in systems with heavy atoms with effects such as Rashba/Dresselhaus splittings~\cite{Dresselhaus1955,Bychkov1984}. 
Other relativistic effects, such as Fermi contact and spin-dipole interactions~\cite{Helgaker2000,Melo2005} are often neglected in solid-state DFT simulations, but can be of importance in quantum information science applications~\cite{Gugler2018}. 
\Inq provides a framework where these effects can be implemented in an incremental and unintrusive way. 
As an example application, two dimensional layered van der Waals systems with 5d heavy transition metals are a materials class where electron-electron interactions and the spin-orbit coupling are important energy scales. 
These materials are known to exhibit new broken symmetry or topological phases, including charge density waves and  metastable metallic states \cite{Stojchevska177,Lee2019,Siddiqui2020,Ligges2018,Storeck2019}.
Real-time TDDFT will clarify details of the mechanisms behind the evolution of charge instabilities, periodic lattice distortions, and coherent phonon generation in these 2D layered systems. 
In systems with antiferromagnetic order, magnon modes may emerge as collective excitations. 
Twisted boundary conditions are often used to simulate such systems and those with incommensurate spiral physics in general~\cite{Starykh2015}. 
These capabilities can be included in \inq by modifying the currently available boundary conditions~\ref{sec:poisson}.

\inq supports Ehrenfest dynamics, and can be extended to other forms of non-adiabatic molecular dynamics~\cite{Tapavicza2013,Curchod2018}. 
A variety of proposed dynamics departing from Born-Oppenheimer (BO) and beyond Ehrenfest would be faster to experiment with.
For example, it has been noted recently that friction terms can be added \emph{ab initio} to model electron-phonon coupling with classical ions.
These dissipative terms can be calculated on-the-fly by launching \emph{tangential} short TDDFT simulations (in parallel to the main trajectory evolution).
The dissipative contributions can then be added to the Newton's equation of motion by terms in the form of \(M \ddot{\vec{R}} = \vec{F}_\text{BO} - \beta\dot{\vec{R}} + \vec{\xi}\), where is \(\beta\) and \(\xi\) are a friction tensor and correlated noise respectively~\cite{Tamm2018}.

These functionalities are required for applications that involve any form of coupled electron-ion dynamics, and become especially useful in challenging materials systems that also require advanced electronic structure methods such as those above. 
For instance, the bulk photovoltaic effect involves a number of nonlinear photo-current generation effects driven by different mechanisms~\cite{Belinicher1980,Sipe2000,Tan2016,Fregoso2018,Andrade2018,Parker2019}. 
Recent tight binding model simulations have predicted an enhanced bulk photovoltaic effect due to coupled spin and phonon dynamics in a strongly correlated manganite perovskite~\cite{Rajpurohit2021}.
First-principles exploration and comparison of this enhancement effect across a variety of materials systems will result in materials design insights for maximizing its magnitude.

\Inq is written with the intention that the user can take an active role in developing electronic structure calculations according to their needs. 
While the above examples provide some possible future directions, the user is in fact limited only by their imagination when it comes to new feature development.

\section{Conclusion}

We have presented a new framework for the computational simulation of electronic systems.
It has several new design characteristics that offer unique advantages over existing legacy DFT codes.
Using a modern approach to coding, based on the \CC language, has allowed us to write a very compact code that directly expresses the formulation of the problem and not the implementation details.
When combined with an extensive use of testing of the code, we can develop and implement new features very fast.
The code was designed from scratch to work with GPUs and \textsc{MPI} parallelization, which means it can make use of modern supercomputing platforms and be quickly adapted for future ones.
Within the scope of basic DFT, DFT-MD, and TDDFT the code is production ready.

This collection of features make \inq an ideal platform to apply TDDFT to a range of physical problems that are hard to approach with current software.
Of course, this will require further development of the code and theoretical tools by \inq developers and the broader electronic structure community.
In this spirit, we look to make \inq an open platform that other researchers can use and adapt for their research, and that can interact with other software components.

\begin{acknowledgments}

We thank M. Morales, M. Dewing, E. Draeger, F. Gygi, and D. Strubbe, for insightful interactions.
We also acknowledge M.\,A.\,L. Marques and S. Lehtola for their advice in implementing GPU support in \textsc{libxc}.

The work was supported by the Center for Non-Perturbative Studies of Functional Materials Under Non-Equilibrium Conditions (NPNEQ) funded by the Computational Materials Sciences Program of the US Department of Energy, Office of Science, Basic Energy Sciences, Materials Sciences and Engineering Division. 
Work by X.A, T.O and A.C was performed under the auspices of the U.S. Department of Energy by Lawrence Livermore National Laboratory under Contract DE-AC52-07NA27344. C.D.P, A.K, J.X and A.L were supported by the U.S. Department of Energy, Office of Basic Energy Sciences, Division of Materials Sciences and Engineering, under Contract No. DE-AC02-76SF00515 at SLAC. 
L.Z.T. was supported by the Molecular Foundry, a DOE Office of Science User Facility supported by the Office of Science of the U.S. Department of Energy under Contract No. DE-AC02-05CH11231. 
Computing support for this work came from the Lawrence Livermore National Laboratory Institutional Computing Grand Challenge program.

\end{acknowledgments}

\bibliographystyle{achemso}
\bibliography{\jobname}

\providecommand{\latin}[1]{#1}
\makeatletter
\providecommand{\doi}
  {\begingroup\let\do\@makeother\dospecials
  \catcode`\{=1 \catcode`\}=2 \doi@aux}
\providecommand{\doi@aux}[1]{\endgroup\texttt{#1}}
\makeatother
\providecommand*\mcitethebibliography{\thebibliography}
\csname @ifundefined\endcsname{endmcitethebibliography}
  {\let\endmcitethebibliography\endthebibliography}{}
\begin{mcitethebibliography}{190}
\providecommand*\natexlab[1]{#1}
\providecommand*\mciteSetBstSublistMode[1]{}
\providecommand*\mciteSetBstMaxWidthForm[2]{}
\providecommand*\mciteBstWouldAddEndPuncttrue
  {\def\EndOfBibitem{\unskip.}}
\providecommand*\mciteBstWouldAddEndPunctfalse
  {\let\EndOfBibitem\relax}
\providecommand*\mciteSetBstMidEndSepPunct[3]{}
\providecommand*\mciteSetBstSublistLabelBeginEnd[3]{}
\providecommand*\EndOfBibitem{}
\mciteSetBstSublistMode{f}
\mciteSetBstMaxWidthForm{subitem}{(\alph{mcitesubitemcount})}
\mciteSetBstSublistLabelBeginEnd
  {\mcitemaxwidthsubitemform\space}
  {\relax}
  {\relax}

\bibitem[Hohenberg and Kohn(1964)Hohenberg, and Kohn]{Hohenberg_1964}
Hohenberg,~P.; Kohn,~W. Inhomogeneous Electron Gas. \emph{Phys. Rev.}
  \textbf{1964}, \emph{136}, B864--B871\relax
\mciteBstWouldAddEndPuncttrue
\mciteSetBstMidEndSepPunct{\mcitedefaultmidpunct}
{\mcitedefaultendpunct}{\mcitedefaultseppunct}\relax
\EndOfBibitem
\bibitem[Kohn and Sham(1965)Kohn, and Sham]{Kohn_1965}
Kohn,~W.; Sham,~L.~J. Self-Consistent Equations Including Exchange and
  Correlation Effects. \emph{Phys. Rev.} \textbf{1965}, \emph{140},
  A1133--A1138\relax
\mciteBstWouldAddEndPuncttrue
\mciteSetBstMidEndSepPunct{\mcitedefaultmidpunct}
{\mcitedefaultendpunct}{\mcitedefaultseppunct}\relax
\EndOfBibitem
\bibitem[Kohn(1999)]{Kohn_1999}
Kohn,~W. Nobel Lecture: Electronic structure of matter---wave functions and
  density functionals. \emph{Rev. Mod. Phys.} \textbf{1999}, \emph{71},
  1253--1266\relax
\mciteBstWouldAddEndPuncttrue
\mciteSetBstMidEndSepPunct{\mcitedefaultmidpunct}
{\mcitedefaultendpunct}{\mcitedefaultseppunct}\relax
\EndOfBibitem
\bibitem[Runge and Gross(1984)Runge, and Gross]{Runge_1984}
Runge,~E.; Gross,~E. K.~U. Density-Functional Theory for Time-Dependent
  Systems. \emph{Phys. Rev. Lett.} \textbf{1984}, \emph{52}, 997--1000\relax
\mciteBstWouldAddEndPuncttrue
\mciteSetBstMidEndSepPunct{\mcitedefaultmidpunct}
{\mcitedefaultendpunct}{\mcitedefaultseppunct}\relax
\EndOfBibitem
\bibitem[Hafner \latin{et~al.}(2006)Hafner, Wolverton, and Ceder]{Hafner2006}
Hafner,~J.; Wolverton,~C.; Ceder,~G. Toward computational materials design: the
  impact of density functional theory on materials research. \emph{MRS
  bulletin} \textbf{2006}, \emph{31}, 659--668\relax
\mciteBstWouldAddEndPuncttrue
\mciteSetBstMidEndSepPunct{\mcitedefaultmidpunct}
{\mcitedefaultendpunct}{\mcitedefaultseppunct}\relax
\EndOfBibitem
\bibitem[Burke(2012)]{Burke2012}
Burke,~K. Perspective on density functional theory. \emph{J. Chem. Phys.}
  \textbf{2012}, \emph{136}, 150901\relax
\mciteBstWouldAddEndPuncttrue
\mciteSetBstMidEndSepPunct{\mcitedefaultmidpunct}
{\mcitedefaultendpunct}{\mcitedefaultseppunct}\relax
\EndOfBibitem
\bibitem[Mardirossian and Head-Gordon(2017)Mardirossian, and
  Head-Gordon]{Mardirossian2017}
Mardirossian,~N.; Head-Gordon,~M. Thirty years of density functional theory in
  computational chemistry: an overview and extensive assessment of 200 density
  functionals. \emph{Mol. Phys.} \textbf{2017}, \emph{115}, 2315--2372\relax
\mciteBstWouldAddEndPuncttrue
\mciteSetBstMidEndSepPunct{\mcitedefaultmidpunct}
{\mcitedefaultendpunct}{\mcitedefaultseppunct}\relax
\EndOfBibitem
\bibitem[Hasnip \latin{et~al.}(2014)Hasnip, Refson, Probert, Yates, Clark, and
  Pickard]{Hasnip2014}
Hasnip,~P.~J.; Refson,~K.; Probert,~M.~I.; Yates,~J.~R.; Clark,~S.~J.;
  Pickard,~C.~J. Density functional theory in the solid state. \emph{Philos.
  Trans. R. Soc. A} \textbf{2014}, \emph{372}, 20130270\relax
\mciteBstWouldAddEndPuncttrue
\mciteSetBstMidEndSepPunct{\mcitedefaultmidpunct}
{\mcitedefaultendpunct}{\mcitedefaultseppunct}\relax
\EndOfBibitem
\bibitem[Hehre \latin{et~al.}(1969)Hehre, Stewart, and Pople]{Hehre1969}
Hehre,~W.~J.; Stewart,~R.~F.; Pople,~J.~A. Self-consistent molecular-orbital
  methods. I. Use of Gaussian expansions of Slater-type atomic orbitals.
  \emph{J. Chem. Phys.} \textbf{1969}, \emph{51}, 2657--2664\relax
\mciteBstWouldAddEndPuncttrue
\mciteSetBstMidEndSepPunct{\mcitedefaultmidpunct}
{\mcitedefaultendpunct}{\mcitedefaultseppunct}\relax
\EndOfBibitem
\bibitem[Chelikowsky \latin{et~al.}(1994)Chelikowsky, Troullier, and
  Saad]{Chelikowsky1994}
Chelikowsky,~J.~R.; Troullier,~N.; Saad,~Y. Finite-difference-pseudopotential
  method: Electronic structure calculations without a basis. \emph{Phys. Rev.
  Lett.} \textbf{1994}, \emph{72}, 1240\relax
\mciteBstWouldAddEndPuncttrue
\mciteSetBstMidEndSepPunct{\mcitedefaultmidpunct}
{\mcitedefaultendpunct}{\mcitedefaultseppunct}\relax
\EndOfBibitem
\bibitem[Briggs \latin{et~al.}(1996)Briggs, Sullivan, and Bernholc]{Briggs1996}
Briggs,~E.; Sullivan,~D.; Bernholc,~J. Real-space multigrid-based approach to
  large-scale electronic structure calculations. \emph{Phys. Rev. B}
  \textbf{1996}, \emph{54}, 14362\relax
\mciteBstWouldAddEndPuncttrue
\mciteSetBstMidEndSepPunct{\mcitedefaultmidpunct}
{\mcitedefaultendpunct}{\mcitedefaultseppunct}\relax
\EndOfBibitem
\bibitem[Kresse and Furthm{\"u}ller(1996)Kresse, and
  Furthm{\"u}ller]{Kresse1996}
Kresse,~G.; Furthm{\"u}ller,~J. Efficient iterative schemes for ab initio
  total-energy calculations using a plane-wave basis set. \emph{Phys. Rev. B}
  \textbf{1996}, \emph{54}, 11169\relax
\mciteBstWouldAddEndPuncttrue
\mciteSetBstMidEndSepPunct{\mcitedefaultmidpunct}
{\mcitedefaultendpunct}{\mcitedefaultseppunct}\relax
\EndOfBibitem
\bibitem[Soler \latin{et~al.}(2002)Soler, Artacho, Gale, Garc{\'\i}a, Junquera,
  Ordej{\'o}n, and S{\'a}nchez-Portal]{Soler2002}
Soler,~J.~M.; Artacho,~E.; Gale,~J.~D.; Garc{\'\i}a,~A.; Junquera,~J.;
  Ordej{\'o}n,~P.; S{\'a}nchez-Portal,~D. The SIESTA method for ab initio
  order-N materials simulation. \emph{J. Phys.: Cond. Matt.} \textbf{2002},
  \emph{14}, 2745\relax
\mciteBstWouldAddEndPuncttrue
\mciteSetBstMidEndSepPunct{\mcitedefaultmidpunct}
{\mcitedefaultendpunct}{\mcitedefaultseppunct}\relax
\EndOfBibitem
\bibitem[Castro \latin{et~al.}(2006)Castro, Appel, Oliveira, Rozzi, Andrade,
  Lorenzen, Marques, Gross, and Rubio]{Castro2006}
Castro,~A.; Appel,~H.; Oliveira,~M.; Rozzi,~C.~A.; Andrade,~X.; Lorenzen,~F.;
  Marques,~M.~A.; Gross,~E.; Rubio,~A. Octopus: a tool for the application of
  time-dependent density functional theory. \emph{Phys. Stat. Sol. (b)}
  \textbf{2006}, \emph{243}, 2465--2488\relax
\mciteBstWouldAddEndPuncttrue
\mciteSetBstMidEndSepPunct{\mcitedefaultmidpunct}
{\mcitedefaultendpunct}{\mcitedefaultseppunct}\relax
\EndOfBibitem
\bibitem[Gygi(2008)]{Gygi2008}
Gygi,~F. Architecture of Qbox: A scalable first-principles molecular dynamics
  code. \emph{{IBM} J. Res. Dev.} \textbf{2008}, \emph{52}, 137--144\relax
\mciteBstWouldAddEndPuncttrue
\mciteSetBstMidEndSepPunct{\mcitedefaultmidpunct}
{\mcitedefaultendpunct}{\mcitedefaultseppunct}\relax
\EndOfBibitem
\bibitem[Blum \latin{et~al.}(2009)Blum, Gehrke, Hanke, Havu, Havu, Ren, Reuter,
  and Scheffler]{Blum2009}
Blum,~V.; Gehrke,~R.; Hanke,~F.; Havu,~P.; Havu,~V.; Ren,~X.; Reuter,~K.;
  Scheffler,~M. Ab initio molecular simulations with numeric atom-centered
  orbitals. \emph{Comput. Phys. Comm.} \textbf{2009}, \emph{180},
  2175--2196\relax
\mciteBstWouldAddEndPuncttrue
\mciteSetBstMidEndSepPunct{\mcitedefaultmidpunct}
{\mcitedefaultendpunct}{\mcitedefaultseppunct}\relax
\EndOfBibitem
\bibitem[Enkovaara \latin{et~al.}(2010)Enkovaara, Rostgaard, Mortensen, Chen,
  Du{\l}ak, Ferrighi, Gavnholt, Glinsvad, Haikola, Hansen, \latin{et~al.}
  others]{Enkovaara2010}
Enkovaara,~J.; Rostgaard,~C.; Mortensen,~J.~J.; Chen,~J.; Du{\l}ak,~M.;
  Ferrighi,~L.; Gavnholt,~J.; Glinsvad,~C.; Haikola,~V.; Hansen,~H.,
  \latin{et~al.}  Electronic structure calculations with GPAW: a real-space
  implementation of the projector augmented-wave method. \emph{J. Phys.: Cond.
  Matt.} \textbf{2010}, \emph{22}, 253202\relax
\mciteBstWouldAddEndPuncttrue
\mciteSetBstMidEndSepPunct{\mcitedefaultmidpunct}
{\mcitedefaultendpunct}{\mcitedefaultseppunct}\relax
\EndOfBibitem
\bibitem[Shao \latin{et~al.}(2015)Shao, Gan, Epifanovsky, Gilbert, Wormit,
  Kussmann, Lange, Behn, Deng, Feng, \latin{et~al.} others]{Shao2015}
Shao,~Y.; Gan,~Z.; Epifanovsky,~E.; Gilbert,~A.~T.; Wormit,~M.; Kussmann,~J.;
  Lange,~A.~W.; Behn,~A.; Deng,~J.; Feng,~X., \latin{et~al.}  Advances in
  molecular quantum chemistry contained in the Q-Chem 4 program package.
  \emph{Mol. Phys.} \textbf{2015}, \emph{113}, 184--215\relax
\mciteBstWouldAddEndPuncttrue
\mciteSetBstMidEndSepPunct{\mcitedefaultmidpunct}
{\mcitedefaultendpunct}{\mcitedefaultseppunct}\relax
\EndOfBibitem
\bibitem[Genova \latin{et~al.}(2017)Genova, Ceresoli, Krishtal, Andreussi,
  DiStasio~Jr, and Pavanello]{Genova2017}
Genova,~A.; Ceresoli,~D.; Krishtal,~A.; Andreussi,~O.; DiStasio~Jr,~R.~A.;
  Pavanello,~M. eQE: An open-source density functional embedding theory code
  for the condensed phase. \emph{Int. J. Quant. Chem.} \textbf{2017},
  \emph{117}, e25401\relax
\mciteBstWouldAddEndPuncttrue
\mciteSetBstMidEndSepPunct{\mcitedefaultmidpunct}
{\mcitedefaultendpunct}{\mcitedefaultseppunct}\relax
\EndOfBibitem
\bibitem[Draeger \latin{et~al.}(2017)Draeger, Andrade, Gunnels, Bhatele,
  Schleife, and Correa]{Draeger2017}
Draeger,~E.~W.; Andrade,~X.; Gunnels,~J.~A.; Bhatele,~A.; Schleife,~A.;
  Correa,~A.~A. Massively parallel first-principles simulation of electron
  dynamics in materials. \emph{J. Parallel Distrib. Comput.} \textbf{2017},
  \emph{106}, 205--214\relax
\mciteBstWouldAddEndPuncttrue
\mciteSetBstMidEndSepPunct{\mcitedefaultmidpunct}
{\mcitedefaultendpunct}{\mcitedefaultseppunct}\relax
\EndOfBibitem
\bibitem[Giannozzi \latin{et~al.}(2017)Giannozzi, Andreussi, Brumme, Bunau,
  Nardelli, Calandra, Car, Cavazzoni, Ceresoli, Cococcioni, Colonna, Carnimeo,
  Corso, de~Gironcoli, Delugas, DiStasio, Ferretti, Floris, Fratesi, Fugallo,
  Gebauer, Gerstmann, Giustino, Gorni, Jia, Kawamura, Ko, Kokalj,
  Kü{\c{c}}ükbenli, Lazzeri, Marsili, Marzari, Mauri, Nguyen, Nguyen, de-la
  Roza, Paulatto, Ponc{\'{e}}, Rocca, Sabatini, Santra, Schlipf, Seitsonen,
  Smogunov, Timrov, Thonhauser, Umari, Vast, Wu, and Baroni]{Giannozzi2017}
Giannozzi,~P. \latin{et~al.}  Advanced capabilities for materials modelling
  with Quantum {ESPRESSO}. \emph{Journal of Physics: Condensed Matter}
  \textbf{2017}, \emph{29}, 465901\relax
\mciteBstWouldAddEndPuncttrue
\mciteSetBstMidEndSepPunct{\mcitedefaultmidpunct}
{\mcitedefaultendpunct}{\mcitedefaultseppunct}\relax
\EndOfBibitem
\bibitem[Noda \latin{et~al.}(2019)Noda, Sato, Hirokawa, Uemoto, Takeuchi,
  Yamada, Yamada, Shinohara, Yamaguchi, Iida, \latin{et~al.} others]{Noda2019}
Noda,~M.; Sato,~S.~A.; Hirokawa,~Y.; Uemoto,~M.; Takeuchi,~T.; Yamada,~S.;
  Yamada,~A.; Shinohara,~Y.; Yamaguchi,~M.; Iida,~K., \latin{et~al.}  Salmon:
  Scalable ab-initio light--matter simulator for optics and nanoscience.
  \emph{Comput. Phys. Comm.} \textbf{2019}, \emph{235}, 356--365\relax
\mciteBstWouldAddEndPuncttrue
\mciteSetBstMidEndSepPunct{\mcitedefaultmidpunct}
{\mcitedefaultendpunct}{\mcitedefaultseppunct}\relax
\EndOfBibitem
\bibitem[Apr\'a \latin{et~al.}(2020)Apr\'a, Bylaska, de~Jong, Govind, Kowalski,
  Straatsma, Valiev, van Dam, Alexeev, Anchell, Anisimov, Aquino, Atta-Fynn,
  Autschbach, Bauman, Becca, Bernholdt, Bhaskaran-Nair, Bogatko, Borowski,
  Boschen, Brabec, Bruner, E., Chen, Chuev, Cramer, Daily, Deegan, Dunning,
  Dupuis, Dyall, Fann, Fischer, Fonari, Fr\"uchtl, Gagliardi, Garza, Gawande,
  Ghosh, Glaesemann, G\"otz, Hammond, Helms, Hermes, Hirao, Hirata, Jacquelin,
  Jensen, Johnson, J\'onsson, Kendall, Klemm, Kobayashi, Konkov,
  Krishnamoorthy, Krishnan, Lin, Lins, Littlefield, Logsdail, Lopata, Ma,
  Marenich, Martin~del Campo, Mejia-Rodriguez, Moore, Mullin, Nakajima,
  Nascimento, Nichols, Nichols, Nieplocha, Otero-de-la Roza, Palmer, Panyala,
  Pirojsirikul, Peng, Peverati, Pittner, Pollack, Richard, Sadayappan, Schatz,
  Shelton, Silverstein, Smith, Soares, Song, Swart, Taylor, Thomas, Tipparaju,
  Truhlar, Tsemekhman, Van~Voorhis, V\'azquez-Mayagoitia, Verma, Villa, Vishnu,
  Vogiatzis, Wang, Weare, Williamson, Windus, Woli\'nski, Wong, Wu, Yang, Yu,
  Zacharias, Zhang, Zhao, and Harrison]{Apra2020}
Apr\'a,~E. \latin{et~al.}  NWChem: Past, present, and future. \emph{J. Chem.
  Phys.} \textbf{2020}, \emph{152}, 184102\relax
\mciteBstWouldAddEndPuncttrue
\mciteSetBstMidEndSepPunct{\mcitedefaultmidpunct}
{\mcitedefaultendpunct}{\mcitedefaultseppunct}\relax
\EndOfBibitem
\bibitem[K{\"u}hne \latin{et~al.}(2020)K{\"u}hne, Iannuzzi, Del~Ben, Rybkin,
  Seewald, Stein, Laino, Khaliullin, Sch{\"u}tt, Schiffmann, \latin{et~al.}
  others]{Kuhne2020}
K{\"u}hne,~T.~D.; Iannuzzi,~M.; Del~Ben,~M.; Rybkin,~V.~V.; Seewald,~P.;
  Stein,~F.; Laino,~T.; Khaliullin,~R.~Z.; Sch{\"u}tt,~O.; Schiffmann,~F.,
  \latin{et~al.}  CP2K: An electronic structure and molecular dynamics software
  package-Quickstep: Efficient and accurate electronic structure calculations.
  \emph{J. Chem. Phys.} \textbf{2020}, \emph{152}, 194103\relax
\mciteBstWouldAddEndPuncttrue
\mciteSetBstMidEndSepPunct{\mcitedefaultmidpunct}
{\mcitedefaultendpunct}{\mcitedefaultseppunct}\relax
\EndOfBibitem
\bibitem[Gonze \latin{et~al.}(2020)Gonze, Amadon, Antonius, Arnardi, Baguet,
  Beuken, Bieder, Bottin, Bouchet, Bousquet, Brouwer, Bruneval, Brunin,
  Cavignac, Charraud, Chen, Côté, Cottenier, Denier, Geneste, Ghosez,
  Giantomassi, Gillet, Gingras, Hamann, Hautier, He, Helbig, Holzwarth, Jia,
  Jollet, Lafargue-Dit-Hauret, Lejaeghere, Marques, Martin, Martins, Miranda,
  Naccarato, Persson, Petretto, Planes, Pouillon, Prokhorenko, Ricci,
  Rignanese, Romero, Schmitt, Torrent, van Setten, Troeye, Verstraete, Zérah,
  and Zwanziger]{Gonze2020}
Gonze,~X. \latin{et~al.}  The Abinit project: Impact, environment and recent
  developments. \emph{Comput. Phys. Commun.} \textbf{2020}, \emph{248},
  107042\relax
\mciteBstWouldAddEndPuncttrue
\mciteSetBstMidEndSepPunct{\mcitedefaultmidpunct}
{\mcitedefaultendpunct}{\mcitedefaultseppunct}\relax
\EndOfBibitem
\bibitem[Seritan \latin{et~al.}(2021)Seritan, Bannwarth, Fales, Hohenstein,
  Isborn, Kokkila-Schumacher, Li, Liu, Luehr, Snyder~Jr, \latin{et~al.}
  others]{Seritan2021}
Seritan,~S.; Bannwarth,~C.; Fales,~B.~S.; Hohenstein,~E.~G.; Isborn,~C.~M.;
  Kokkila-Schumacher,~S.~I.; Li,~X.; Liu,~F.; Luehr,~N.; Snyder~Jr,~J.~W.,
  \latin{et~al.}  TeraChem: A graphical processing unit-accelerated electronic
  structure package for large-scale ab initio molecular dynamics. \emph{Wiley
  Interdiscip. Rev. Comput. Mol. Sci.} \textbf{2021}, \emph{11}, e1494\relax
\mciteBstWouldAddEndPuncttrue
\mciteSetBstMidEndSepPunct{\mcitedefaultmidpunct}
{\mcitedefaultendpunct}{\mcitedefaultseppunct}\relax
\EndOfBibitem
\bibitem[Verma and Truhlar(2020)Verma, and Truhlar]{Verma2020}
Verma,~P.; Truhlar,~D.~G. Status and challenges of density functional theory.
  \emph{Trends Chem.} \textbf{2020}, \emph{2}, 302--318\relax
\mciteBstWouldAddEndPuncttrue
\mciteSetBstMidEndSepPunct{\mcitedefaultmidpunct}
{\mcitedefaultendpunct}{\mcitedefaultseppunct}\relax
\EndOfBibitem
\bibitem[Curtarolo \latin{et~al.}(2013)Curtarolo, Hart, Nardelli, Mingo,
  Sanvito, and Levy]{Curtarolo2013}
Curtarolo,~S.; Hart,~G.~L.; Nardelli,~M.~B.; Mingo,~N.; Sanvito,~S.; Levy,~O.
  The high-throughput highway to computational materials design. \emph{Nat.
  Mat.} \textbf{2013}, \emph{12}, 191--201\relax
\mciteBstWouldAddEndPuncttrue
\mciteSetBstMidEndSepPunct{\mcitedefaultmidpunct}
{\mcitedefaultendpunct}{\mcitedefaultseppunct}\relax
\EndOfBibitem
\bibitem[Oba and Kumagai(2018)Oba, and Kumagai]{Oba2018}
Oba,~F.; Kumagai,~Y. Design and exploration of semiconductors from first
  principles: A review of recent advances. \emph{App. Phys. Express}
  \textbf{2018}, \emph{11}, 060101\relax
\mciteBstWouldAddEndPuncttrue
\mciteSetBstMidEndSepPunct{\mcitedefaultmidpunct}
{\mcitedefaultendpunct}{\mcitedefaultseppunct}\relax
\EndOfBibitem
\bibitem[Marzari \latin{et~al.}(2021)Marzari, Ferretti, and
  Wolverton]{Marzari2021}
Marzari,~N.; Ferretti,~A.; Wolverton,~C. Electronic-structure methods for
  materials design. \emph{Nat. Mater.} \textbf{2021}, \emph{20}, 736--749\relax
\mciteBstWouldAddEndPuncttrue
\mciteSetBstMidEndSepPunct{\mcitedefaultmidpunct}
{\mcitedefaultendpunct}{\mcitedefaultseppunct}\relax
\EndOfBibitem
\bibitem[Oganov and Glass(2006)Oganov, and Glass]{Oganov2006}
Oganov,~A.~R.; Glass,~C.~W. Crystal structure prediction usingab
  initioevolutionary techniques: Principles and applications. \emph{J. Chem.
  Phys.} \textbf{2006}, \emph{124}, 244704\relax
\mciteBstWouldAddEndPuncttrue
\mciteSetBstMidEndSepPunct{\mcitedefaultmidpunct}
{\mcitedefaultendpunct}{\mcitedefaultseppunct}\relax
\EndOfBibitem
\bibitem[Lemonick(2018)]{Lemonick2018}
Lemonick,~S. Is machine learning overhyped? \emph{Chem. Eng. News}
  \textbf{2018}, \emph{96}, 16--20\relax
\mciteBstWouldAddEndPuncttrue
\mciteSetBstMidEndSepPunct{\mcitedefaultmidpunct}
{\mcitedefaultendpunct}{\mcitedefaultseppunct}\relax
\EndOfBibitem
\bibitem[Schmidt \latin{et~al.}(2019)Schmidt, Marques, Botti, and
  Marques]{Schmidt2019}
Schmidt,~J.; Marques,~M.~R.; Botti,~S.; Marques,~M.~A. Recent advances and
  applications of machine learning in solid-state materials science. \emph{npj
  Comput. Mat.} \textbf{2019}, \emph{5}, 1--36\relax
\mciteBstWouldAddEndPuncttrue
\mciteSetBstMidEndSepPunct{\mcitedefaultmidpunct}
{\mcitedefaultendpunct}{\mcitedefaultseppunct}\relax
\EndOfBibitem
\bibitem[Tkatchenko(2020)]{Tkatchenko2020}
Tkatchenko,~A. Machine learning for chemical discovery. \emph{Nat. Comm.}
  \textbf{2020}, \emph{11}, 1--4\relax
\mciteBstWouldAddEndPuncttrue
\mciteSetBstMidEndSepPunct{\mcitedefaultmidpunct}
{\mcitedefaultendpunct}{\mcitedefaultseppunct}\relax
\EndOfBibitem
\bibitem[Li \latin{et~al.}(2021)Li, Hoyer, Pederson, Sun, Cubuk, Riley, Burke,
  \latin{et~al.} others]{Li2021}
Li,~L.; Hoyer,~S.; Pederson,~R.; Sun,~R.; Cubuk,~E.~D.; Riley,~P.; Burke,~K.,
  \latin{et~al.}  Kohn-Sham equations as regularizer: Building prior knowledge
  into machine-learned physics. \emph{Phys. Rev. Lett.} \textbf{2021},
  \emph{126}, 036401\relax
\mciteBstWouldAddEndPuncttrue
\mciteSetBstMidEndSepPunct{\mcitedefaultmidpunct}
{\mcitedefaultendpunct}{\mcitedefaultseppunct}\relax
\EndOfBibitem
\bibitem[Shaidu \latin{et~al.}(2021)Shaidu, K{\"u}{\c{c}}{\"u}kbenli, Lot,
  Pellegrini, Kaxiras, and de~Gironcoli]{Shaidu2021}
Shaidu,~Y.; K{\"u}{\c{c}}{\"u}kbenli,~E.; Lot,~R.; Pellegrini,~F.; Kaxiras,~E.;
  de~Gironcoli,~S. A systematic approach to generating accurate neural network
  potentials: the case of carbon. \emph{npj Comput. Mat.} \textbf{2021},
  \emph{7}, 1--13\relax
\mciteBstWouldAddEndPuncttrue
\mciteSetBstMidEndSepPunct{\mcitedefaultmidpunct}
{\mcitedefaultendpunct}{\mcitedefaultseppunct}\relax
\EndOfBibitem
\bibitem[Hachmann \latin{et~al.}(2014)Hachmann, Olivares-Amaya, Jinich,
  Appleton, Blood-Forsythe, Seress, Rom{\'a}n-Salgado, Trepte, Atahan-Evrenk,
  Er, \latin{et~al.} others]{Hachmann2014}
Hachmann,~J.; Olivares-Amaya,~R.; Jinich,~A.; Appleton,~A.~L.;
  Blood-Forsythe,~M.~A.; Seress,~L.~R.; Rom{\'a}n-Salgado,~C.; Trepte,~K.;
  Atahan-Evrenk,~S.; Er,~S., \latin{et~al.}  Lead candidates for
  high-performance organic photovoltaics from high-throughput quantum
  chemistry--the Harvard Clean Energy Project. \emph{Energy \& Environmental
  Science} \textbf{2014}, \emph{7}, 698--704\relax
\mciteBstWouldAddEndPuncttrue
\mciteSetBstMidEndSepPunct{\mcitedefaultmidpunct}
{\mcitedefaultendpunct}{\mcitedefaultseppunct}\relax
\EndOfBibitem
\bibitem[Cohen \latin{et~al.}(2008)Cohen, Mori-S{\'a}nchez, and
  Yang]{Cohen2008}
Cohen,~A.~J.; Mori-S{\'a}nchez,~P.; Yang,~W. Insights into current limitations
  of density functional theory. \emph{Sci.} \textbf{2008}, \emph{321},
  792--794\relax
\mciteBstWouldAddEndPuncttrue
\mciteSetBstMidEndSepPunct{\mcitedefaultmidpunct}
{\mcitedefaultendpunct}{\mcitedefaultseppunct}\relax
\EndOfBibitem
\bibitem[Gori-Giorgi and Seidl(2010)Gori-Giorgi, and Seidl]{Gori2010}
Gori-Giorgi,~P.; Seidl,~M. Density functional theory for strongly-interacting
  electrons: perspectives for physics and chemistry. \emph{Phys. Chem. Chem.
  Phys.} \textbf{2010}, \emph{12}, 14405--14419\relax
\mciteBstWouldAddEndPuncttrue
\mciteSetBstMidEndSepPunct{\mcitedefaultmidpunct}
{\mcitedefaultendpunct}{\mcitedefaultseppunct}\relax
\EndOfBibitem
\bibitem[Jain \latin{et~al.}(2016)Jain, Shin, and Persson]{Jain2016}
Jain,~A.; Shin,~Y.; Persson,~K.~A. Computational predictions of energy
  materials using density functional theory. \emph{Nat. Rev. Mat.}
  \textbf{2016}, \emph{1}, 1--13\relax
\mciteBstWouldAddEndPuncttrue
\mciteSetBstMidEndSepPunct{\mcitedefaultmidpunct}
{\mcitedefaultendpunct}{\mcitedefaultseppunct}\relax
\EndOfBibitem
\bibitem[Ullrich and Yang(2014)Ullrich, and Yang]{Ullrich2014}
Ullrich,~C.~A.; Yang,~Z.-h. A brief compendium of time-dependent density
  functional theory. \emph{Braz. J. Phys.} \textbf{2014}, \emph{44},
  154--188\relax
\mciteBstWouldAddEndPuncttrue
\mciteSetBstMidEndSepPunct{\mcitedefaultmidpunct}
{\mcitedefaultendpunct}{\mcitedefaultseppunct}\relax
\EndOfBibitem
\bibitem[Onida \latin{et~al.}(2002)Onida, Reining, and Rubio]{Onida2002}
Onida,~G.; Reining,~L.; Rubio,~A. Electronic excitations: density-functional
  versus many-body Green's-function approaches. \emph{Rev. Mod. Phys.}
  \textbf{2002}, \emph{74}, 601\relax
\mciteBstWouldAddEndPuncttrue
\mciteSetBstMidEndSepPunct{\mcitedefaultmidpunct}
{\mcitedefaultendpunct}{\mcitedefaultseppunct}\relax
\EndOfBibitem
\bibitem[Castro \latin{et~al.}(2009)Castro, Marques, Varsano, Sottile, and
  Rubio]{Castro2009}
Castro,~A.; Marques,~M.~A.; Varsano,~D.; Sottile,~F.; Rubio,~A. The challenge
  of predicting optical properties of biomolecules: what can we learn from
  time-dependent density-functional theory? \emph{Comptes Rendus Physique}
  \textbf{2009}, \emph{10}, 469--490\relax
\mciteBstWouldAddEndPuncttrue
\mciteSetBstMidEndSepPunct{\mcitedefaultmidpunct}
{\mcitedefaultendpunct}{\mcitedefaultseppunct}\relax
\EndOfBibitem
\bibitem[Elliott \latin{et~al.}(2011)Elliott, Goldson, Canahui, and
  Maitra]{Elliott2011}
Elliott,~P.; Goldson,~S.; Canahui,~C.; Maitra,~N.~T. Perspectives on
  double-excitations in TDDFT. \emph{Chem. Phys.} \textbf{2011}, \emph{391},
  110--119\relax
\mciteBstWouldAddEndPuncttrue
\mciteSetBstMidEndSepPunct{\mcitedefaultmidpunct}
{\mcitedefaultendpunct}{\mcitedefaultseppunct}\relax
\EndOfBibitem
\bibitem[Lian \latin{et~al.}(2018)Lian, Guan, Hu, Zhang, and Meng]{Lian2018}
Lian,~C.; Guan,~M.; Hu,~S.; Zhang,~J.; Meng,~S. Photoexcitation in Solids:
  First-Principles Quantum Simulations by Real-Time TDDFT. \emph{Adv. Theory
  Simul.} \textbf{2018}, \emph{1}, 1800055\relax
\mciteBstWouldAddEndPuncttrue
\mciteSetBstMidEndSepPunct{\mcitedefaultmidpunct}
{\mcitedefaultendpunct}{\mcitedefaultseppunct}\relax
\EndOfBibitem
\bibitem[Izmaylov and Scuseria(2008)Izmaylov, and Scuseria]{Izmaylov2008}
Izmaylov,~A.~F.; Scuseria,~G.~E. Why are time-dependent density functional
  theory excitations in solids equal to band structure energy gaps for
  semilocal functionals, and how does nonlocal Hartree--Fock-type exchange
  introduce excitonic effects? \emph{J. Chem. Phys.} \textbf{2008}, \emph{129},
  034101\relax
\mciteBstWouldAddEndPuncttrue
\mciteSetBstMidEndSepPunct{\mcitedefaultmidpunct}
{\mcitedefaultendpunct}{\mcitedefaultseppunct}\relax
\EndOfBibitem
\bibitem[Ullrich and Yang(2016)Ullrich, and Yang]{Ullrich2016}
Ullrich,~C.~A.; Yang,~Z.-h. In \emph{Density-Functional Methods for Excited
  States}; Ferr{\'e},~N., Filatov,~M., Huix-Rotllant,~M., Eds.; Springer
  International Publishing: Cham, 2016; pp 185--217\relax
\mciteBstWouldAddEndPuncttrue
\mciteSetBstMidEndSepPunct{\mcitedefaultmidpunct}
{\mcitedefaultendpunct}{\mcitedefaultseppunct}\relax
\EndOfBibitem
\bibitem[Refaely-Abramson \latin{et~al.}(2015)Refaely-Abramson, Jain,
  Sharifzadeh, Neaton, and Kronik]{Refaely-Abramson2015a}
Refaely-Abramson,~S.; Jain,~M.; Sharifzadeh,~S.; Neaton,~J.~B.; Kronik,~L.
  {Solid-state optical absorption from optimally tuned time-dependent
  range-separated hybrid density functional theory}. \emph{Phys. Rev. B}
  \textbf{2015}, \emph{92}, 081204\relax
\mciteBstWouldAddEndPuncttrue
\mciteSetBstMidEndSepPunct{\mcitedefaultmidpunct}
{\mcitedefaultendpunct}{\mcitedefaultseppunct}\relax
\EndOfBibitem
\bibitem[K{\"u}mmel(2017)]{Kummel2017}
K{\"u}mmel,~S. Charge-Transfer Excitations: a challenge for time-dependent
  density functional theory that has been met. \emph{Advanced Energy Materials}
  \textbf{2017}, \emph{7}, 1700440\relax
\mciteBstWouldAddEndPuncttrue
\mciteSetBstMidEndSepPunct{\mcitedefaultmidpunct}
{\mcitedefaultendpunct}{\mcitedefaultseppunct}\relax
\EndOfBibitem
\bibitem[Pemmaraju(2019)]{Pemmaraju2018b}
Pemmaraju,~C.~D. {Valence and core excitons in solids from velocity-gauge
  real-time TDDFT with range-separated hybrid functionals: An LCAO approach}.
  \emph{Comput. Cond. Matt.} \textbf{2019}, \emph{18}, e00348\relax
\mciteBstWouldAddEndPuncttrue
\mciteSetBstMidEndSepPunct{\mcitedefaultmidpunct}
{\mcitedefaultendpunct}{\mcitedefaultseppunct}\relax
\EndOfBibitem
\bibitem[Sun \latin{et~al.}(2021)Sun, Lee, Kononov, Schleife, and
  Ullrich]{Sun2021}
Sun,~J.; Lee,~C.-W.; Kononov,~A.; Schleife,~A.; Ullrich,~C.~A. Real-time
  exciton dynamics with time-dependent density-functional theory. \emph{arXiv
  preprint arXiv:2102.01796} \textbf{2021}, \relax
\mciteBstWouldAddEndPunctfalse
\mciteSetBstMidEndSepPunct{\mcitedefaultmidpunct}
{}{\mcitedefaultseppunct}\relax
\EndOfBibitem
\bibitem[Kothe \latin{et~al.}(2019)Kothe, Lee, and Qualters]{Kothe2019}
Kothe,~D.; Lee,~S.; Qualters,~I. Exascale Computing in the United States.
  \emph{Comput. Sci. Eng.} \textbf{2019}, \emph{21}, 17--29\relax
\mciteBstWouldAddEndPuncttrue
\mciteSetBstMidEndSepPunct{\mcitedefaultmidpunct}
{\mcitedefaultendpunct}{\mcitedefaultseppunct}\relax
\EndOfBibitem
\bibitem[Stone \latin{et~al.}(2007)Stone, Phillips, Freddolino, Hardy, Trabuco,
  and Schulten]{Stone2007}
Stone,~J.~E.; Phillips,~J.~C.; Freddolino,~P.~L.; Hardy,~D.~J.; Trabuco,~L.~G.;
  Schulten,~K. Accelerating molecular modeling applications with graphics
  processors. \emph{J. Comput. Chem.} \textbf{2007}, \emph{28},
  2618--2640\relax
\mciteBstWouldAddEndPuncttrue
\mciteSetBstMidEndSepPunct{\mcitedefaultmidpunct}
{\mcitedefaultendpunct}{\mcitedefaultseppunct}\relax
\EndOfBibitem
\bibitem[Biagini \latin{et~al.}(2019)Biagini, Petrizzelli, Truglio, Cespa,
  Barbieri, Capocefalo, Castellana, Tevy, Carella, and Mazza]{Biagini2019}
Biagini,~T.; Petrizzelli,~F.; Truglio,~M.; Cespa,~R.; Barbieri,~A.;
  Capocefalo,~D.; Castellana,~S.; Tevy,~M.~F.; Carella,~M.; Mazza,~T. Are
  Gaming-Enabled Graphic Processing Unit Cards Convenient for Molecular
  Dynamics Simulation? \emph{Evol. Bioinform.} \textbf{2019}, \emph{15},
  1176934319850144\relax
\mciteBstWouldAddEndPuncttrue
\mciteSetBstMidEndSepPunct{\mcitedefaultmidpunct}
{\mcitedefaultendpunct}{\mcitedefaultseppunct}\relax
\EndOfBibitem
\bibitem[Genovese \latin{et~al.}(2009)Genovese, Ospici, Deutsch, M{\'e}haut,
  Neelov, and Goedecker]{Genovese2009}
Genovese,~L.; Ospici,~M.; Deutsch,~T.; M{\'e}haut,~J.-F.; Neelov,~A.;
  Goedecker,~S. Density functional theory calculation on many-cores hybrid
  central processing unit-graphic processing unit architectures. \emph{J. Chem.
  Phys.} \textbf{2009}, \emph{131}, 034103\relax
\mciteBstWouldAddEndPuncttrue
\mciteSetBstMidEndSepPunct{\mcitedefaultmidpunct}
{\mcitedefaultendpunct}{\mcitedefaultseppunct}\relax
\EndOfBibitem
\bibitem[Andrade and Aspuru-Guzik(2013)Andrade, and Aspuru-Guzik]{Andrade2013}
Andrade,~X.; Aspuru-Guzik,~A. Real-space density functional theory on graphical
  processing units: computational approach and comparison to gaussian basis set
  methods. \emph{J. Chem. Theo. Comput.} \textbf{2013}, \emph{9},
  4360--4373\relax
\mciteBstWouldAddEndPuncttrue
\mciteSetBstMidEndSepPunct{\mcitedefaultmidpunct}
{\mcitedefaultendpunct}{\mcitedefaultseppunct}\relax
\EndOfBibitem
\bibitem[Hacene \latin{et~al.}(2012)Hacene, Anciaux-Sedrakian, Rozanska, Klahr,
  Guignon, and Fleurat-Lessard]{Hacene2012}
Hacene,~M.; Anciaux-Sedrakian,~A.; Rozanska,~X.; Klahr,~D.; Guignon,~T.;
  Fleurat-Lessard,~P. Accelerating VASP electronic structure calculations using
  graphic processing units. \emph{J. Comput. Chem.} \textbf{2012}, \emph{33},
  2581--2589\relax
\mciteBstWouldAddEndPuncttrue
\mciteSetBstMidEndSepPunct{\mcitedefaultmidpunct}
{\mcitedefaultendpunct}{\mcitedefaultseppunct}\relax
\EndOfBibitem
\bibitem[Romero \latin{et~al.}(2017)Romero, Phillips, Ruetsch, Fatica, Spiga,
  and Giannozzi]{Romero2017}
Romero,~J.; Phillips,~E.; Ruetsch,~G.; Fatica,~M.; Spiga,~F.; Giannozzi,~P. A
  performance study of Quantum ESPRESSO's PWscf code on multi-core and GPU
  systems. International Workshop on Performance Modeling, Benchmarking and
  Simulation of High Performance Computer Systems. 2017; pp 67--87\relax
\mciteBstWouldAddEndPuncttrue
\mciteSetBstMidEndSepPunct{\mcitedefaultmidpunct}
{\mcitedefaultendpunct}{\mcitedefaultseppunct}\relax
\EndOfBibitem
\bibitem[Huhn \latin{et~al.}(2020)Huhn, Lange, Yu, Yoon, and Blum]{Huhn2020}
Huhn,~W.~P.; Lange,~B.; Yu,~V. W.-z.; Yoon,~M.; Blum,~V. GPU acceleration of
  all-electron electronic structure theory using localized numeric
  atom-centered basis functions. \emph{Comput. Phys. Comm.} \textbf{2020},
  \emph{254}, 107314\relax
\mciteBstWouldAddEndPuncttrue
\mciteSetBstMidEndSepPunct{\mcitedefaultmidpunct}
{\mcitedefaultendpunct}{\mcitedefaultseppunct}\relax
\EndOfBibitem
\bibitem[Andrade(2010)]{Andrade2010thesis}
Andrade,~X. Linear and non-linear response phenomena of molecular systems
  within time-dependent density functional theory. \textbf{2010}, \relax
\mciteBstWouldAddEndPunctfalse
\mciteSetBstMidEndSepPunct{\mcitedefaultmidpunct}
{}{\mcitedefaultseppunct}\relax
\EndOfBibitem
\bibitem[Andrade \latin{et~al.}(2012)Andrade, Alberdi-Rodriguez, Strubbe,
  Oliveira, Nogueira, Castro, Muguerza, Arruabarrena, Louie, Aspuru-Guzik,
  \latin{et~al.} others]{Andrade2012}
Andrade,~X.; Alberdi-Rodriguez,~J.; Strubbe,~D.~A.; Oliveira,~M.~J.;
  Nogueira,~F.; Castro,~A.; Muguerza,~J.; Arruabarrena,~A.; Louie,~S.~G.;
  Aspuru-Guzik,~A., \latin{et~al.}  Time-dependent density-functional theory in
  massively parallel computer architectures: the octopus project. \emph{J.
  Phys.: Cond. Matt.} \textbf{2012}, \emph{24}, 233202\relax
\mciteBstWouldAddEndPuncttrue
\mciteSetBstMidEndSepPunct{\mcitedefaultmidpunct}
{\mcitedefaultendpunct}{\mcitedefaultseppunct}\relax
\EndOfBibitem
\bibitem[Andrade \latin{et~al.}(2015)Andrade, Strubbe, De~Giovannini, Larsen,
  Oliveira, Alberdi-Rodriguez, Varas, Theophilou, Helbig, Verstraete,
  \latin{et~al.} others]{Andrade2015}
Andrade,~X.; Strubbe,~D.; De~Giovannini,~U.; Larsen,~A.~H.; Oliveira,~M.~J.;
  Alberdi-Rodriguez,~J.; Varas,~A.; Theophilou,~I.; Helbig,~N.;
  Verstraete,~M.~J., \latin{et~al.}  Real-space grids and the Octopus code as
  tools for the development of new simulation approaches for electronic
  systems. \emph{Phys. Chem. Chem. Phys.} \textbf{2015}, \emph{17},
  31371--31396\relax
\mciteBstWouldAddEndPuncttrue
\mciteSetBstMidEndSepPunct{\mcitedefaultmidpunct}
{\mcitedefaultendpunct}{\mcitedefaultseppunct}\relax
\EndOfBibitem
\bibitem[Tancogne-Dejean \latin{et~al.}(2020)Tancogne-Dejean, Oliveira,
  Andrade, Appel, Borca, Le~Breton, Buchholz, Castro, Corni, Correa,
  \latin{et~al.} others]{Tancogne2020}
Tancogne-Dejean,~N.; Oliveira,~M.~J.; Andrade,~X.; Appel,~H.; Borca,~C.~H.;
  Le~Breton,~G.; Buchholz,~F.; Castro,~A.; Corni,~S.; Correa,~A.~A.,
  \latin{et~al.}  Octopus, a computational framework for exploring light-driven
  phenomena and quantum dynamics in extended and finite systems. \emph{J. Chem.
  Phys.} \textbf{2020}, \emph{152}, 124119\relax
\mciteBstWouldAddEndPuncttrue
\mciteSetBstMidEndSepPunct{\mcitedefaultmidpunct}
{\mcitedefaultendpunct}{\mcitedefaultseppunct}\relax
\EndOfBibitem
\bibitem[Schleife \latin{et~al.}(2012)Schleife, Draeger, Kanai, and
  Correa]{Schleife2012}
Schleife,~A.; Draeger,~E.~W.; Kanai,~Y.; Correa,~A.~A. Plane-wave
  pseudopotential implementation of explicit integrators for time-dependent
  Kohn-Sham equations in large-scale simulations. \emph{J. Chem. Phys.}
  \textbf{2012}, \emph{137}, 22A546\relax
\mciteBstWouldAddEndPuncttrue
\mciteSetBstMidEndSepPunct{\mcitedefaultmidpunct}
{\mcitedefaultendpunct}{\mcitedefaultseppunct}\relax
\EndOfBibitem
\bibitem[Schleife \latin{et~al.}(2015)Schleife, Kanai, and
  Correa]{Schleife2015}
Schleife,~A.; Kanai,~Y.; Correa,~A.~A. Accurate atomistic first-principles
  calculations of electronic stopping. \emph{Phys. Rev. B} \textbf{2015},
  \emph{91}, 014306\relax
\mciteBstWouldAddEndPuncttrue
\mciteSetBstMidEndSepPunct{\mcitedefaultmidpunct}
{\mcitedefaultendpunct}{\mcitedefaultseppunct}\relax
\EndOfBibitem
\bibitem[Gygi \latin{et~al.}(2005)Gygi, Yates, Lorenz, Draeger, Franchetti,
  Ueberhuber, de~Supinski, Kral, Gunnels, and Sexton]{Gygi2005}
Gygi,~F.; Yates,~R.~K.; Lorenz,~J.; Draeger,~E.~W.; Franchetti,~F.;
  Ueberhuber,~C.~W.; de~Supinski,~B.~R.; Kral,~S.; Gunnels,~J.~A.;
  Sexton,~J.~C. Large-scale first-principles molecular dynamics simulations on
  the bluegene/l platform using the qbox code. SC'05: Proceedings of the 2005
  ACM/IEEE conference on Supercomputing. 2005; pp 24--24\relax
\mciteBstWouldAddEndPuncttrue
\mciteSetBstMidEndSepPunct{\mcitedefaultmidpunct}
{\mcitedefaultendpunct}{\mcitedefaultseppunct}\relax
\EndOfBibitem
\bibitem[Martin(2009)]{Martin2009}
Martin,~R.~C. \emph{Clean code: a handbook of agile software craftsmanship};
  Pearson Education, 2009\relax
\mciteBstWouldAddEndPuncttrue
\mciteSetBstMidEndSepPunct{\mcitedefaultmidpunct}
{\mcitedefaultendpunct}{\mcitedefaultseppunct}\relax
\EndOfBibitem
\bibitem[Loeliger and McCullough(2012)Loeliger, and McCullough]{Loeliger2012}
Loeliger,~J.; McCullough,~M. \emph{Version Control with Git: Powerful tools and
  techniques for collaborative software development}; O'Reilly Media,
  2012\relax
\mciteBstWouldAddEndPuncttrue
\mciteSetBstMidEndSepPunct{\mcitedefaultmidpunct}
{\mcitedefaultendpunct}{\mcitedefaultseppunct}\relax
\EndOfBibitem
\bibitem[Larsen \latin{et~al.}(2017)Larsen, Mortensen, Blomqvist, Castelli,
  Christensen, Du{\l}ak, Friis, Groves, Hammer, Hargus, Hermes, Jennings,
  Jensen, Kermode, Kitchin, Kolsbjerg, Kubal, Kaasbjerg, Lysgaard, Maronsson,
  Maxson, Olsen, Pastewka, Peterson, Rostgaard, Schi{\o}tz, Sch\"{u}tt,
  Strange, Thygesen, Vegge, Vilhelmsen, Walter, Zeng, and
  Jacobsen]{HjorthLarsen2017}
Larsen,~A.~H. \latin{et~al.}  The atomic simulation environment{\textemdash}a
  Python library for working with atoms. \emph{Journal of Physics: Condensed
  Matter} \textbf{2017}, \emph{29}, 273002\relax
\mciteBstWouldAddEndPuncttrue
\mciteSetBstMidEndSepPunct{\mcitedefaultmidpunct}
{\mcitedefaultendpunct}{\mcitedefaultseppunct}\relax
\EndOfBibitem
\bibitem[Jain \latin{et~al.}(2013)Jain, Ong, Hautier, Chen, Richards, Dacek,
  Cholia, Gunter, Skinner, Ceder, \latin{et~al.} others]{Jain2013}
Jain,~A.; Ong,~S.~P.; Hautier,~G.; Chen,~W.; Richards,~W.~D.; Dacek,~S.;
  Cholia,~S.; Gunter,~D.; Skinner,~D.; Ceder,~G., \latin{et~al.}  Commentary:
  The Materials Project: A materials genome approach to accelerating materials
  innovation. \emph{APL materials} \textbf{2013}, \emph{1}, 011002\relax
\mciteBstWouldAddEndPuncttrue
\mciteSetBstMidEndSepPunct{\mcitedefaultmidpunct}
{\mcitedefaultendpunct}{\mcitedefaultseppunct}\relax
\EndOfBibitem
\bibitem[Ince \latin{et~al.}(2012)Ince, Hatton, and Graham-Cumming]{Ince2012}
Ince,~D.~C.; Hatton,~L.; Graham-Cumming,~J. The case for open computer
  programs. \emph{Nat.} \textbf{2012}, \emph{482}, 485--488\relax
\mciteBstWouldAddEndPuncttrue
\mciteSetBstMidEndSepPunct{\mcitedefaultmidpunct}
{\mcitedefaultendpunct}{\mcitedefaultseppunct}\relax
\EndOfBibitem
\bibitem[Smart(2018)]{Smart2018}
Smart,~A.~G. The war over supercooled water. \emph{Phys. Today} \textbf{2018},
  \emph{22}\relax
\mciteBstWouldAddEndPuncttrue
\mciteSetBstMidEndSepPunct{\mcitedefaultmidpunct}
{\mcitedefaultendpunct}{\mcitedefaultseppunct}\relax
\EndOfBibitem
\bibitem[Stepanov and McJones(2009)Stepanov, and McJones]{Stepanov2009}
Stepanov,~A.; McJones,~P. \emph{Elements of Programming}; Addison-Wesley,
  2009\relax
\mciteBstWouldAddEndPuncttrue
\mciteSetBstMidEndSepPunct{\mcitedefaultmidpunct}
{\mcitedefaultendpunct}{\mcitedefaultseppunct}\relax
\EndOfBibitem
\bibitem[Oliveira \latin{et~al.}(2020)Oliveira, Papior, Pouillon, Blum,
  Artacho, Caliste, Corsetti, de~Gironcoli, Elena, Garc{\'{\i}}a,
  Garc{\'{\i}}a-Su{\'{a}}rez, Genovese, Huhn, Huhs, Kokott,
  K\"{u}{\c{c}}\"{u}kbenli, Larsen, Lazzaro, Lebedeva, Li,
  L{\'{o}}pez-Dur{\'{a}}n, L{\'{o}}pez-Tarifa, L\"{u}ders, Marques, Minar,
  Mohr, Mostofi, O'Cais, Payne, Ruh, Smith, Soler, Strubbe, Tancogne-Dejean,
  Tildesley, Torrent, and zhe Yu]{Oliveira2020}
Oliveira,~M. J.~T. \latin{et~al.}  The {CECAM} electronic structure library and
  the modular software development paradigm. \emph{J. Chem. Phys.}
  \textbf{2020}, \emph{153}, 024117\relax
\mciteBstWouldAddEndPuncttrue
\mciteSetBstMidEndSepPunct{\mcitedefaultmidpunct}
{\mcitedefaultendpunct}{\mcitedefaultseppunct}\relax
\EndOfBibitem
\bibitem[Togo and Tanaka(2018)Togo, and Tanaka]{Togo2018}
Togo,~A.; Tanaka,~I. Spglib: a software library for crystal symmetry search.
  \emph{arXiv preprint arXiv:1808.01590} \textbf{2018}, \relax
\mciteBstWouldAddEndPunctfalse
\mciteSetBstMidEndSepPunct{\mcitedefaultmidpunct}
{}{\mcitedefaultseppunct}\relax
\EndOfBibitem
\bibitem[Cooley and Tukey(1965)Cooley, and Tukey]{Cooley1965}
Cooley,~J.~W.; Tukey,~J.~W. An algorithm for the machine calculation of complex
  Fourier series. \emph{Math. Comput.} \textbf{1965}, \emph{19}, 297--301\relax
\mciteBstWouldAddEndPuncttrue
\mciteSetBstMidEndSepPunct{\mcitedefaultmidpunct}
{\mcitedefaultendpunct}{\mcitedefaultseppunct}\relax
\EndOfBibitem
\bibitem[Kim \latin{et~al.}(2018)Kim, Baczewski, Beaudet, Benali, Bennett,
  Berrill, Blunt, Borda, Casula, Ceperley, Chiesa, Clark, Clay, Delaney,
  Dewing, Esler, Hao, Heinonen, Kent, Krogel, Kyl\"anp\"a\"a, Li, Lopez, Luo,
  Malone, Martin, Mathuriya, McMinis, Melton, Mitas, Morales, Neuscamman,
  Parker, Flores, Romero, Rubenstein, Shea, Shin, Shulenburger, Tillack,
  Townsend, Tubman, Goetz, Vincent, Yang, Yang, Zhang, and Zhao]{Kim2018}
Kim,~J. \latin{et~al.}  {QMCPACK}: an open sourceab initioquantum Monte Carlo
  package for the electronic structure of atoms, molecules and solids. \emph{J.
  Phys.: Cond. Matt.} \textbf{2018}, \emph{30}, 195901\relax
\mciteBstWouldAddEndPuncttrue
\mciteSetBstMidEndSepPunct{\mcitedefaultmidpunct}
{\mcitedefaultendpunct}{\mcitedefaultseppunct}\relax
\EndOfBibitem
\bibitem[Garc{\'\i}a \latin{et~al.}(2018)Garc{\'\i}a, Verstraete, Pouillon, and
  Junquera]{Garcia2018}
Garc{\'\i}a,~A.; Verstraete,~M.~J.; Pouillon,~Y.; Junquera,~J. The psml format
  and library for norm-conserving pseudopotential data curation and
  interoperability. \emph{Comput. Phys. Comm.} \textbf{2018}, \emph{227},
  51--71\relax
\mciteBstWouldAddEndPuncttrue
\mciteSetBstMidEndSepPunct{\mcitedefaultmidpunct}
{\mcitedefaultendpunct}{\mcitedefaultseppunct}\relax
\EndOfBibitem
\bibitem[Hamann(2013)]{Hamann2013}
Hamann,~D. Optimized norm-conserving Vanderbilt pseudopotentials. \emph{Phys.
  Rev. B} \textbf{2013}, \emph{88}, 085117\relax
\mciteBstWouldAddEndPuncttrue
\mciteSetBstMidEndSepPunct{\mcitedefaultmidpunct}
{\mcitedefaultendpunct}{\mcitedefaultseppunct}\relax
\EndOfBibitem
\bibitem[Tafipolsky and Schmid(2006)Tafipolsky, and Schmid]{Tafipolsky2006}
Tafipolsky,~M.; Schmid,~R. A general and efficient pseudopotential Fourier
  filtering scheme for real space methods using mask functions. \emph{J. Chem.
  Phys.} \textbf{2006}, \emph{124}, 174102\relax
\mciteBstWouldAddEndPuncttrue
\mciteSetBstMidEndSepPunct{\mcitedefaultmidpunct}
{\mcitedefaultendpunct}{\mcitedefaultseppunct}\relax
\EndOfBibitem
\bibitem[Willand \latin{et~al.}(2013)Willand, Kvashnin, Genovese,
  V{\'a}zquez-Mayagoitia, Deb, Sadeghi, Deutsch, and Goedecker]{Willand2013}
Willand,~A.; Kvashnin,~Y.~O.; Genovese,~L.; V{\'a}zquez-Mayagoitia,~{\'A}.;
  Deb,~A.~K.; Sadeghi,~A.; Deutsch,~T.; Goedecker,~S. Norm-conserving
  pseudopotentials with chemical accuracy compared to all-electron
  calculations. \emph{J. Chem. Phys.} \textbf{2013}, \emph{138}, 104109\relax
\mciteBstWouldAddEndPuncttrue
\mciteSetBstMidEndSepPunct{\mcitedefaultmidpunct}
{\mcitedefaultendpunct}{\mcitedefaultseppunct}\relax
\EndOfBibitem
\bibitem[Dal~Corso(2014)]{DalCorso2014}
Dal~Corso,~A. Pseudopotentials periodic table: From H to Pu. \emph{Comput. Mat.
  Sci.} \textbf{2014}, \emph{95}, 337--350\relax
\mciteBstWouldAddEndPuncttrue
\mciteSetBstMidEndSepPunct{\mcitedefaultmidpunct}
{\mcitedefaultendpunct}{\mcitedefaultseppunct}\relax
\EndOfBibitem
\bibitem[Garrity \latin{et~al.}(2014)Garrity, Bennett, Rabe, and
  Vanderbilt]{Garrity2014}
Garrity,~K.~F.; Bennett,~J.~W.; Rabe,~K.~M.; Vanderbilt,~D. Pseudopotentials
  for high-throughput DFT calculations. \emph{Comput. Mat. Sci.} \textbf{2014},
  \emph{81}, 446--452\relax
\mciteBstWouldAddEndPuncttrue
\mciteSetBstMidEndSepPunct{\mcitedefaultmidpunct}
{\mcitedefaultendpunct}{\mcitedefaultseppunct}\relax
\EndOfBibitem
\bibitem[Kucukbenli \latin{et~al.}(2014)Kucukbenli, Monni, Adetunji, Ge,
  Adebayo, Marzari, De~Gironcoli, and Corso]{Kucukbenli2014}
Kucukbenli,~E.; Monni,~M.; Adetunji,~B.; Ge,~X.; Adebayo,~G.; Marzari,~N.;
  De~Gironcoli,~S.; Corso,~A.~D. Projector augmented-wave and all-electron
  calculations across the periodic table: a comparison of structural and
  energetic properties. \emph{arXiv preprint arXiv:1404.3015} \textbf{2014},
  \relax
\mciteBstWouldAddEndPunctfalse
\mciteSetBstMidEndSepPunct{\mcitedefaultmidpunct}
{}{\mcitedefaultseppunct}\relax
\EndOfBibitem
\bibitem[Topsakal and Wentzcovitch(2014)Topsakal, and
  Wentzcovitch]{Topsakal2014}
Topsakal,~M.; Wentzcovitch,~R. Accurate projected augmented wave (PAW) datasets
  for rare-earth elements (RE= La--Lu). \emph{Comput. Mat. Sci.} \textbf{2014},
  \emph{95}, 263--270\relax
\mciteBstWouldAddEndPuncttrue
\mciteSetBstMidEndSepPunct{\mcitedefaultmidpunct}
{\mcitedefaultendpunct}{\mcitedefaultseppunct}\relax
\EndOfBibitem
\bibitem[Schlipf and Gygi(2015)Schlipf, and Gygi]{Schlipf2015}
Schlipf,~M.; Gygi,~F. Optimization algorithm for the generation of ONCV
  pseudopotentials. \emph{Comput. Phys. Comm.} \textbf{2015}, \emph{196},
  36--44\relax
\mciteBstWouldAddEndPuncttrue
\mciteSetBstMidEndSepPunct{\mcitedefaultmidpunct}
{\mcitedefaultendpunct}{\mcitedefaultseppunct}\relax
\EndOfBibitem
\bibitem[Prandini \latin{et~al.}(2018)Prandini, Marrazzo, Castelli, Mounet, and
  Marzari]{Prandini2018}
Prandini,~G.; Marrazzo,~A.; Castelli,~I.~E.; Mounet,~N.; Marzari,~N. Precision
  and efficiency in solid-state pseudopotential calculations. \emph{npj Comput.
  Mat.} \textbf{2018}, \emph{4}, 1--13\relax
\mciteBstWouldAddEndPuncttrue
\mciteSetBstMidEndSepPunct{\mcitedefaultmidpunct}
{\mcitedefaultendpunct}{\mcitedefaultseppunct}\relax
\EndOfBibitem
\bibitem[Van~Setten \latin{et~al.}(2018)Van~Setten, Giantomassi, Bousquet,
  Verstraete, Hamann, Gonze, and Rignanese]{Van2018}
Van~Setten,~M.; Giantomassi,~M.; Bousquet,~E.; Verstraete,~M.~J.;
  Hamann,~D.~R.; Gonze,~X.; Rignanese,~G.-M. The PseudoDojo: Training and
  grading a 85 element optimized norm-conserving pseudopotential table.
  \emph{Comput. Phys. Comm.} \textbf{2018}, \emph{226}, 39--54\relax
\mciteBstWouldAddEndPuncttrue
\mciteSetBstMidEndSepPunct{\mcitedefaultmidpunct}
{\mcitedefaultendpunct}{\mcitedefaultseppunct}\relax
\EndOfBibitem
\bibitem[Marques \latin{et~al.}(2012)Marques, Oliveira, and
  Burnus]{Marques2012}
Marques,~M.~A.; Oliveira,~M.~J.; Burnus,~T. Libxc: A library of exchange and
  correlation functionals for density functional theory. \emph{Comput. Phys.
  Comm.} \textbf{2012}, \emph{183}, 2272--2281\relax
\mciteBstWouldAddEndPuncttrue
\mciteSetBstMidEndSepPunct{\mcitedefaultmidpunct}
{\mcitedefaultendpunct}{\mcitedefaultseppunct}\relax
\EndOfBibitem
\bibitem[Lehtola \latin{et~al.}(2018)Lehtola, Steigemann, Oliveira, and
  Marques]{Lehtola2018}
Lehtola,~S.; Steigemann,~C.; Oliveira,~M.~J.; Marques,~M.~A. Recent
  developments in libxc -- A comprehensive library of functionals for density
  functional theory. \emph{SoftwareX} \textbf{2018}, \emph{7}, 1--5\relax
\mciteBstWouldAddEndPuncttrue
\mciteSetBstMidEndSepPunct{\mcitedefaultmidpunct}
{\mcitedefaultendpunct}{\mcitedefaultseppunct}\relax
\EndOfBibitem
\bibitem[Stepanov and Rose(2014)Stepanov, and Rose]{Stepanov2014}
Stepanov,~A.; Rose,~D. \emph{From Mathematics to Generic Programming};
  Addison-Wesley, 2014\relax
\mciteBstWouldAddEndPuncttrue
\mciteSetBstMidEndSepPunct{\mcitedefaultmidpunct}
{\mcitedefaultendpunct}{\mcitedefaultseppunct}\relax
\EndOfBibitem
\bibitem[Andrade and Genovese(2012)Andrade, and Genovese]{Andrade2012c}
Andrade,~X.; Genovese,~L. \emph{Fundamentals of Time-Dependent Density
  Functional Theory}; Springer, 2012; pp 401--413\relax
\mciteBstWouldAddEndPuncttrue
\mciteSetBstMidEndSepPunct{\mcitedefaultmidpunct}
{\mcitedefaultendpunct}{\mcitedefaultseppunct}\relax
\EndOfBibitem
\bibitem[Beckingsale \latin{et~al.}(2019)Beckingsale, Burmark, Hornung, Jones,
  Killian, Kunen, Pearce, Robinson, Ryujin, and Scogland]{Beckingsale2019}
Beckingsale,~D.~A.; Burmark,~J.; Hornung,~R.; Jones,~H.; Killian,~W.;
  Kunen,~A.~J.; Pearce,~O.; Robinson,~P.; Ryujin,~B.~S.; Scogland,~T.~R. RAJA:
  Portable performance for large-scale scientific applications. 2019 ieee/acm
  international workshop on performance, portability and productivity in {HPC}
  ({P3HPC}). 2019; pp 71--81\relax
\mciteBstWouldAddEndPuncttrue
\mciteSetBstMidEndSepPunct{\mcitedefaultmidpunct}
{\mcitedefaultendpunct}{\mcitedefaultseppunct}\relax
\EndOfBibitem
\bibitem[Edwards \latin{et~al.}(2014)Edwards, Trott, and
  Sunderland]{CarterEdwards2014}
Edwards,~H.~C.; Trott,~C.~R.; Sunderland,~D. Kokkos: Enabling manycore
  performance portability through polymorphic memory access patterns. \emph{J.
  Parallel Distrib. Comput.} \textbf{2014}, \emph{74}, 3202 -- 3216\relax
\mciteBstWouldAddEndPuncttrue
\mciteSetBstMidEndSepPunct{\mcitedefaultmidpunct}
{\mcitedefaultendpunct}{\mcitedefaultseppunct}\relax
\EndOfBibitem
\bibitem[Lee and Eigenmann(2010)Lee, and Eigenmann]{Lee2010}
Lee,~S.; Eigenmann,~R. OpenMPC: Extended OpenMP programming and tuning for
  GPUs. SC'10: Proceedings of the 2010 ACM/IEEE International Conference for
  High Performance Computing, Networking, Storage and Analysis. 2010; pp
  1--11\relax
\mciteBstWouldAddEndPuncttrue
\mciteSetBstMidEndSepPunct{\mcitedefaultmidpunct}
{\mcitedefaultendpunct}{\mcitedefaultseppunct}\relax
\EndOfBibitem
\bibitem[Alpay and Heuveline(2020)Alpay, and Heuveline]{Alpay2020}
Alpay,~A.; Heuveline,~V. SYCL beyond OpenCL: The architecture, current state
  and future direction of hipSYCL. Proceedings of the International Workshop on
  OpenCL. 2020; pp 1--1\relax
\mciteBstWouldAddEndPuncttrue
\mciteSetBstMidEndSepPunct{\mcitedefaultmidpunct}
{\mcitedefaultendpunct}{\mcitedefaultseppunct}\relax
\EndOfBibitem
\bibitem[Chien \latin{et~al.}(2019)Chien, Peng, and Markidis]{Chien2019}
Chien,~S.; Peng,~I.; Markidis,~S. Performance Evaluation of Advanced Features
  in CUDA Unified Memory. 2019 IEEE/ACM Workshop on Memory Centric High
  Performance Computing (MCHPC). 2019; pp 50--57\relax
\mciteBstWouldAddEndPuncttrue
\mciteSetBstMidEndSepPunct{\mcitedefaultmidpunct}
{\mcitedefaultendpunct}{\mcitedefaultseppunct}\relax
\EndOfBibitem
\bibitem[Harris \latin{et~al.}(2007)Harris, \latin{et~al.} others]{Harris2007}
Harris,~M., \latin{et~al.}  Optimizing parallel reduction in CUDA. \emph{Nvidia
  Dev. Tech.} \textbf{2007}, \emph{2}, 1--39\relax
\mciteBstWouldAddEndPuncttrue
\mciteSetBstMidEndSepPunct{\mcitedefaultmidpunct}
{\mcitedefaultendpunct}{\mcitedefaultseppunct}\relax
\EndOfBibitem
\bibitem[Pacheco(1997)]{Pacheco1997}
Pacheco,~P. \emph{Parallel Programming with MPI}; Elsevier Science, 1997\relax
\mciteBstWouldAddEndPuncttrue
\mciteSetBstMidEndSepPunct{\mcitedefaultmidpunct}
{\mcitedefaultendpunct}{\mcitedefaultseppunct}\relax
\EndOfBibitem
\bibitem[Jornet-Somoza \latin{et~al.}(2015)Jornet-Somoza, Alberdi-Rodriguez,
  Milne, Andrade, Marques, Nogueira, Oliveira, Stewart, and Rubio]{Jornet2015}
Jornet-Somoza,~J.; Alberdi-Rodriguez,~J.; Milne,~B.~F.; Andrade,~X.;
  Marques,~M.~A.; Nogueira,~F.; Oliveira,~M.~J.; Stewart,~J.~J.; Rubio,~A.
  Insights into colour-tuning of chlorophyll optical response in green plants.
  \emph{Phys. Chem. Chem. Phys.} \textbf{2015}, \emph{17}, 26599--26606\relax
\mciteBstWouldAddEndPuncttrue
\mciteSetBstMidEndSepPunct{\mcitedefaultmidpunct}
{\mcitedefaultendpunct}{\mcitedefaultseppunct}\relax
\EndOfBibitem
\bibitem[Hasegawa \latin{et~al.}(2011)Hasegawa, Iwata, Tsuji, Takahashi,
  Oshiyama, Minami, Boku, Shoji, Uno, Kurokawa, \latin{et~al.}
  others]{Hasegawa2011}
Hasegawa,~Y.; Iwata,~J.-I.; Tsuji,~M.; Takahashi,~D.; Oshiyama,~A.; Minami,~K.;
  Boku,~T.; Shoji,~F.; Uno,~A.; Kurokawa,~M., \latin{et~al.}  First-principles
  calculations of electron states of a silicon nanowire with 100,000 atoms on
  the K computer. Proceedings of 2011 International Conference for High
  Performance Computing, Networking, Storage and Analysis. 2011; pp 1--11\relax
\mciteBstWouldAddEndPuncttrue
\mciteSetBstMidEndSepPunct{\mcitedefaultmidpunct}
{\mcitedefaultendpunct}{\mcitedefaultseppunct}\relax
\EndOfBibitem
\bibitem[Ehrenfest(1927)]{Ehrenfest1927}
Ehrenfest,~P. Bemerkung {\"u}ber die angen{\"a}herte G{\"u}ltigkeit der
  klassischen Mechanik innerhalb der Quantenmechanik. \emph{Z. Phys.}
  \textbf{1927}, \emph{45}, 455--457\relax
\mciteBstWouldAddEndPuncttrue
\mciteSetBstMidEndSepPunct{\mcitedefaultmidpunct}
{\mcitedefaultendpunct}{\mcitedefaultseppunct}\relax
\EndOfBibitem
\bibitem[Andrade \latin{et~al.}(2009)Andrade, Castro, Zueco, Alonso, Echenique,
  Falceto, and Rubio]{Andrade2009}
Andrade,~X.; Castro,~A.; Zueco,~D.; Alonso,~J.; Echenique,~P.; Falceto,~F.;
  Rubio,~A. Modified Ehrenfest formalism for efficient large-scale ab initio
  molecular dynamics. \emph{J. Chem. Theo. Comput.} \textbf{2009}, \emph{5},
  728--742\relax
\mciteBstWouldAddEndPuncttrue
\mciteSetBstMidEndSepPunct{\mcitedefaultmidpunct}
{\mcitedefaultendpunct}{\mcitedefaultseppunct}\relax
\EndOfBibitem
\bibitem[Szabo and Ostlund(2012)Szabo, and Ostlund]{Szabo2012}
Szabo,~A.; Ostlund,~N. \emph{Modern Quantum Chemistry: Introduction to Advanced
  Electronic Structure Theory}; Dover Books on Chemistry; Dover Publications,
  2012\relax
\mciteBstWouldAddEndPuncttrue
\mciteSetBstMidEndSepPunct{\mcitedefaultmidpunct}
{\mcitedefaultendpunct}{\mcitedefaultseppunct}\relax
\EndOfBibitem
\bibitem[Olsen(2021)]{Olsen2021}
Olsen,~J. \emph{Basis Sets in Computational Chemistry}; Springer International
  Publishing: Cham, 2021; pp 1--16\relax
\mciteBstWouldAddEndPuncttrue
\mciteSetBstMidEndSepPunct{\mcitedefaultmidpunct}
{\mcitedefaultendpunct}{\mcitedefaultseppunct}\relax
\EndOfBibitem
\bibitem[Broyden(1965)]{Broyden1965}
Broyden,~C.~G. A class of methods for solving nonlinear simultaneous equations.
  \emph{Math. Comput.} \textbf{1965}, \emph{19}, 577--593\relax
\mciteBstWouldAddEndPuncttrue
\mciteSetBstMidEndSepPunct{\mcitedefaultmidpunct}
{\mcitedefaultendpunct}{\mcitedefaultseppunct}\relax
\EndOfBibitem
\bibitem[Pulay(1980)]{Pulay1980}
Pulay,~P. Convergence acceleration of iterative sequences. The case of SCF
  iteration. \emph{Chem. Phys. Lett.} \textbf{1980}, \emph{73}, 393--398\relax
\mciteBstWouldAddEndPuncttrue
\mciteSetBstMidEndSepPunct{\mcitedefaultmidpunct}
{\mcitedefaultendpunct}{\mcitedefaultseppunct}\relax
\EndOfBibitem
\bibitem[Saad(2011)]{Saad2011}
Saad,~Y. \emph{Numerical Methods for Large Eigenvalue Problems}; Society for
  Industrial and Applied Mathematics, 2011\relax
\mciteBstWouldAddEndPuncttrue
\mciteSetBstMidEndSepPunct{\mcitedefaultmidpunct}
{\mcitedefaultendpunct}{\mcitedefaultseppunct}\relax
\EndOfBibitem
\bibitem[Axler(2014)]{Axler2014}
Axler,~S. \emph{Linear Algebra Done Right}; Undergraduate Texts in Mathematics;
  Springer International Publishing, 2014; p 151\relax
\mciteBstWouldAddEndPuncttrue
\mciteSetBstMidEndSepPunct{\mcitedefaultmidpunct}
{\mcitedefaultendpunct}{\mcitedefaultseppunct}\relax
\EndOfBibitem
\bibitem[Payne \latin{et~al.}(1992)Payne, Teter, Allan, Arias, and
  Joannopoulos]{Payne1992}
Payne,~M.~C.; Teter,~M.~P.; Allan,~D.~C.; Arias,~T.; Joannopoulos,~a.~J.
  Iterative minimization techniques for ab initio total-energy calculations:
  molecular dynamics and conjugate gradients. \emph{Rev. Mod. Phys.}
  \textbf{1992}, \emph{64}, 1045\relax
\mciteBstWouldAddEndPuncttrue
\mciteSetBstMidEndSepPunct{\mcitedefaultmidpunct}
{\mcitedefaultendpunct}{\mcitedefaultseppunct}\relax
\EndOfBibitem
\bibitem[Davidson(1975)]{Davidson1975}
Davidson,~E. Theiterative calculation of a few ofthe lowest eigenvalues and
  corresponding eigenvectors of large real-symmetric matrices. \emph{J. Comput.
  Phys} \textbf{1975}, \emph{17}, 87--94\relax
\mciteBstWouldAddEndPuncttrue
\mciteSetBstMidEndSepPunct{\mcitedefaultmidpunct}
{\mcitedefaultendpunct}{\mcitedefaultseppunct}\relax
\EndOfBibitem
\bibitem[Teter \latin{et~al.}(1989)Teter, Payne, and Allan]{Teter1989}
Teter,~M.~P.; Payne,~M.~C.; Allan,~D.~C. Solution of Schr{\"o}dinger's equation
  for large systems. \emph{Phys. Rev. B} \textbf{1989}, \emph{40}, 12255\relax
\mciteBstWouldAddEndPuncttrue
\mciteSetBstMidEndSepPunct{\mcitedefaultmidpunct}
{\mcitedefaultendpunct}{\mcitedefaultseppunct}\relax
\EndOfBibitem
\bibitem[Blackford \latin{et~al.}(2002)Blackford, Petitet, Pozo, Remington,
  Whaley, Demmel, Dongarra, Duff, Hammarling, Henry, \latin{et~al.}
  others]{Blackford2002}
Blackford,~L.~S.; Petitet,~A.; Pozo,~R.; Remington,~K.; Whaley,~R.~C.;
  Demmel,~J.; Dongarra,~J.; Duff,~I.; Hammarling,~S.; Henry,~G., \latin{et~al.}
   An updated set of basic linear algebra subprograms (BLAS). \emph{ACM
  Transactions on Mathematical Software} \textbf{2002}, \emph{28},
  135--151\relax
\mciteBstWouldAddEndPuncttrue
\mciteSetBstMidEndSepPunct{\mcitedefaultmidpunct}
{\mcitedefaultendpunct}{\mcitedefaultseppunct}\relax
\EndOfBibitem
\bibitem[Anderson \latin{et~al.}(1999)Anderson, Bai, Bischof, Blackford,
  Demmel, Dongarra, Du~Croz, Greenbaum, Hammarling, McKenney, and
  Sorensen]{Anderson1999}
Anderson,~E.; Bai,~Z.; Bischof,~C.; Blackford,~S.; Demmel,~J.; Dongarra,~J.;
  Du~Croz,~J.; Greenbaum,~A.; Hammarling,~S.; McKenney,~A.; Sorensen,~D.
  \emph{{LAPACK} Users' Guide}, 3rd ed.; Society for Industrial and Applied
  Mathematics: Philadelphia, PA, 1999\relax
\mciteBstWouldAddEndPuncttrue
\mciteSetBstMidEndSepPunct{\mcitedefaultmidpunct}
{\mcitedefaultendpunct}{\mcitedefaultseppunct}\relax
\EndOfBibitem
\bibitem[Blackford \latin{et~al.}(1997)Blackford, Choi, Cleary, D'Azevedo,
  Demmel, Dhillon, Dongarra, Hammarling, Henry, Petitet, Stanley, Walker, and
  Whaley]{Blackford1997}
Blackford,~L.~S.; Choi,~J.; Cleary,~A.; D'Azevedo,~E.; Demmel,~J.; Dhillon,~I.;
  Dongarra,~J.; Hammarling,~S.; Henry,~G.; Petitet,~A.; Stanley,~K.;
  Walker,~D.; Whaley,~R.~C. \emph{{ScaLAPACK} Users' Guide}; Society for
  Industrial and Applied Mathematics: Philadelphia, PA, 1997\relax
\mciteBstWouldAddEndPuncttrue
\mciteSetBstMidEndSepPunct{\mcitedefaultmidpunct}
{\mcitedefaultendpunct}{\mcitedefaultseppunct}\relax
\EndOfBibitem
\bibitem[Gates \latin{et~al.}(2019)Gates, Kurzak, Charara, YarKhan, and
  Dongarra]{Gates2019}
Gates,~M.; Kurzak,~J.; Charara,~A.; YarKhan,~A.; Dongarra,~J. Slate: Design of
  a modern distributed and accelerated linear algebra library. Proceedings of
  the International Conference for High Performance Computing, Networking,
  Storage and Analysis. 2019; pp 1--18\relax
\mciteBstWouldAddEndPuncttrue
\mciteSetBstMidEndSepPunct{\mcitedefaultmidpunct}
{\mcitedefaultendpunct}{\mcitedefaultseppunct}\relax
\EndOfBibitem
\bibitem[O'Neill(2014)]{Oneill2014}
O'Neill,~M.~E. PCG: A family of simple fast space-efficient statistically good
  algorithms for random number generation. \emph{ACM Transactions on
  Mathematical Software} \textbf{2014}, \relax
\mciteBstWouldAddEndPunctfalse
\mciteSetBstMidEndSepPunct{\mcitedefaultmidpunct}
{}{\mcitedefaultseppunct}\relax
\EndOfBibitem
\bibitem[Castro \latin{et~al.}(2004)Castro, Marques, and Rubio]{Castro2004}
Castro,~A.; Marques,~M.~A.; Rubio,~A. Propagators for the time-dependent
  Kohn--Sham equations. \emph{J. Chem. Phys.} \textbf{2004}, \emph{121},
  3425--3433\relax
\mciteBstWouldAddEndPuncttrue
\mciteSetBstMidEndSepPunct{\mcitedefaultmidpunct}
{\mcitedefaultendpunct}{\mcitedefaultseppunct}\relax
\EndOfBibitem
\bibitem[Kidd \latin{et~al.}(2017)Kidd, Covington, and Varga]{Kidd2017}
Kidd,~D.; Covington,~C.; Varga,~K. Exponential integrators in time-dependent
  density-functional calculations. \emph{Phys. Rev. E} \textbf{2017},
  \emph{96}, 063307\relax
\mciteBstWouldAddEndPuncttrue
\mciteSetBstMidEndSepPunct{\mcitedefaultmidpunct}
{\mcitedefaultendpunct}{\mcitedefaultseppunct}\relax
\EndOfBibitem
\bibitem[Gomez~Pueyo \latin{et~al.}(2018)Gomez~Pueyo, Marques, Rubio, and
  Castro]{Gomez2018}
Gomez~Pueyo,~A.; Marques,~M.~A.; Rubio,~A.; Castro,~A. Propagators for the
  time-dependent Kohn--Sham equations: Multistep, Runge--Kutta, exponential
  Runge--Kutta, and commutator free magnus methods. \emph{J Chem. Theo.
  Comput.} \textbf{2018}, \emph{14}, 3040--3052\relax
\mciteBstWouldAddEndPuncttrue
\mciteSetBstMidEndSepPunct{\mcitedefaultmidpunct}
{\mcitedefaultendpunct}{\mcitedefaultseppunct}\relax
\EndOfBibitem
\bibitem[Jia \latin{et~al.}(2018)Jia, An, Wang, and Lin]{Jia2018}
Jia,~W.; An,~D.; Wang,~L.-W.; Lin,~L. Fast real-time time-dependent density
  functional theory calculations with the parallel transport gauge. \emph{J.
  Chem. Theo. Comput.} \textbf{2018}, \emph{14}, 5645--5652\relax
\mciteBstWouldAddEndPuncttrue
\mciteSetBstMidEndSepPunct{\mcitedefaultmidpunct}
{\mcitedefaultendpunct}{\mcitedefaultseppunct}\relax
\EndOfBibitem
\bibitem[Bader \latin{et~al.}(2018)Bader, Blanes, and Kopylov]{Bader2018}
Bader,~P.; Blanes,~S.; Kopylov,~N. Exponential propagators for the
  Schr{\"o}dinger equation with a time-dependent potential. \emph{J. Chem.
  Phys.} \textbf{2018}, \emph{148}, 244109\relax
\mciteBstWouldAddEndPuncttrue
\mciteSetBstMidEndSepPunct{\mcitedefaultmidpunct}
{\mcitedefaultendpunct}{\mcitedefaultseppunct}\relax
\EndOfBibitem
\bibitem[Zhu and Herbert(2018)Zhu, and Herbert]{Zhu2018}
Zhu,~Y.; Herbert,~J.~M. Self-consistent predictor/corrector algorithms for
  stable and efficient integration of the time-dependent Kohn-Sham equation.
  \emph{J. Chem. Phys.} \textbf{2018}, \emph{148}, 044117\relax
\mciteBstWouldAddEndPuncttrue
\mciteSetBstMidEndSepPunct{\mcitedefaultmidpunct}
{\mcitedefaultendpunct}{\mcitedefaultseppunct}\relax
\EndOfBibitem
\bibitem[Verlet(1967)]{Verlet1967}
Verlet,~L. Computer" experiments" on classical fluids. I. Thermodynamical
  properties of Lennard-Jones molecules. \emph{Phys. Rev.} \textbf{1967},
  \emph{159}, 98\relax
\mciteBstWouldAddEndPuncttrue
\mciteSetBstMidEndSepPunct{\mcitedefaultmidpunct}
{\mcitedefaultendpunct}{\mcitedefaultseppunct}\relax
\EndOfBibitem
\bibitem[Ayala \latin{et~al.}(2020)Ayala, Tomov, Haidar, and
  Dongarra]{Ayala2020}
Ayala,~A.; Tomov,~S.; Haidar,~A.; Dongarra,~J. heffte: Highly efficient fft for
  exascale. International Conference on Computational Science. 2020; pp
  262--275\relax
\mciteBstWouldAddEndPuncttrue
\mciteSetBstMidEndSepPunct{\mcitedefaultmidpunct}
{\mcitedefaultendpunct}{\mcitedefaultseppunct}\relax
\EndOfBibitem
\bibitem[Wadleigh and Crawford(2000)Wadleigh, and Crawford]{Wadleigh2000}
Wadleigh,~K.; Crawford,~I. \emph{Software Optimization for High-performance
  Computing}; HP Professional; Prentice Hall PTR, 2000\relax
\mciteBstWouldAddEndPuncttrue
\mciteSetBstMidEndSepPunct{\mcitedefaultmidpunct}
{\mcitedefaultendpunct}{\mcitedefaultseppunct}\relax
\EndOfBibitem
\bibitem[Vanderbilt(1990)]{Vanderbilt1990}
Vanderbilt,~D. Soft self-consistent pseudopotentials in a generalized
  eigenvalue formalism. \emph{Phys. Rev. B} \textbf{1990}, \emph{41},
  7892\relax
\mciteBstWouldAddEndPuncttrue
\mciteSetBstMidEndSepPunct{\mcitedefaultmidpunct}
{\mcitedefaultendpunct}{\mcitedefaultseppunct}\relax
\EndOfBibitem
\bibitem[Bl{\"o}chl(1994)]{Blochl1994}
Bl{\"o}chl,~P.~E. Projector augmented-wave method. \emph{Phys. Rev. B}
  \textbf{1994}, \emph{50}, 17953\relax
\mciteBstWouldAddEndPuncttrue
\mciteSetBstMidEndSepPunct{\mcitedefaultmidpunct}
{\mcitedefaultendpunct}{\mcitedefaultseppunct}\relax
\EndOfBibitem
\bibitem[Lejaeghere \latin{et~al.}(2016)Lejaeghere, Bihlmayer, Bj{\"o}rkman,
  Blaha, Bl{\"u}gel, Blum, Caliste, Castelli, Clark, Dal~Corso, \latin{et~al.}
  others]{Lejaeghere2016}
Lejaeghere,~K.; Bihlmayer,~G.; Bj{\"o}rkman,~T.; Blaha,~P.; Bl{\"u}gel,~S.;
  Blum,~V.; Caliste,~D.; Castelli,~I.~E.; Clark,~S.~J.; Dal~Corso,~A.,
  \latin{et~al.}  Reproducibility in density functional theory calculations of
  solids. \emph{Sci.} \textbf{2016}, \emph{351}\relax
\mciteBstWouldAddEndPuncttrue
\mciteSetBstMidEndSepPunct{\mcitedefaultmidpunct}
{\mcitedefaultendpunct}{\mcitedefaultseppunct}\relax
\EndOfBibitem
\bibitem[Varsano \latin{et~al.}(2009)Varsano, Espinosa-Leal, Andrade, Marques,
  Di~Felice, and Rubio]{Varsano2009}
Varsano,~D.; Espinosa-Leal,~L.~A.; Andrade,~X.; Marques,~M.~A.; Di~Felice,~R.;
  Rubio,~A. Towards a gauge invariant method for molecular chiroptical
  properties in TDDFT. \emph{Phys. Chem. Chem. Phys.} \textbf{2009}, \emph{11},
  4481--4489\relax
\mciteBstWouldAddEndPuncttrue
\mciteSetBstMidEndSepPunct{\mcitedefaultmidpunct}
{\mcitedefaultendpunct}{\mcitedefaultseppunct}\relax
\EndOfBibitem
\bibitem[Rozzi \latin{et~al.}(2006)Rozzi, Varsano, Marini, Gross, and
  Rubio]{Rozzi2006}
Rozzi,~C.~A.; Varsano,~D.; Marini,~A.; Gross,~E.~K.; Rubio,~A. Exact Coulomb
  cutoff technique for supercell calculations. \emph{Phys. Rev. B}
  \textbf{2006}, \emph{73}, 205119\relax
\mciteBstWouldAddEndPuncttrue
\mciteSetBstMidEndSepPunct{\mcitedefaultmidpunct}
{\mcitedefaultendpunct}{\mcitedefaultseppunct}\relax
\EndOfBibitem
\bibitem[Garc{\'\i}a-Risue{\~n}o \latin{et~al.}(2014)Garc{\'\i}a-Risue{\~n}o,
  Alberdi-Rodriguez, Oliveira, Andrade, Pippig, Muguerza, Arruabarrena, and
  Rubio]{Garcia2014}
Garc{\'\i}a-Risue{\~n}o,~P.; Alberdi-Rodriguez,~J.; Oliveira,~M.~J.;
  Andrade,~X.; Pippig,~M.; Muguerza,~J.; Arruabarrena,~A.; Rubio,~A. A survey
  of the parallel performance and accuracy of Poisson solvers for electronic
  structure calculations. \emph{J. Comput. Chem.} \textbf{2014}, \emph{35},
  427--444\relax
\mciteBstWouldAddEndPuncttrue
\mciteSetBstMidEndSepPunct{\mcitedefaultmidpunct}
{\mcitedefaultendpunct}{\mcitedefaultseppunct}\relax
\EndOfBibitem
\bibitem[Hirose(2005)]{Hirose2005}
Hirose,~K. \emph{First-principles Calculations in Real-space Formalism:
  Electronic Configurations and Transport Properties of Nanostructures};
  Imperial College Press, 2005; p~12\relax
\mciteBstWouldAddEndPuncttrue
\mciteSetBstMidEndSepPunct{\mcitedefaultmidpunct}
{\mcitedefaultendpunct}{\mcitedefaultseppunct}\relax
\EndOfBibitem
\bibitem[Baroni \latin{et~al.}(2001)Baroni, De~Gironcoli, Dal~Corso, and
  Giannozzi]{Baroni2001}
Baroni,~S.; De~Gironcoli,~S.; Dal~Corso,~A.; Giannozzi,~P. Phonons and related
  crystal properties from density-functional perturbation theory. \emph{Rev.
  Mod. Phys.} \textbf{2001}, \emph{73}, 515\relax
\mciteBstWouldAddEndPuncttrue
\mciteSetBstMidEndSepPunct{\mcitedefaultmidpunct}
{\mcitedefaultendpunct}{\mcitedefaultseppunct}\relax
\EndOfBibitem
\bibitem[Yabana \latin{et~al.}(2012)Yabana, Sugiyama, Shinohara, Otobe, and
  Bertsch]{Yabana2012}
Yabana,~K.; Sugiyama,~T.; Shinohara,~Y.; Otobe,~T.; Bertsch,~G.~F.
  {Time-dependent density functional theory for strong electromagnetic fields
  in crystalline solids}. \emph{Phys. Rev. B} \textbf{2012}, \emph{85},
  045134\relax
\mciteBstWouldAddEndPuncttrue
\mciteSetBstMidEndSepPunct{\mcitedefaultmidpunct}
{\mcitedefaultendpunct}{\mcitedefaultseppunct}\relax
\EndOfBibitem
\bibitem[Sigmund(2006)]{Sigmund2006}
Sigmund,~P. \emph{Particle Penetration and Radiation Effects}; Springer,
  2006\relax
\mciteBstWouldAddEndPuncttrue
\mciteSetBstMidEndSepPunct{\mcitedefaultmidpunct}
{\mcitedefaultendpunct}{\mcitedefaultseppunct}\relax
\EndOfBibitem
\bibitem[Ziegler \latin{et~al.}(1985)Ziegler, Littmark, and
  Biersack]{Ziegler1985}
Ziegler,~J.~F.; Littmark,~U.; Biersack,~J.~P. \emph{The stopping and range of
  ions in solids}; New York : Pergamon, 1985, 1985; p 321 p.:\relax
\mciteBstWouldAddEndPuncttrue
\mciteSetBstMidEndSepPunct{\mcitedefaultmidpunct}
{\mcitedefaultendpunct}{\mcitedefaultseppunct}\relax
\EndOfBibitem
\bibitem[Zhang \latin{et~al.}(2016)Zhang, Jin, Xue, Lu, Olsen, Beland, Ullah,
  Zhao, Bei, Aidhy, Samolyuk, Wang, Caro, Caro, Stocks, Larson, Robertson,
  Correa, and Weber]{Zhang2016}
Zhang,~Y. \latin{et~al.}  Influence of chemical disorder on energy dissipation
  and defect evolution in advanced alloys. \emph{J. Mat. Res.} \textbf{2016},
  \emph{31}, 2363--2375\relax
\mciteBstWouldAddEndPuncttrue
\mciteSetBstMidEndSepPunct{\mcitedefaultmidpunct}
{\mcitedefaultendpunct}{\mcitedefaultseppunct}\relax
\EndOfBibitem
\bibitem[Haussalo \latin{et~al.}(1996)Haussalo, Nordlund, and
  Keinonen]{Haussalo1996}
Haussalo,~P.; Nordlund,~K.; Keinonen,~J. Stopping of 5{\textendash}100 {keV}
  helium in tantalum, niobium, tungsten, and {AISI} 316L steel. \emph{Nucl.
  Instrum. Meth. B} \textbf{1996}, \emph{111}, 1--6\relax
\mciteBstWouldAddEndPuncttrue
\mciteSetBstMidEndSepPunct{\mcitedefaultmidpunct}
{\mcitedefaultendpunct}{\mcitedefaultseppunct}\relax
\EndOfBibitem
\bibitem[Caporaso \latin{et~al.}(2009)Caporaso, Chen, and
  Sampayan]{Caporaso2009}
Caporaso,~G.~J.; Chen,~Y.-J.; Sampayan,~S.~E. The Dielectric Wall Accelerator.
  \emph{Rev. Accel. Sci. Tech.} \textbf{2009}, \emph{02}, 253--263\relax
\mciteBstWouldAddEndPuncttrue
\mciteSetBstMidEndSepPunct{\mcitedefaultmidpunct}
{\mcitedefaultendpunct}{\mcitedefaultseppunct}\relax
\EndOfBibitem
\bibitem[Thomson(1912)]{Thomson1912}
Thomson,~J.~J. XLII. Ionization by moving electrified particles. \emph{Philos.
  Mag.} \textbf{1912}, \emph{23}, 449--457\relax
\mciteBstWouldAddEndPuncttrue
\mciteSetBstMidEndSepPunct{\mcitedefaultmidpunct}
{\mcitedefaultendpunct}{\mcitedefaultseppunct}\relax
\EndOfBibitem
\bibitem[Darwin(1912)]{Darwin1912}
Darwin,~C. XC. A theory of the absorption and scattering of the alpha rays.
  \emph{Philos. Mag.} \textbf{1912}, \emph{23}, 901--920\relax
\mciteBstWouldAddEndPuncttrue
\mciteSetBstMidEndSepPunct{\mcitedefaultmidpunct}
{\mcitedefaultendpunct}{\mcitedefaultseppunct}\relax
\EndOfBibitem
\bibitem[Bohr(1913)]{Bohr1913}
Bohr,~N. I. On the constitution of atoms and molecules. \emph{Philos. Mag.}
  \textbf{1913}, \emph{26}, 1--25\relax
\mciteBstWouldAddEndPuncttrue
\mciteSetBstMidEndSepPunct{\mcitedefaultmidpunct}
{\mcitedefaultendpunct}{\mcitedefaultseppunct}\relax
\EndOfBibitem
\bibitem[Bethe(1930)]{Bethe1930}
Bethe,~H. Zur Theorie des Durchgangs schneller Korpuskularstrahlen durch
  Materie. \emph{Ann. Phys.} \textbf{1930}, \emph{397}, 325--400\relax
\mciteBstWouldAddEndPuncttrue
\mciteSetBstMidEndSepPunct{\mcitedefaultmidpunct}
{\mcitedefaultendpunct}{\mcitedefaultseppunct}\relax
\EndOfBibitem
\bibitem[Fermi and Teller(1947)Fermi, and Teller]{Fermi1947}
Fermi,~E.; Teller,~E. The Capture of Negative Mesotrons in Matter. \emph{Phys.
  Rev.} \textbf{1947}, \emph{72}, 399--408\relax
\mciteBstWouldAddEndPuncttrue
\mciteSetBstMidEndSepPunct{\mcitedefaultmidpunct}
{\mcitedefaultendpunct}{\mcitedefaultseppunct}\relax
\EndOfBibitem
\bibitem[Lindhard and Winther(1964)Lindhard, and Winther]{Lindhard1964}
Lindhard,~J.; Winther,~A. Stopping Power of Electron Gas and Equipartition
  Rule. \emph{Mat. Fys. Medd. Dan. Vid. Selsk.} \textbf{1964}, \emph{34},
  1--24\relax
\mciteBstWouldAddEndPuncttrue
\mciteSetBstMidEndSepPunct{\mcitedefaultmidpunct}
{\mcitedefaultendpunct}{\mcitedefaultseppunct}\relax
\EndOfBibitem
\bibitem[Correa \latin{et~al.}(2012)Correa, Kohanoff, Artacho,
  S\'anchez-Portal, and Caro]{Correa2012}
Correa,~A.~A.; Kohanoff,~J.; Artacho,~E.; S\'anchez-Portal,~D.; Caro,~A.
  Nonadiabatic Forces in Ion-Solid Interactions: The Initial Stages of
  Radiation Damage. \emph{Phys. Rev. Lett.} \textbf{2012}, \emph{108},
  213201\relax
\mciteBstWouldAddEndPuncttrue
\mciteSetBstMidEndSepPunct{\mcitedefaultmidpunct}
{\mcitedefaultendpunct}{\mcitedefaultseppunct}\relax
\EndOfBibitem
\bibitem[Quashie \latin{et~al.}(2016)Quashie, Saha, and Correa]{Quashie2016}
Quashie,~E.~E.; Saha,~B.~C.; Correa,~A.~A. Electronic band structure effects in
  the stopping of protons in copper. \emph{Phys. Rev. B} \textbf{2016},
  \emph{94}\relax
\mciteBstWouldAddEndPuncttrue
\mciteSetBstMidEndSepPunct{\mcitedefaultmidpunct}
{\mcitedefaultendpunct}{\mcitedefaultseppunct}\relax
\EndOfBibitem
\bibitem[Ullah \latin{et~al.}(2018)Ullah, Artacho, and Correa]{Ullah2018}
Ullah,~R.; Artacho,~E.; Correa,~A.~A. Core Electrons in the Electronic Stopping
  of Heavy Ions. \emph{Phys. Rev. Lett.} \textbf{2018}, \emph{121}\relax
\mciteBstWouldAddEndPuncttrue
\mciteSetBstMidEndSepPunct{\mcitedefaultmidpunct}
{\mcitedefaultendpunct}{\mcitedefaultseppunct}\relax
\EndOfBibitem
\bibitem[Correa(2018)]{Correa2018}
Correa,~A.~A. Calculating electronic stopping power in materials from first
  principles. \emph{Comput. Mat. Sci.} \textbf{2018}, \emph{150},
  291--303\relax
\mciteBstWouldAddEndPuncttrue
\mciteSetBstMidEndSepPunct{\mcitedefaultmidpunct}
{\mcitedefaultendpunct}{\mcitedefaultseppunct}\relax
\EndOfBibitem
\bibitem[Pruneda \latin{et~al.}(2007)Pruneda, S\'anchez-Portal, Arnau,
  Juaristi, and Artacho]{Pruneda2007}
Pruneda,~J.~M.; S\'anchez-Portal,~D.; Arnau,~A.; Juaristi,~J.~I.; Artacho,~E.
  Electronic Stopping Power in LiF from First Principles. \emph{Phys. Rev.
  Lett.} \textbf{2007}, \emph{99}, 235501\relax
\mciteBstWouldAddEndPuncttrue
\mciteSetBstMidEndSepPunct{\mcitedefaultmidpunct}
{\mcitedefaultendpunct}{\mcitedefaultseppunct}\relax
\EndOfBibitem
\bibitem[Lin(2016)]{Lin2016}
Lin,~L. Adaptively compressed exchange operator. \emph{J. Chem. Theo. Comput.}
  \textbf{2016}, \emph{12}, 2242--2249\relax
\mciteBstWouldAddEndPuncttrue
\mciteSetBstMidEndSepPunct{\mcitedefaultmidpunct}
{\mcitedefaultendpunct}{\mcitedefaultseppunct}\relax
\EndOfBibitem
\bibitem[Jia and Lin(2019)Jia, and Lin]{Jia2019}
Jia,~W.; Lin,~L. Fast real-time time-dependent hybrid functional calculations
  with the parallel transport gauge and the adaptively compressed exchange
  formulation. \emph{Comput. Phys. Comm.} \textbf{2019}, \emph{240},
  21--29\relax
\mciteBstWouldAddEndPuncttrue
\mciteSetBstMidEndSepPunct{\mcitedefaultmidpunct}
{\mcitedefaultendpunct}{\mcitedefaultseppunct}\relax
\EndOfBibitem
\bibitem[Carnimeo \latin{et~al.}(2019)Carnimeo, Baroni, and
  Giannozzi]{Carnimeo2019}
Carnimeo,~I.; Baroni,~S.; Giannozzi,~P. Fast hybrid density-functional
  computations using plane-wave basis sets. \emph{Electronic Structure}
  \textbf{2019}, \emph{1}, 015009\relax
\mciteBstWouldAddEndPuncttrue
\mciteSetBstMidEndSepPunct{\mcitedefaultmidpunct}
{\mcitedefaultendpunct}{\mcitedefaultseppunct}\relax
\EndOfBibitem
\bibitem[Ehrke \latin{et~al.}(2011)Ehrke, Tobey, Wall, Cavill, F\"orst, Khanna,
  Garl, Stojanovic, Prabhakaran, Boothroyd, Gensch, Mirone, Reutler,
  Revcolevschi, Dhesi, and Cavalleri]{Ehrke2011}
Ehrke,~H. \latin{et~al.}  Photoinduced Melting of Antiferromagnetic Order in
  ${\mathrm{La}}_{0.5}{\mathrm{Sr}}_{1.5}{\mathrm{MnO}}_{4}$ Measured Using
  Ultrafast Resonant Soft X-Ray Diffraction. \emph{Phys. Rev. Lett.}
  \textbf{2011}, \emph{106}, 217401\relax
\mciteBstWouldAddEndPuncttrue
\mciteSetBstMidEndSepPunct{\mcitedefaultmidpunct}
{\mcitedefaultendpunct}{\mcitedefaultseppunct}\relax
\EndOfBibitem
\bibitem[Li \latin{et~al.}(2013)Li, Patz, Mouchliadis, Yan, Lograsso, Perakis,
  and Wang]{Li2013}
Li,~T.; Patz,~A.; Mouchliadis,~L.; Yan,~J.; Lograsso,~T.~A.; Perakis,~I.~E.;
  Wang,~J. {Femtosecond switching of magnetism via strongly correlated
  spin--charge quantum excitations}. \emph{Nat.} \textbf{2013}, \emph{496},
  69--73\relax
\mciteBstWouldAddEndPuncttrue
\mciteSetBstMidEndSepPunct{\mcitedefaultmidpunct}
{\mcitedefaultendpunct}{\mcitedefaultseppunct}\relax
\EndOfBibitem
\bibitem[{Rajpurohit} \latin{et~al.}(2020){Rajpurohit}, {Jooss}, and
  {Bl{\"o}chl}]{Rajpurohit2020}
{Rajpurohit},~S.; {Jooss},~C.; {Bl{\"o}chl},~P.~E. {Evolution of the magnetic
  and polaronic order of $\rm{Pr_{1/2}Ca_{1/2}MnO_3}$ following an ultrashort
  light pulse}. \emph{Phys. Rev. B} \textbf{2020}, \emph{102}, 014302\relax
\mciteBstWouldAddEndPuncttrue
\mciteSetBstMidEndSepPunct{\mcitedefaultmidpunct}
{\mcitedefaultendpunct}{\mcitedefaultseppunct}\relax
\EndOfBibitem
\bibitem[Morrison \latin{et~al.}(2014)Morrison, Chatelain, Tiwari, Hendaoui,
  Bruh{\'a}cs, Chaker, and Siwick]{Morrison2014}
Morrison,~V.~R.; Chatelain,~R.~P.; Tiwari,~K.~L.; Hendaoui,~A.;
  Bruh{\'a}cs,~A.; Chaker,~M.; Siwick,~B.~J. A photoinduced metal-like phase of
  monoclinic VO2 revealed by ultrafast electron diffraction. \emph{Sci.}
  \textbf{2014}, \emph{346}, 445--448\relax
\mciteBstWouldAddEndPuncttrue
\mciteSetBstMidEndSepPunct{\mcitedefaultmidpunct}
{\mcitedefaultendpunct}{\mcitedefaultseppunct}\relax
\EndOfBibitem
\bibitem[Otto \latin{et~al.}(2019)Otto, Ren{\'e}~de Cotret, Valverde-Chavez,
  Tiwari, {\'E}mond, Chaker, Cooke, and Siwick]{Otto2019}
Otto,~M.~R.; Ren{\'e}~de Cotret,~L.~P.; Valverde-Chavez,~D.~A.; Tiwari,~K.~L.;
  {\'E}mond,~N.; Chaker,~M.; Cooke,~D.~G.; Siwick,~B.~J. How optical excitation
  controls the structure and properties of vanadium dioxide. \emph{Proc. Natl.
  Acad. Sci. U.S.A.} \textbf{2019}, \emph{116}, 450--455\relax
\mciteBstWouldAddEndPuncttrue
\mciteSetBstMidEndSepPunct{\mcitedefaultmidpunct}
{\mcitedefaultendpunct}{\mcitedefaultseppunct}\relax
\EndOfBibitem
\bibitem[Caviglia \latin{et~al.}(2013)Caviglia, F\"orst, Scherwitzl, Khanna,
  Bromberger, Mankowsky, Singla, Chuang, Lee, Krupin, Schlotter, Turner,
  Dakovski, Minitti, Robinson, Scagnoli, Wilkins, Cavill, Gibert, Gariglio,
  Zubko, Triscone, Hill, Dhesi, and Cavalleri]{Caviglia2013}
Caviglia,~A.~D. \latin{et~al.}  Photoinduced melting of magnetic order in the
  correlated electron insulator NdNiO${}_{3}$. \emph{Phys. Rev. B}
  \textbf{2013}, \emph{88}, 220401\relax
\mciteBstWouldAddEndPuncttrue
\mciteSetBstMidEndSepPunct{\mcitedefaultmidpunct}
{\mcitedefaultendpunct}{\mcitedefaultseppunct}\relax
\EndOfBibitem
\bibitem[De~Filippis \latin{et~al.}(2012)De~Filippis, Cataudella, Nowadnick,
  Devereaux, Mishchenko, and Nagaosa]{Filippis2012}
De~Filippis,~G.; Cataudella,~V.; Nowadnick,~E.~A.; Devereaux,~T.~P.;
  Mishchenko,~A.~S.; Nagaosa,~N. Quantum Dynamics of the Hubbard-Holstein Model
  in Equilibrium and Nonequilibrium: Application to Pump-Probe Phenomena.
  \emph{Phys. Rev. Lett.} \textbf{2012}, \emph{109}, 176402\relax
\mciteBstWouldAddEndPuncttrue
\mciteSetBstMidEndSepPunct{\mcitedefaultmidpunct}
{\mcitedefaultendpunct}{\mcitedefaultseppunct}\relax
\EndOfBibitem
\bibitem[Werner and Eckstein(2015)Werner, and Eckstein]{Werner2015}
Werner,~P.; Eckstein,~M. Field-induced polaron formation in the
  Holstein-Hubbard model. \emph{EPL} \textbf{2015}, \emph{109}, 37002\relax
\mciteBstWouldAddEndPuncttrue
\mciteSetBstMidEndSepPunct{\mcitedefaultmidpunct}
{\mcitedefaultendpunct}{\mcitedefaultseppunct}\relax
\EndOfBibitem
\bibitem[K\"ohler \latin{et~al.}(2018)K\"ohler, Rajpurohit, Schumann, Paeckel,
  Biebl, Sotoudeh, Kramer, Bl\"ochl, Kehrein, and Manmana]{Kohler2018}
K\"ohler,~T.; Rajpurohit,~S.; Schumann,~O.; Paeckel,~S.; Biebl,~F. R.~A.;
  Sotoudeh,~M.; Kramer,~S.~C.; Bl\"ochl,~P.~E.; Kehrein,~S.; Manmana,~S.~R.
  Relaxation of photoexcitations in polaron-induced magnetic microstructures.
  \emph{Phys. Rev. B} \textbf{2018}, \emph{97}, 235120\relax
\mciteBstWouldAddEndPuncttrue
\mciteSetBstMidEndSepPunct{\mcitedefaultmidpunct}
{\mcitedefaultendpunct}{\mcitedefaultseppunct}\relax
\EndOfBibitem
\bibitem[Stolpp \latin{et~al.}(2020)Stolpp, Herbrych, Dorfner, Dagotto, and
  Heidrich-Meisner]{Stolpp2020}
Stolpp,~J.; Herbrych,~J.; Dorfner,~F.; Dagotto,~E.; Heidrich-Meisner,~F.
  Charge-density-wave melting in the one-dimensional Holstein model.
  \emph{Phys. Rev. B} \textbf{2020}, \emph{101}, 035134\relax
\mciteBstWouldAddEndPuncttrue
\mciteSetBstMidEndSepPunct{\mcitedefaultmidpunct}
{\mcitedefaultendpunct}{\mcitedefaultseppunct}\relax
\EndOfBibitem
\bibitem[Rajpurohit \latin{et~al.}(2020)Rajpurohit, Tan, Jooss, and
  Blöchl]{Rajpurohit2020a}
Rajpurohit,~S.; Tan,~L.~Z.; Jooss,~C.; Blöchl,~P.~E. Ultrafast spin-nematic
  and ferroelectric phase transitions induced by femtosecond light pulses.
  \emph{Phys. Rev. B} \textbf{2020}, \emph{102}, 174430\relax
\mciteBstWouldAddEndPuncttrue
\mciteSetBstMidEndSepPunct{\mcitedefaultmidpunct}
{\mcitedefaultendpunct}{\mcitedefaultseppunct}\relax
\EndOfBibitem
\bibitem[Stoica \latin{et~al.}(2019)Stoica, Laanait, Dai, Hong, Yuan, Zhang,
  Lei, Mccarter, Yadav, Damodaran, and et~al.]{Stoica2019}
Stoica,~V.~A.; Laanait,~N.; Dai,~C.; Hong,~Z.; Yuan,~Y.; Zhang,~Z.; Lei,~S.;
  Mccarter,~M.~R.; Yadav,~A.; Damodaran,~A.~R.; et~al., Optical creation of a
  supercrystal with three-dimensional nanoscale periodicity. \emph{Nature
  Materials} \textbf{2019}, \emph{18}, 377–383\relax
\mciteBstWouldAddEndPuncttrue
\mciteSetBstMidEndSepPunct{\mcitedefaultmidpunct}
{\mcitedefaultendpunct}{\mcitedefaultseppunct}\relax
\EndOfBibitem
\bibitem[Guzelturk \latin{et~al.}(2020)Guzelturk, Mei, Zhang, Tan, Donahue,
  Singh, Schlom, Martin, and Lindenberg]{Guzelturk2020}
Guzelturk,~B.; Mei,~A.~B.; Zhang,~L.; Tan,~L.~Z.; Donahue,~P.; Singh,~A.~G.;
  Schlom,~D.~G.; Martin,~L.~W.; Lindenberg,~A.~M. Light-{Induced} {Currents} at
  {Domain} {Walls} in {Multiferroic} {BiFeO} $_{\textrm{3}}$. \emph{Nano Lett.}
  \textbf{2020}, \emph{20}, 145--151\relax
\mciteBstWouldAddEndPuncttrue
\mciteSetBstMidEndSepPunct{\mcitedefaultmidpunct}
{\mcitedefaultendpunct}{\mcitedefaultseppunct}\relax
\EndOfBibitem
\bibitem[Trigo \latin{et~al.}(2013)Trigo, Fuchs, Chen, Jiang, Cammarata, Fahy,
  Fritz, Gaffney, Ghimire, Higginbotham, Johnson, Kozina, Larsson, Lemke,
  Lindenberg, Ndabashimiye, Quirin, Sokolowski-Tinten, Uher, Wang, Wark, Zhu,
  and Reis]{Trigo2013}
Trigo,~M. \latin{et~al.}  Fourier-transform inelastic {X}-ray scattering from
  time- and momentum-dependent phonon–phonon correlations. \emph{Nat. Phys.}
  \textbf{2013}, \emph{9}, 790--794\relax
\mciteBstWouldAddEndPuncttrue
\mciteSetBstMidEndSepPunct{\mcitedefaultmidpunct}
{\mcitedefaultendpunct}{\mcitedefaultseppunct}\relax
\EndOfBibitem
\bibitem[Sie \latin{et~al.}(2019)Sie, Nyby, Pemmaraju, Park, Shen, Yang,
  Hoffmann, Ofori-Okai, Li, Reid, and et~al.]{Sie2019}
Sie,~E.~J.; Nyby,~C.~M.; Pemmaraju,~C.~D.; Park,~S.~J.; Shen,~X.; Yang,~J.;
  Hoffmann,~M.~C.; Ofori-Okai,~B.~K.; Li,~R.; Reid,~A.~H.; et~al., An ultrafast
  symmetry switch in a {W}eyl semimetal. \emph{Nature} \textbf{2019},
  \emph{565}, 61–66\relax
\mciteBstWouldAddEndPuncttrue
\mciteSetBstMidEndSepPunct{\mcitedefaultmidpunct}
{\mcitedefaultendpunct}{\mcitedefaultseppunct}\relax
\EndOfBibitem
\bibitem[Dresselhaus(1955)]{Dresselhaus1955}
Dresselhaus,~G. Spin-{Orbit} {Coupling} {Effects} in {Zinc} {Blende}
  {Structures}. \emph{Phys. Rev.} \textbf{1955}, \emph{100}, 580--586\relax
\mciteBstWouldAddEndPuncttrue
\mciteSetBstMidEndSepPunct{\mcitedefaultmidpunct}
{\mcitedefaultendpunct}{\mcitedefaultseppunct}\relax
\EndOfBibitem
\bibitem[Bychkov and Rashba(1984)Bychkov, and Rashba]{Bychkov1984}
Bychkov,~Y.~A.; Rashba,~E.~I. Properties of a {2D} electron gas with lifted
  spectral degeneracy. \emph{Sov. J. Exp. Theo. Phys. Lett.} \textbf{1984},
  \emph{39}, 78\relax
\mciteBstWouldAddEndPuncttrue
\mciteSetBstMidEndSepPunct{\mcitedefaultmidpunct}
{\mcitedefaultendpunct}{\mcitedefaultseppunct}\relax
\EndOfBibitem
\bibitem[Helgaker \latin{et~al.}(2000)Helgaker, Watson, and
  Handy]{Helgaker2000}
Helgaker,~T.; Watson,~M.; Handy,~N.~C. Analytical calculation of nuclear
  magnetic resonance indirect spin–spin coupling constants at the generalized
  gradient approximation and hybrid levels of density-functional theory.
  \emph{J. Chem. Phys.} \textbf{2000}, \emph{113}, 9402--9409\relax
\mciteBstWouldAddEndPuncttrue
\mciteSetBstMidEndSepPunct{\mcitedefaultmidpunct}
{\mcitedefaultendpunct}{\mcitedefaultseppunct}\relax
\EndOfBibitem
\bibitem[Melo \latin{et~al.}(2005)Melo, Ruiz~de Azúa, Peralta, and
  Scuseria]{Melo2005}
Melo,~J.~I.; Ruiz~de Azúa,~M.~C.; Peralta,~J.~E.; Scuseria,~G.~E. Relativistic
  calculation of indirect {NMR} spin-spin couplings using the
  {Douglas}-{Kroll}-{Hess} approximation. \emph{J. Chem. Phys.} \textbf{2005},
  \emph{123}, 204112\relax
\mciteBstWouldAddEndPuncttrue
\mciteSetBstMidEndSepPunct{\mcitedefaultmidpunct}
{\mcitedefaultendpunct}{\mcitedefaultseppunct}\relax
\EndOfBibitem
\bibitem[Gugler \latin{et~al.}(2018)Gugler, Astner, Angerer, Schmiedmayer,
  Majer, and Mohn]{Gugler2018}
Gugler,~J.; Astner,~T.; Angerer,~A.; Schmiedmayer,~J.; Majer,~J.; Mohn,~P. Ab
  initio calculation of the spin lattice relaxation time \(T_1\) for
  nitrogen-vacancy centers in diamond. \emph{Phys. Rev. B} \textbf{2018},
  \emph{98}, 214442\relax
\mciteBstWouldAddEndPuncttrue
\mciteSetBstMidEndSepPunct{\mcitedefaultmidpunct}
{\mcitedefaultendpunct}{\mcitedefaultseppunct}\relax
\EndOfBibitem
\bibitem[Stojchevska \latin{et~al.}(2014)Stojchevska, Vaskivskyi, Mertelj,
  Kusar, Svetin, Brazovskii, and Mihailovic]{Stojchevska177}
Stojchevska,~L.; Vaskivskyi,~I.; Mertelj,~T.; Kusar,~P.; Svetin,~D.;
  Brazovskii,~S.; Mihailovic,~D. Ultrafast Switching to a Stable Hidden Quantum
  State in an Electronic Crystal. \emph{Sci.} \textbf{2014}, \emph{344},
  177--180\relax
\mciteBstWouldAddEndPuncttrue
\mciteSetBstMidEndSepPunct{\mcitedefaultmidpunct}
{\mcitedefaultendpunct}{\mcitedefaultseppunct}\relax
\EndOfBibitem
\bibitem[Lee \latin{et~al.}(2019)Lee, Goh, and Cho]{Lee2019}
Lee,~S.-H.; Goh,~J.~S.; Cho,~D. Origin of the Insulating Phase and First-Order
  Metal-Insulator Transition in $1T\text{\ensuremath{-}}{\mathrm{TaS}}_{2}$.
  \emph{Phys. Rev. Lett.} \textbf{2019}, \emph{122}, 106404\relax
\mciteBstWouldAddEndPuncttrue
\mciteSetBstMidEndSepPunct{\mcitedefaultmidpunct}
{\mcitedefaultendpunct}{\mcitedefaultseppunct}\relax
\EndOfBibitem
\bibitem[Siddiqui \latin{et~al.}(2020)Siddiqui, Durham, Cropp, Ophus,
  Rajpurohit, Zhu, Carlstr\"om, Stavrakas, Mao, Raja, Musumeci, Tan, Minor,
  Filippetto, and Kaindl]{Siddiqui2020}
Siddiqui,~K.~M.; Durham,~D.~B.; Cropp,~F.; Ophus,~C.; Rajpurohit,~S.; Zhu,~Y.;
  Carlstr\"om,~J.~D.; Stavrakas,~C.; Mao,~Z.; Raja,~A.; Musumeci,~P.;
  Tan,~L.~Z.; Minor,~A.~M.; Filippetto,~D.; Kaindl,~R.~A. Ultrafast optical
  melting of trimer superstructure in layered {1T}'-{TaTe2}.
  \emph{arXiv:2009.02891 [cond-mat]} \textbf{2020}, arXiv: 2009.02891\relax
\mciteBstWouldAddEndPuncttrue
\mciteSetBstMidEndSepPunct{\mcitedefaultmidpunct}
{\mcitedefaultendpunct}{\mcitedefaultseppunct}\relax
\EndOfBibitem
\bibitem[Ligges \latin{et~al.}(2018)Ligges, Avigo, Gole\ifmmode~\check{z}\else
  \v{z}\fi{}, Strand, Beyazit, Hanff, Diekmann, Stojchevska, Kall\"ane, Zhou,
  Rossnagel, Eckstein, Werner, and Bovensiepen]{Ligges2018}
Ligges,~M.; Avigo,~I.; Gole\ifmmode~\check{z}\else \v{z}\fi{},~D.; Strand,~H.
  U.~R.; Beyazit,~Y.; Hanff,~K.; Diekmann,~F.; Stojchevska,~L.; Kall\"ane,~M.;
  Zhou,~P.; Rossnagel,~K.; Eckstein,~M.; Werner,~P.; Bovensiepen,~U. Ultrafast
  Doublon Dynamics in Photoexcited $1T$-${\mathrm{TaS}}_{2}$. \emph{Phys. Rev.
  Lett.} \textbf{2018}, \emph{120}, 166401\relax
\mciteBstWouldAddEndPuncttrue
\mciteSetBstMidEndSepPunct{\mcitedefaultmidpunct}
{\mcitedefaultendpunct}{\mcitedefaultseppunct}\relax
\EndOfBibitem
\bibitem[{Storeck} \latin{et~al.}(2019){Storeck}, {Gerrit Horstmann},
  {Diekmann}, {Vogelgesang}, {von Witte}, {Yalunin}, {Rossnagel}, and
  {Ropers}]{Storeck2019}
{Storeck},~G.; {Gerrit Horstmann},~J.; {Diekmann},~T.; {Vogelgesang},~S.; {von
  Witte},~G.; {Yalunin},~S.; {Rossnagel},~K.; {Ropers},~C. {Structural Dynamics
  of incommensurate Charge-Density Waves tracked by Ultrafast Low-Energy
  Electron Diffraction}. \emph{arXiv e-prints} \textbf{2019},
  arXiv:1909.10793\relax
\mciteBstWouldAddEndPuncttrue
\mciteSetBstMidEndSepPunct{\mcitedefaultmidpunct}
{\mcitedefaultendpunct}{\mcitedefaultseppunct}\relax
\EndOfBibitem
\bibitem[Starykh(2015)]{Starykh2015}
Starykh,~O.~A. Unusual ordered phases of highly frustrated magnets: a review.
  \emph{Rep. Prog. Phys.} \textbf{2015}, \emph{78}, 052502\relax
\mciteBstWouldAddEndPuncttrue
\mciteSetBstMidEndSepPunct{\mcitedefaultmidpunct}
{\mcitedefaultendpunct}{\mcitedefaultseppunct}\relax
\EndOfBibitem
\bibitem[Tapavicza \latin{et~al.}(2013)Tapavicza, Bellchambers, Vincent, and
  Furche]{Tapavicza2013}
Tapavicza,~E.; Bellchambers,~G.~D.; Vincent,~J.~C.; Furche,~F. Ab initio
  non-adiabatic molecular dynamics. \emph{Phys. Chem. Chem. Phys.}
  \textbf{2013}, \emph{15}, 18336--18348\relax
\mciteBstWouldAddEndPuncttrue
\mciteSetBstMidEndSepPunct{\mcitedefaultmidpunct}
{\mcitedefaultendpunct}{\mcitedefaultseppunct}\relax
\EndOfBibitem
\bibitem[Curchod and Mart\'{\i}nez(2018)Curchod, and
  Mart\'{\i}nez]{Curchod2018}
Curchod,~B. F.~E.; Mart\'{\i}nez,~T.~J. Ab {Initio} {Nonadiabatic} {Quantum}
  {Molecular} {Dynamics}. \emph{Chem. Rev.} \textbf{2018}, \emph{118},
  3305--3336\relax
\mciteBstWouldAddEndPuncttrue
\mciteSetBstMidEndSepPunct{\mcitedefaultmidpunct}
{\mcitedefaultendpunct}{\mcitedefaultseppunct}\relax
\EndOfBibitem
\bibitem[Tamm \latin{et~al.}(2018)Tamm, Caro, Caro, Samolyuk, Klintenberg, and
  Correa]{Tamm2018}
Tamm,~A.; Caro,~M.; Caro,~A.; Samolyuk,~G.; Klintenberg,~M.; Correa,~A.~A.
  Langevin Dynamics with Spatial Correlations as a Model for Electron-Phonon
  Coupling. \emph{Phys. Rev. Lett.} \textbf{2018}, \emph{120}\relax
\mciteBstWouldAddEndPuncttrue
\mciteSetBstMidEndSepPunct{\mcitedefaultmidpunct}
{\mcitedefaultendpunct}{\mcitedefaultseppunct}\relax
\EndOfBibitem
\bibitem[Belinicher and Sturman(1980)Belinicher, and Sturman]{Belinicher1980}
Belinicher,~V.~I.; Sturman,~B.~I. The photogalvanic effect in media lacking a
  center of symmetry. \emph{Sov. Phys. Uspekhi} \textbf{1980}, \emph{23},
  199--223\relax
\mciteBstWouldAddEndPuncttrue
\mciteSetBstMidEndSepPunct{\mcitedefaultmidpunct}
{\mcitedefaultendpunct}{\mcitedefaultseppunct}\relax
\EndOfBibitem
\bibitem[Sipe and Shkrebtii(2000)Sipe, and Shkrebtii]{Sipe2000}
Sipe,~J.~E.; Shkrebtii,~A.~I. Second-order optical response in semiconductors.
  \emph{Phys. Rev. B} \textbf{2000}, \emph{61}, 5337--5352\relax
\mciteBstWouldAddEndPuncttrue
\mciteSetBstMidEndSepPunct{\mcitedefaultmidpunct}
{\mcitedefaultendpunct}{\mcitedefaultseppunct}\relax
\EndOfBibitem
\bibitem[Tan \latin{et~al.}(2016)Tan, Zheng, Young, Wang, Liu, and
  Rappe]{Tan2016}
Tan,~L.~Z.; Zheng,~F.; Young,~S.~M.; Wang,~F.; Liu,~S.; Rappe,~A.~M. Shift
  current bulk photovoltaic effect in polar materials--hybrid and oxide
  perovskites and beyond. \emph{npj Comput. Mat.} \textbf{2016}, \emph{2},
  16026\relax
\mciteBstWouldAddEndPuncttrue
\mciteSetBstMidEndSepPunct{\mcitedefaultmidpunct}
{\mcitedefaultendpunct}{\mcitedefaultseppunct}\relax
\EndOfBibitem
\bibitem[Fregoso \latin{et~al.}(2018)Fregoso, Muniz, and Sipe]{Fregoso2018}
Fregoso,~B.~M.; Muniz,~R.~A.; Sipe,~J.~E. Jerk {Current}: {A} {Novel} {Bulk}
  {Photovoltaic} {Effect}. \emph{Phys. Rev. Lett.} \textbf{2018}, \emph{121},
  176604\relax
\mciteBstWouldAddEndPuncttrue
\mciteSetBstMidEndSepPunct{\mcitedefaultmidpunct}
{\mcitedefaultendpunct}{\mcitedefaultseppunct}\relax
\EndOfBibitem
\bibitem[Andrade \latin{et~al.}(2018)Andrade, Hamel, and Correa]{Andrade2018}
Andrade,~X.; Hamel,~S.; Correa,~A.~A. Negative differential conductivity in
  liquid aluminum from real-time quantum simulations. \emph{Eur. Phys. J. B}
  \textbf{2018}, \emph{91}, 1--7\relax
\mciteBstWouldAddEndPuncttrue
\mciteSetBstMidEndSepPunct{\mcitedefaultmidpunct}
{\mcitedefaultendpunct}{\mcitedefaultseppunct}\relax
\EndOfBibitem
\bibitem[Parker \latin{et~al.}(2019)Parker, Morimoto, Orenstein, and
  Moore]{Parker2019}
Parker,~D.~E.; Morimoto,~T.; Orenstein,~J.; Moore,~J.~E. Diagrammatic approach
  to nonlinear optical response with application to {Weyl} semimetals.
  \emph{Phys. Rev. B} \textbf{2019}, \emph{99}, 045121\relax
\mciteBstWouldAddEndPuncttrue
\mciteSetBstMidEndSepPunct{\mcitedefaultmidpunct}
{\mcitedefaultendpunct}{\mcitedefaultseppunct}\relax
\EndOfBibitem
\bibitem[{Rajpurohit} \latin{et~al.}(2021){Rajpurohit}, {Das Pemmaraju},
  {Ogitsu}, and {Tan}]{Rajpurohit2021}
{Rajpurohit},~S.; {Das Pemmaraju},~C.; {Ogitsu},~T.; {Tan},~L.~Z. {A
  non-perturbative study of bulk photovoltaic effect enhanced by an optically
  induced phase transition}. \emph{arXiv e-prints} \textbf{2021},
  arXiv:2105.11310\relax
\mciteBstWouldAddEndPuncttrue
\mciteSetBstMidEndSepPunct{\mcitedefaultmidpunct}
{\mcitedefaultendpunct}{\mcitedefaultseppunct}\relax
\EndOfBibitem
\end{mcitethebibliography}

\end{document}
